\documentclass[twocolumn,twocolappendix]{aastex631}
\usepackage{newtxtext,newtxmath}

\usepackage{threeparttable}

\renewcommand{\vec}[1]{\pmb{#1}}

\begin{document}

\title{On the Physical Origins of Long Period Radio Transients}

\author[0000-0003-4721-4869]{Yuanhong Qu}\thanks{E-mail: yuanhong.qu@unlv.edu}
\affiliation{Nevada Center for Astrophysics, University of Nevada, Las Vegas, NV 89154}
\affiliation{Department of Physics and Astronomy, University of Nevada Las Vegas, Las Vegas, NV 89154, USA}
\affiliation{Department of Physics, University of Helsinki, P.O. Box 64, University of Helsinki, FI-00014, Finland}

\author[0000-0002-9725-2524]{Bing Zhang}\thanks{E-mail: bzhang1@hku.hk}
\affiliation{The Hong Kong Institute for Astronomy and Astrophysics, The University of Hong Kong, Pokfulam Road, Hong Kong, China}
\affiliation{Department of Physics, Department of Physics, The University of Hong Kong, Pokfulam Road, Hong Kong, China}

\begin{abstract}
Long-period radio transients (LPRTs) are a rapidly growing class of coherent radio sources with periods ranging from minutes to hours, whose central engines and emission mechanisms remain unclear. 
Motivated by the detection of red dwarf (RD) companions in several LPRTs and by generic period constraints from the Roche limit and the mass transfer limit, we argue that LPRTs naturally separate into two broad classes: shorter-period sources that are likely isolated compact objects and longer-period sources that are compact objects in binary systems that are likely detached.
For isolated objects, we find that isolated white dwarfs (WDs) generally have difficulty sustaining pair production and coherent radio emission unless the surface temperature is extremely high, while slow rotating neutron stars (NSs) can remain marginally active through inverse-Compton-driven pair cascades. 
For binary systems, asynchronous WD / NS + RD systems can power coherent radio emission through unipolar induction when the WD / NS magnetic field dominates the companion surface field, with relativistic electron cyclotron maser emission as the radiation mechanism, while at larger separations the system enters the magnetospheric interaction regime, possibly powered by magnetic reconnection. 
Bright X-ray counterparts favor magnetar-related systems and undetected X-ray emission is expected from WD-related channels. 
We propose a diagnostic flow chart that uses observational criteria to classify LPRTs and identify their central engines.
These criteria lead to a physically motivated classification framework for LPRTs.
\end{abstract}


\section{Introduction}
Long period radio transients (LPRTs)\footnote{These sources are also known in the literature as ultra-long-period objects (ULPOs) or long-period transients (LPTs).} are a newly recognized class of astrophysical radio sources characterized by long duration periodic radio emissions \citep{Hyman2005,Marsh2016,Caleb2022,Hurley-Walker2022,Hurley-Walker2023,Hurley-Walker2024,Pelisoli2023,Caleb2024,DongFQ2025_421s,LiD2024,Ruiter2025,WangZT2025,Lee2025,Segura2025,DongFQ2025,Bloot2025,Anumarlapudi2025,McSweeney2025,Rose2026}. See the recent review by \cite{Rea2026}.
Their repetition periods typically range from minutes to hours, while the individual bursts last from seconds to minutes, implying very small duty cycles.
These objects are referred to as LPRTs due to their unusually long periods. 
Most LPRTs exhibit bright, coherent radio emission; however, three sources show relatively fainter radio luminosities and are believed to be associated with white dwarf (WD) pulsars in binary systems \citep{Marsh2016, Pelisoli2023, Segura2025}.
The recent discovery of an increasing number of such sources, exhibiting various characteristics, suggests a wide variety of central engines and emission mechanisms.
For completeness, we list known LPRTs and their observational properties to date in Table~\ref{table}.
Several sources show evidence of a possible WD + red dwarf (RD) binary system, and X-ray emission has been detected from four of them.
The nature of the central engine remains uncertain for most sources.

Two classes of theoretical models have been proposed as likely progenitors for LPRTs: isolated magnetic WDs or neutron stars (NSs), and binary systems containing a WD or NS:
\begin{itemize}
\item In the isolated scenario, both WDs and NSs are considered plausible central engines, as they are the only two types of compact objects known to produce coherent radio emission, i.e., WD pulsars and NS pulsars. 
Slowly rotating magnetars have been proposed to remain radio-active through twist-initiated pair cascades in charge-starved gaps \citep{Cooper&Wadiasingh2024}.
Due to the fact that the WD’s moment of inertia is several orders of magnitude greater than that of an NS, the WD pulsar model has been invoked to explain certain active LPRTs, such as GCRT J1745-3009 \citep{ZhangGil2005} and GLEAM-X J1627-52 \citep{Katz2022}. 
If this interpretation is correct, the observed emission imposes stringent constraints on the pair production conditions of WDs (see Section~\ref{subsec:WD pair production} for a discussion).
\item For binary systems, several models have been proposed. 
In the case of the AR Scorpii-like sources: 
one suggests an orthogonal WD rotator whose open field-line beams periodically sweep through the magnetized wind of a companion M dwarf (MD), with a bow shock forming and accelerating electrons within the stellar wind \citep{Geng2016}. 
The observed emission may also be driven by rapid spin synchronization and dissipation of the WD’s rotational energy \citep{Katz2017}.
In both theoretical scenarios, synchrotron radiation plays an important role as an incoherent emission mechanism, contributing to the broadband spectral energy distribution (from radio to X-rays) of WD pulsar systems \citep{Marsh2016,Buckley2017,Pelisoli2023,Segura2025}.
\item For other coherent LPRTs distinct from AR Scorpii-like sources, an asynchronous WD + RD binary has been proposed to produce coherent radio bursts via relativistic electron cyclotron maser emission (ECME) \citep{Qu&Zhang2025,Yang2026,Zhong&Most2026}.
In particular, observational evidence already supports the WD + RD binary scenario: 
two LPRTs -- GLEAM-X J 0704-37 and ILT J1101+5521 -- have been confirmed as binary systems containing a WD with an M dwarf companion by optical spectroscopy \citep{Hurley-Walker2024,Ruiter2025,Rodriguez2025}.

\end{itemize}

\begin{figure}
	\includegraphics[width=\columnwidth]{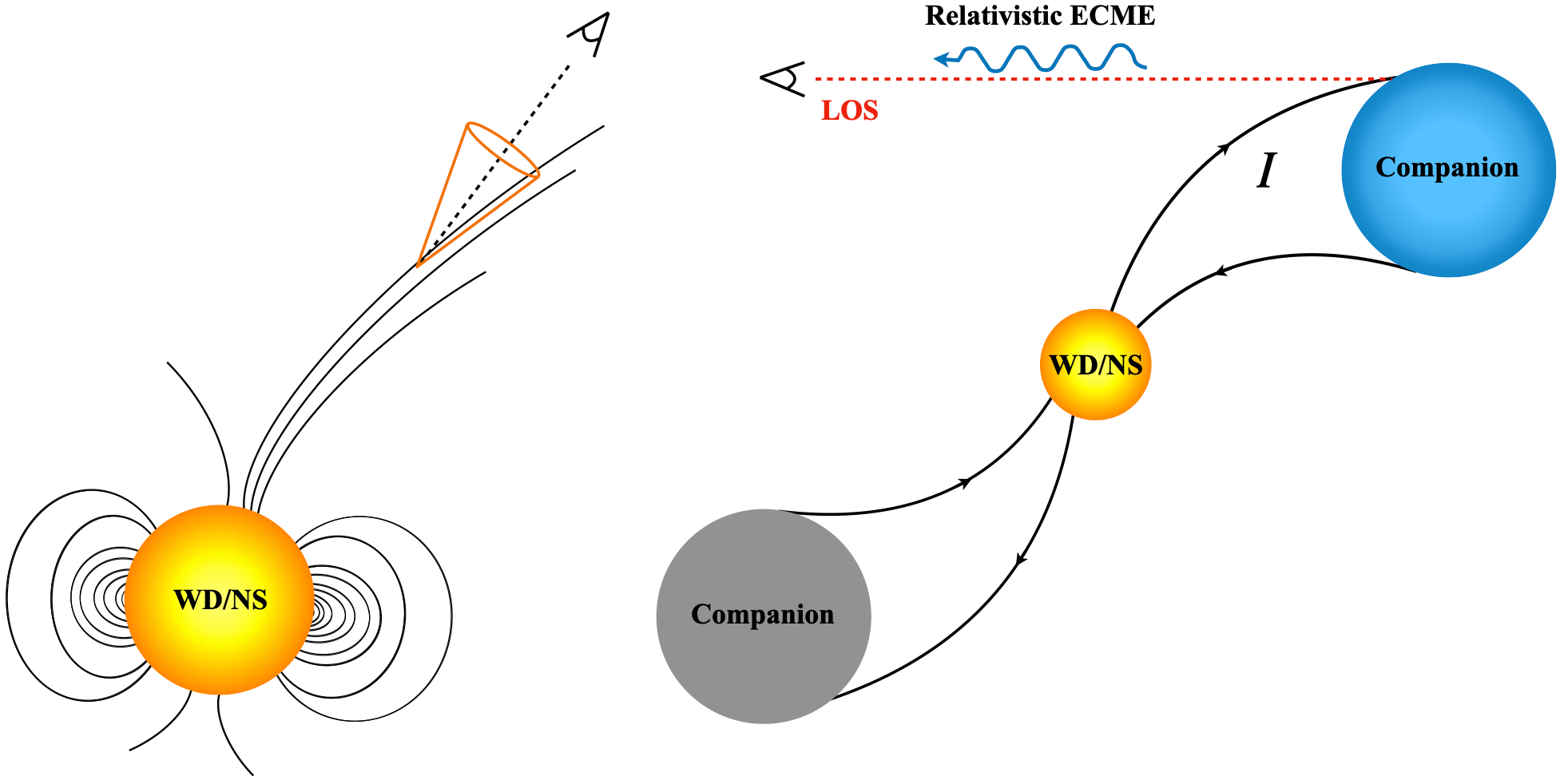}
    \caption{The geometric sketch of the isolated WD or NS (left panel) and WD / NS + RD unipolar induction magnetic interaction model (right panel). 
    In the right panel, the blue and black companions denote the configurations with and without observed radio emission, respectively. 
    The radio emission (blue wiggler) is produced via relativistic ECME in the magnetic loop.}
    \label{fig:cartoon}
\end{figure}

In this paper, we propose a set of generic observational constraints and systematically investigate possible central engines -- including isolated compact stars (WDs and NSs) and binary systems (WD + RD and NS + RD) -- as well as potential intrinsic coherent emission mechanisms and propagation effects of radio waves. 
A classification of LPRTs is proposed based on theoretical considerations.
The relevant central engine scenarios discussed in this work are summarized in Figure~\ref{fig:cartoon}.

This paper is organized as follows.
In Section~\ref{sec:observational results}, we summarize the general observational features of LPRTs and propose physical constraints in Section~\ref{sec:physical constraints}.
We investigate the pair production conditions, radio emission mechanisms and corresponding high-energy counterparts within the context of isolated WDs and NSs in Sections~\ref{sec:Isolated magnetic white dwarf} and \ref{sec:Isolated neutron star}, respectively.
In Section~\ref{sec:high_energy}, we study X-ray counterparts of LPRTs.
In Section~\ref{sec:WD / NS-Red dwarf binary system}, we investigate WD / NS + RD systems and confront observations with theoretical predictions and physical modeling.
In Section~\ref{sec:propagation effects}, we discuss three propagation effects (resonant cyclotron absorption, Faraday conversion, and scintillation).
In Section~\ref{sec:application}, we discuss the classification of LPRTs and their possible connections to other types of radio sources, such as cataclysmic variables (CVs) and WD pulsars.
The main conclusions and discussions are summarized in Section~\ref{sec:conclusions}.
Throughout the paper, the convention $Q=10^nQ_n$ in cgs units is adopted.

\begin{table*}[htbp]
\centering\caption{Published LPRTs with distances, periods, linear and circular polarization degrees (in the radio band), possible optical counterparts and high-energy (HE) counterparts. 
Corresponding references are listed below: [1]\cite{Hyman2005}, 
[2]\cite{Kaplan2008}, 
[3]\cite{Rea2022}, 
[4]\cite{Hurley-Walker2022}, [5]\cite{Hurley-Walker2023}, [6]\cite{Men2025}, 
[7]\cite{Horvath2026},
[8]\cite{Caleb2024}, 
[9]\cite{Ruiter2025}, 
[10]\cite{Hurley-Walker2024}, [11]\cite{WangZT2025}, 
[12]\cite{LiD2024}, 
[13]\cite{Marsh2016},  [14]\cite{Buckley2017}, [15]\cite{Stanway2018}, [16]\cite{Pelisoli2023},
[17]\cite{Lee2025},
[18]\cite{Segura2025}, 
[19]\cite{DongFQ2025},
[20]\cite{Bloot2025},
[21]\cite{Anumarlapudi2025},
[22]\cite{Dobie2024},
[23]\cite{McSweeney2025},
[24]\cite{DongFQ2025_421s},
[25]\cite{Rose2026}.
}
\setlength{\tabcolsep}{1pt}
\begin{tabular}{c|cccccccc}
\hline
Source & $D$ (kpc) & $ P \ ({\rm min})$ & $P_{\rm beat} \ ({\rm min})$ & $\Pi_L \ (\%)$ & $\Pi_V \ (\%)$ & {\rm Optical counterparts} & HE counterparts & Reference\\
\hline
GCRT J1745–3009  & $<0.07$  &77 & / & / & / & No detection & / & [1],[2] \\
\hline
GLEAM-X J1627-52 & $1.3\pm0.5$  &18.18 & / & $88\pm 1$ & / & No detection & / & [3],[4] \\
\hline
GPM J1839-10 & $5.7\pm2.9$ &525 & 21  & 10--100 & $\lesssim 10$ & KD or MD? & / & [5],[6],[7] \\
\hline
ASKAPJ1935+2148 & 4.85 &53.8 & / & $>90$ & $>70$ & near IR source ? & / & [8]\\
\hline
ILT J1101 + 5521 & 0.504 &125.5 & /  & $51\pm 6$ & $<1.6$ & WD+RD & / & [9]\\
\hline
GLEAM-X J 0704-37 & $1.5\pm0.5$  &174 & / &20--50 & 10-30 & RD & / & [10]\\
\hline
ASKAP/DART J1832–0911 & 4.5  & 44 & / & 50-100 &50-100 & NS? & X-ray & [11],[12] \\
\hline
AR Scorpii$^*$ & $0.116\pm0.016$  &213.6 & 1.97 & $\sim 1$ & 30 & WD+RD & {X-ray} & [13],{[14],[15]}\\
\hline
J1912-4410$^*$ &{$0.237\pm 0.005$} &241.8 &5.3 &{4-12} & / &{WD+RD} &{X-ray} &{[16]} \\
\hline
ASKAP J1839-0756 & $4.0\pm1.2$ & 387 & / & 60-90 & 30-60 & No detection & / & [17] \\
\hline
SDSSJ2306 & 1.25 & 209.4 & 1.53 & / & / & WD+RD & / & [18]  \\
\hline
CHIME/ILTJ163430+44501 &1.0-4.3 & $14(70.1 ?)$ & $/$ & 100 & 100 &No detection & / & [19],[20] \\
\hline
ASKAP
J1448-6856 & / & 90 & / &$<9$ -- $82\pm15$ & 26--100 & No detection & X-ray & [21] \\
\hline
ASKAP J1755-2527 &4.7 &69.6 & / & / & / & No detection & / & [22],[23] \\
\hline
CHIME J0630+25 & $0.17^{+0.31}_{-0.1}$ & 7.0 & / & & & No detection & / & [24] \\
\hline
ASKAP J1745-5051$^*$ & 0.4–9.1 & 78 & / & 23-97 & 0-56 & WD + RD & X-ray & [25]\\ 
\hline
\end{tabular}
\parbox{\textwidth}{
\footnotesize
$^{*}$ Three sources marked with an asterisk are discussed separately.
AR Scorpii and J191213.72-441045.1 are included because of their possible similar origin, although their radio brightness temperatures are lower than those of typical LPRTs.
For ASKAP J1745-5051, the brightness temperature reported by \cite{Rose2026} is estimated by assuming an emission region with a radius of one solar radius. Adopting the light-crossing size $\sim c\Delta t$ would lead to an even lower brightness temperature, making it similarly distinct from the majority of known LPRTs.
}
\label{table}
\end{table*}

\section{Observational Features of LPRTs}\label{sec:observational results}

The key observational characteristics of LPRTs may be summarized as follows.

\begin{itemize}

\item The periods of LPRTs range from several minutes \citep{Hurley-Walker2022} to the longest 8.75 hours \citep{Horvath2026}. 
LPRTs exhibit both similarities to and differences from typical long-period rotating pulsars, whose spin periods span from milliseconds to a few minutes \citep{Caleb2022,WangYM2025,DongFQ2025_421s}.
The time duration of each pulse of LPRTs typically ranges from seconds to minutes and the 
LPRT spectrum spans a broad frequency range from $\sim100 \ \rm MHz$ up to a few GHz frequencies.

\item Some LPRTs with a quite long period (roughly longer than one hour) are observed to be in binary systems with a WD and an RD \citep{Marsh2016,Pelisoli2023,Hurley-Walker2024,Ruiter2025,Rodriguez2025}.
Some other LPRTs with a relatively shorter period (roughly shorter than one hour) do not show evidence of a companion, and may belong to an isolated compact star, a WD or an NS \citep{DongFQ2025_421s,WangZT2025,LiD2024}.

\item Both orbital and beat periods have been observed in some LPRTs, which can be naturally attributed to a central engine located in a binary system\footnote{Precession of an isolated WD or NS may also give rise to two characteristic periods: the longer period may be attributed to the precession of the central engine, while the shorter period corresponds to its spin. However, observations already indicated that at least some LPRTs are associated with WD + RD binary systems. So the binary option is more natural.}, and the radiation is modulated by the interaction of the WD / NS and its companion.
Three special LPRTs (or related systems), AR Scorpii, J1912-4410 and SDSS J2306, have been detected to have both orbital and beat periods and they are confirmed to originate from WDs in binary systems \citep{Marsh2016,Pelisoli2023,Segura2025}.
The broad band (from radio to X-ray, except SDSS J2306) radiation is hypothesized to originate from WD pulsars via synchrotron radiation.
Two LPRTs -- CHIME/ILTJ163430+44501 and GPM J1839-10 -- have been detected only through bright coherent radio emission, and they also exhibit both orbital and beat periods \citep{DongFQ2025,Bloot2025,Horvath2026}.

\item We define the ratio of $\Delta\nu$ (full width at half-maximum, hereafter FWHM) to the central frequency $\nu_0$ to describe the narrowness of a radiation spectra. 
The narrow and broad spectra correspond to $\Delta\nu/\nu_0<1$ and $\Delta\nu/\nu_0>1$, respectively.
Some LPRTs (e.g., ASKAP/DART J1832-0911, see Extended Data in \cite{WangZT2025}) have relatively broad spectra with $\Delta\nu/\nu_0>1$.
Interestingly, one burst from GPM J1839-10 shows the down drifting substructure within several tens of millisecond timescale and the spectra of each component are quite narrow with $\Delta\nu/\nu_0\ll1$ \citep{Men2025}.

\item LPRTs often show strong polarization. 
Some LPRTs mainly exhibit high linear polarization \citep{Hurley-Walker2022,Caleb2024}; 
some bursts can also produce significant circular polarization, indicating complex emission geometry or plasma propagation effects.
ASKAP J1935+2148 transitions between a state of highly linearly polarized pulses and another state of highly circularly polarized pulses \citep{Caleb2024}.
For ASKAP/DART J1832-0911, the radio emission polarization properties are also reported to be highly linearly polarized with a total polarization degree $\Pi_p\simeq92\pm3\%$ \citep{WangZT2025,LiD2024}. 
The degree of linear polarization is $\Pi_L\sim 75\%$ and the degree of circular polarization is $\Pi_V\sim 50\%$ \citep{WangZT2025}.
For ASKAP J1839-0756, the emission is highly polarized with a linear polarization degree $\sim90\%$ and a circular polarization degree $\sim37\%$ \citep{Lee2025}.
For CHIME/ILT J1634+44, the emissions show both 100\% linear and circular polarized emission \citep{DongFQ2025,Bloot2025}.

The optical observations of AR Sco show strong linear polarization with $\Pi_L\sim 40\%$ that varies strongly and periodically on both the spin period of the WD and the beat period between the spin and orbital period, as well as low circular polarization degree with a few per cent \citep{Buckley2017}.
The radio emission shows weak linear polarization and a circular polarization with $\Pi_V\sim30\%$ below 10 GHz \citep{Stanway2018}.
J191213.72-441045.1 shows an averaged linear polarization $\Pi_L\sim4\%$ with a maximum reaching $\sim12\%$. No circular polarization was detected \citep{Pelisoli2023}.

\item Some LPRTs have nearly flat polarization angle (PA) evolution \citep{Hurley-Walker2022,LiD2024,WangZT2025}.
PA swings have been observed in some LPRTs \citep{Hurley-Walker2023,Caleb2024,Men2025,Lee2025,Ruiter2025}.
Interestingly, one LPRT (GPM J1839-10) shows $90^\circ$ polarization angle jumps within a timescale of seconds \citep{Hurley-Walker2023,Men2025}.
Faraday conversion resulting in conversion between linear and circular polarization of emission was also observed in GPM J1839-10 \citep{Men2025}.

\item Four LPRTs have high-energy counterparts. The 
X-ray luminosity of AR Scorpii is $L_{\rm X}\simeq4.9\times10^{30} \ \rm erg \ s^{-1}$ \citep{Marsh2016} and that of J1912-4410 is $L_{\rm X}\simeq 1.4\times10^{30} \ \rm erg \ s^{-1}$, roughly a factor of 3 lower than that of AR Scorpii \citep{Schwope2023}.
For ASKAP J1832-0911, the X-ray luminosity is $L_{\rm X}\simeq7.4\times10^{32} \ \rm erg \ s^{-1}$ \citep{WangZT2025}, which is more than two orders of magnitude higher than AR Scorpii and J1912-4410.
For ASKAPJ1448-6856, the X-ray luminosity is $L_{\rm X}\simeq 10^{29} \ \rm erg \ s^{-1}$ \citep{Anumarlapudi2025}, which is nearly four orders of magnitude lower than ASKAP J1832-0911.

\end{itemize}

\section{Physical Constraints}\label{sec:physical constraints}

In this section, we propose several generic constraints on physical models based on the observed temporal, spectral, and polarization properties (Section~\ref{sec:observational results}).

\subsection{Temporal Properties}

\subsubsection{Period and Binary Central Engines}

\begin{figure*}[]
\begin{center}
\setlength{\tabcolsep}{-11pt}
\begin{tabular}{ll}
\resizebox{103mm}{!}{\includegraphics[]{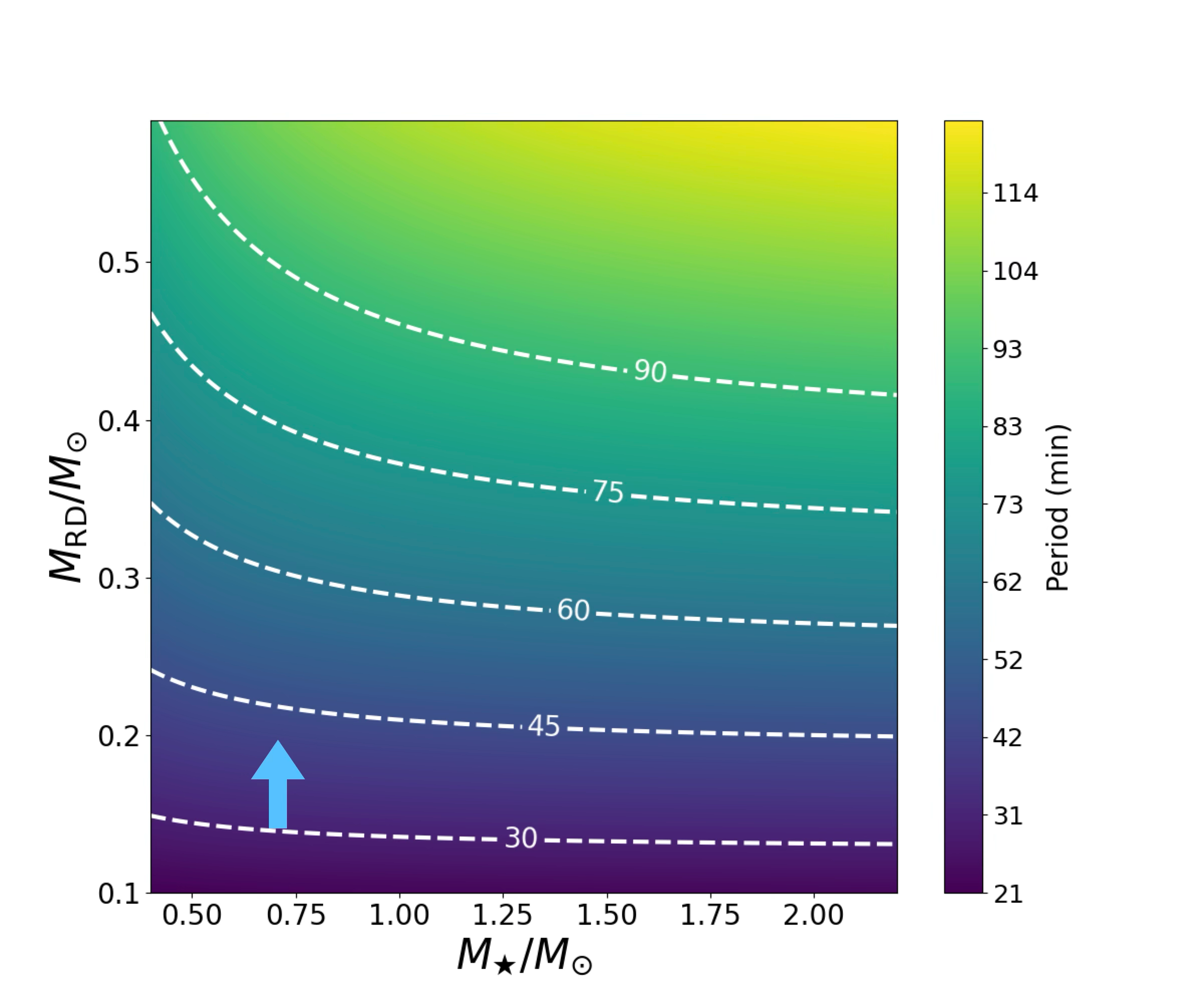}}&
\resizebox{103mm}{!}{\includegraphics[]{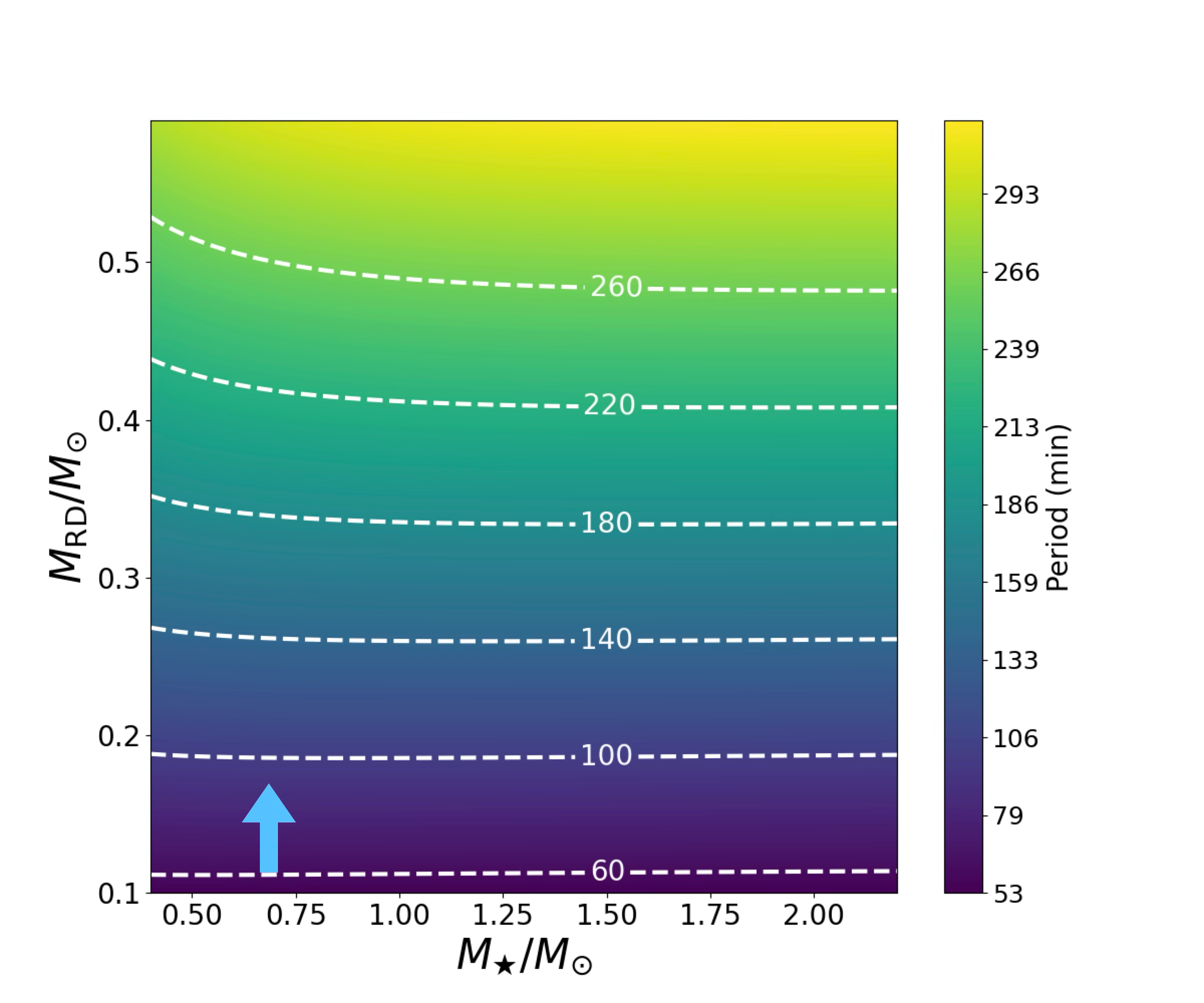}}
\end{tabular}
\caption{The left panel shows the orbital period $P_{\rm Roche}$ for a binary system with an RD companion, as a function of $M_\star/M_\odot$ and $M_{\rm RD}/M_\odot$. 
The right panel shows the orbital period $P_{\rm MT}$ for the same system, as a function of $M_\star/M_\odot$ and $M_{\rm RD}/M_\odot$.
The blue arrows indicate the region of the parameter space where binary systems associated with LPRTs are expected to exist.
}
\label{fig:critical periods}
\end{center}
\end{figure*}

Given that at least some LPRTs are in binary systems, we first investigate what physical constraints can be placed if the observed period is the orbital period of the binary system. 

Physically, two critical periods can be determined if the binary system includes a NS or a WD paired with a RD based on the Roche limit (the condition that the companion is not tidally disrupted) and Roche-lobe overflow (the condition that the mass from the companion is confined to its gravitational potential) conditions. Let us consider a binary system consisting of a WD / NS with mass $M_\star$, radius $R_\star$ and an RD with mass $M_{\rm RD}$ and radius $R_{\rm RD}$.

(i) The Roche limit radius of the WD / NS + RD system is given by
\begin{equation}
\begin{aligned}
R_{\rm Roche}&=R_{\rm RD}\left(\frac{2M_\star}{M_{\rm RD}}\right)^{1/3}\\
&\simeq\left\{
\begin{aligned}
&(2.5\times10^{10} \ {\rm cm}) \ \left(\frac{R_{\rm RD}}{0.2R_\sun}\right)\left(\frac{M_{\rm WD}}{0.6M_\sun}\right)^{1/3}\\
&\times\left(\frac{M_{\rm RD}}{0.2M_\sun}\right)^{-1/3}, \ {\rm WD-RD}, \\
&(3.4\times10^{10} \ {\rm cm})\left(\frac{R_{\rm RD}}{0.2R_\sun}\right)\left(\frac{M_{\rm NS}}{1.4M_\sun}\right)^{1/3}\\
&\times\left(\frac{M_{\rm RD}}{0.2M_\sun}\right)^{-1/3}, \ {\rm NS-RD}.
\end{aligned}
\right.
\end{aligned}
\end{equation}
The RD should be farther away from the WD / NS to be self-bound gravitationally. So a conservative condition is that $R_{\rm Roche}$ should be greater than the $R_{\rm RD}$. This is usually the case, since $M_\star$ is typically greater than $M_{\rm RD}$. The corresponding orbital period defines a critical orbital period due to Roche limit for WD / NS + RD, which can be derived as
\begin{equation}\label{eq:P_Roche}
\begin{aligned}
P_{\rm Roche}&=\frac{2\pi R_{\rm Roche,WD / NS-RD}^{3/2}}{(GM)^{1/2}}\\
&\simeq\left\{
\begin{aligned}
&(41 \ {\rm min}) \ \left(\frac{R_{\rm RD}}{0.2R_\sun}\right)^{3/2}\left(\frac{M_{\rm WD}}{0.6M_\sun}\right)^{1/2}, \\
&\times\left(\frac{M_{\rm RD}}{0.2M_\sun}\right)^{-1/2}\left(\frac{M}{0.8M_{\sun}}\right)^{-1/2}, \ {\rm WD-RD}, \\
&(44 \ {\rm min}) \ \left(\frac{R_{\rm RD}}{0.2R_\sun}\right)^{3/2}\left(\frac{M_{\rm NS}}{1.4M_\sun}\right)^{1/2}\\
&\times\left(\frac{M_{\rm RD}}{0.2M_\sun}\right)^{-1/2}\left(\frac{M}{1.6M_{\sun}}\right)^{-1/2}, \ {\rm NS-RD},
\end{aligned}
\right.
\end{aligned}
\end{equation}
where $M=M_\star+M_{\rm RD}$ denotes the total mass of the binary system and the characteristic values of $M_{\rm WD}=0.6M_\sun$, $M_{\rm NS}=1.4M_\sun$ and $M_{\rm RD}=0.2M_\sun$ are taken for WD / NS and RD, respectively.
Thus, any period of LPRTs shorter than $P_{\rm Roche}$ cannot be a binary WD / NS + RD system, because the RD would break up.

(ii) The second, less stringent constraint on the orbital period is that the matter of the RD is  confined within its Roche lobe. Otherwise, mass flow from the RD onto the WD / NS would suppress coherent radio emission.
The radius of mass transfer (Roche lobe) is given by \citep{Eggleton1983,Sepinsky2007}
\begin{equation}
R_{\rm MT}=r_{\rm L}a(1-e)\simeq\frac{0.49q_r^{2/3}a(1-e)}{0.6q_r^{2/3}+\ln(1+q_r^{1/3})},
\end{equation}
where $r_{\rm L}=R_{\rm MT}/a$ denotes the ratio of the Roche lobe radius to the semimajor axis of the binary, $e$ is the eccentricity and thus $a(1-e)$ is the periastron distance of the binary orbit,
the semimajor axis of the binary system is $a=(GM)^{1/3}\left({P}/{2\pi}\right)^{2/3}$ and $q_r=M_{\rm RD}/M_\star$ is the mass ratio of RD to WD / NS.
For a circular orbit with $e=0$, the condition that matter cannot flow requires $R_{\rm MT}\geq R_{\rm RD}$ and the corresponding orbital period can be calculated as
\begin{equation}\label{eq:P_MT}
\begin{aligned}
P_{\rm MT}&\geq\frac{2\pi}{(GM)^{1/2}}\left(\frac{R_{\rm RD}}{r_{\rm L}}\right)^{3/2}\\
&\simeq\left\{
\begin{aligned}
&(107 \ {\rm min}) \ \left(\frac{M}{0.8M_\sun}\right)^{-1/2}\left(\frac{R_{\rm RD}}{0.2R_\sun}\right)^{3/2}, \ {\rm WD-RD}, \\
&(108 \ {\rm min}) \ \left(\frac{M}{1.6M_\sun}\right)^{-1/2}\left(\frac{R_{\rm RD}}{0.2R_\sun}\right)^{3/2}, \ {\rm NS-RD}.
\end{aligned}
\right.
\end{aligned}
\end{equation}
If the observed LPRT period is interpreted as the orbital period, sources with observed periods shorter than $\sim 110$ minutes are likely not in a WD / NS + RD binary system for the characteristic values of $M_{\rm WD}=0.6M_\sun$, $M_{\rm NS}=1.4M_\sun$ and $M_{\rm RD}=0.2M_\sun$.

We present $P_{\rm Roche}$ and $P_{\rm MT}$ as functions of $M_\star/M_\odot$ and $M_{\rm RD}/M_\odot$ in the left and right panels of Figure~\ref{fig:critical periods}, respectively.
In the left panel, the Roche limit gives a more stringent constraint on period whether a WD / NS + RD binary system can exist.
In the right panel, one can see that the critical orbital period from the Roche lobe radius is nearly constant for a larger mass of the compact objects in the right panel of Figure~\ref{fig:critical periods}.
The orange dashed line, aligned with $\sim 0.2M_{\rm RD}$, denotes $P_{\rm orb}=110 \ \rm min$.
The fitting formula for the lower limit of the orbital period is
\begin{equation}\label{eq:P_lobe}
P_{\rm MT}\gtrsim (550 \ {\rm min}) \ \frac{M_{\rm RD}}{M_\sun}.
\end{equation}
Thus, the observed period of coherent radio emission from LPRTs can put an upper limit on the mass of RD.

We note that the onset of mass transfer can significantly modify the plasma environment in a WD / NS + RD binary.
This is particularly important for coherent radio emission mechanisms which require a relatively low density magnetospheric region for the waves to be generated and to escape.  
If Roche lobe overflow is present, the transferred material is expected to flow along the magnetic field of the WD / NS or to load the interaction region between the two magnetospheres.  
The resulting plasma loading can increase the local plasma frequency and make the conditions for coherent radio emission difficult to satisfy. For elliptical orbits, partial accretion at certain orbital phases is allowed because radio emission can be released when the two stars are far apart (see Section \ref{sec:spin-up-down} for more discussion).

\subsubsection{Pulse duration, central engine size and duty cycles}

\begin{figure}
\hspace*{-3mm}
\includegraphics[width=96mm]{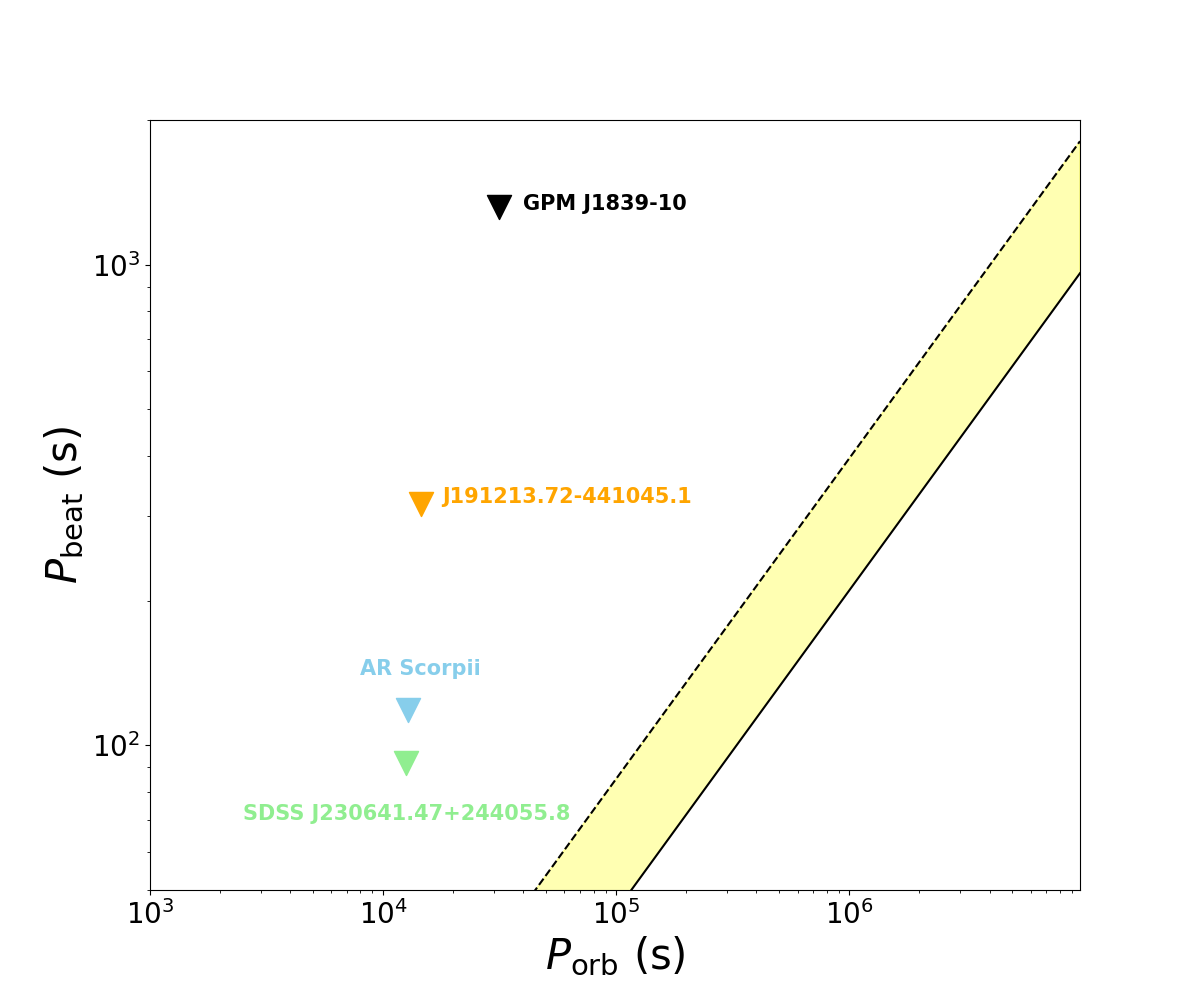}
\caption{Distribution of orbital and spin periods for known LPRTs and AR Scorpii like objects.
For systems with $P_{\rm orb}\gg P_{\rm beat}$, the beat period is approximately equal to the WD / NS spin period.
The shaded yellow region denotes the parameter space in which the light cylinder radius of WD / NS is larger than the binary separation.
The lower boundary (black solid line) corresponds to a low-mass WD + RD system with $M=0.3M_\odot$ and the upper boundary (black dashed line) corresponds to a NS + RD system with $M=2M_\odot$.
}
\label{fig:orbspin}
\end{figure}

The observed LPRTs emit coherent radio pulses that typically last from several tens to hundreds of seconds during each period. 
The intrinsic duration $\Delta t_{\rm LPRT}$ of a single pulse defines a characteristic length scale
\begin{equation}
\begin{aligned}
L_i < c\Delta t_{\rm LPRT}&=(3\times10^{10} \ {\rm cm}) \ \Delta t_{\rm LPRT,2} \\
&\sim 40 \left(\frac{R_{\rm WD}}{0.01R_\sun}\right)\sim 10^4 \left(\frac{R_{\rm NS}}{10^6 \ \rm cm}\right).
\end{aligned}
\end{equation}
The size of the emitter $R_0$ should satisfy $R_0\lesssim L_i$ due to the delay in radio wave propagation between the front end and the rear end of the emission region if the LPRT emitter is moving non-relativistically, as expected.
The light cylinder radius of a WD / NS is given by
\begin{equation}
R_{\rm LC}=\frac{cP_\star}{2\pi}\simeq(2.9\times10^{11} \ {\rm cm}) \ \left(\frac{P_\star}{1 \ \rm min}\right)\sim 10 L_i,
\end{equation}
where $P_\star$ is the spin period of the central engine, normalized to 1 minute.
This suggests that the emission region size is much smaller than that of the WD or NS magnetosphere.

For LPRT sources with both a beat period and an orbital period, we consider that the beat period is approximately equal to the spin period of the WD / NS since the orbital period is much longer than the beat period. For binary interaction models, the light cylinder should be larger than the binary separation ($R_{\rm LC}>a$), i.e.
\begin{equation}
P_{\rm beat}>\frac{2\pi}{c}
\left(GM\right)^{1/3}\left(\frac{P_{\rm orb}}{2\pi}\right)^{2/3},
\label{eq:Pbeat_light_cylinder}
\end{equation}
where $M$ denotes the total mass of the WD / NS + RD binary system, $G$ is the gravitational constant. 
Figure~\ref{fig:orbspin} shows the observed $P_{\rm beat}$ as a function of $P_{\rm orb}$ for sources with both measured periods. 
The shaded yellow band indicates the range of critical beat periods obtained by setting $R_{\rm LC}=a$ for two representative total masses, $M=0.3M_\odot$ and $M=2M_\odot$, corresponding to typical WD + RD and NS + RD systems, respectively.
One can see that LPRTs lie above the yellow region, suggesting that their companions are located within the light cylinder of the WD / NS, thus magnetospheric interaction between the two objects is possible.

\begin{figure*}
\centering
\includegraphics[width=17 cm,height=17 cm]{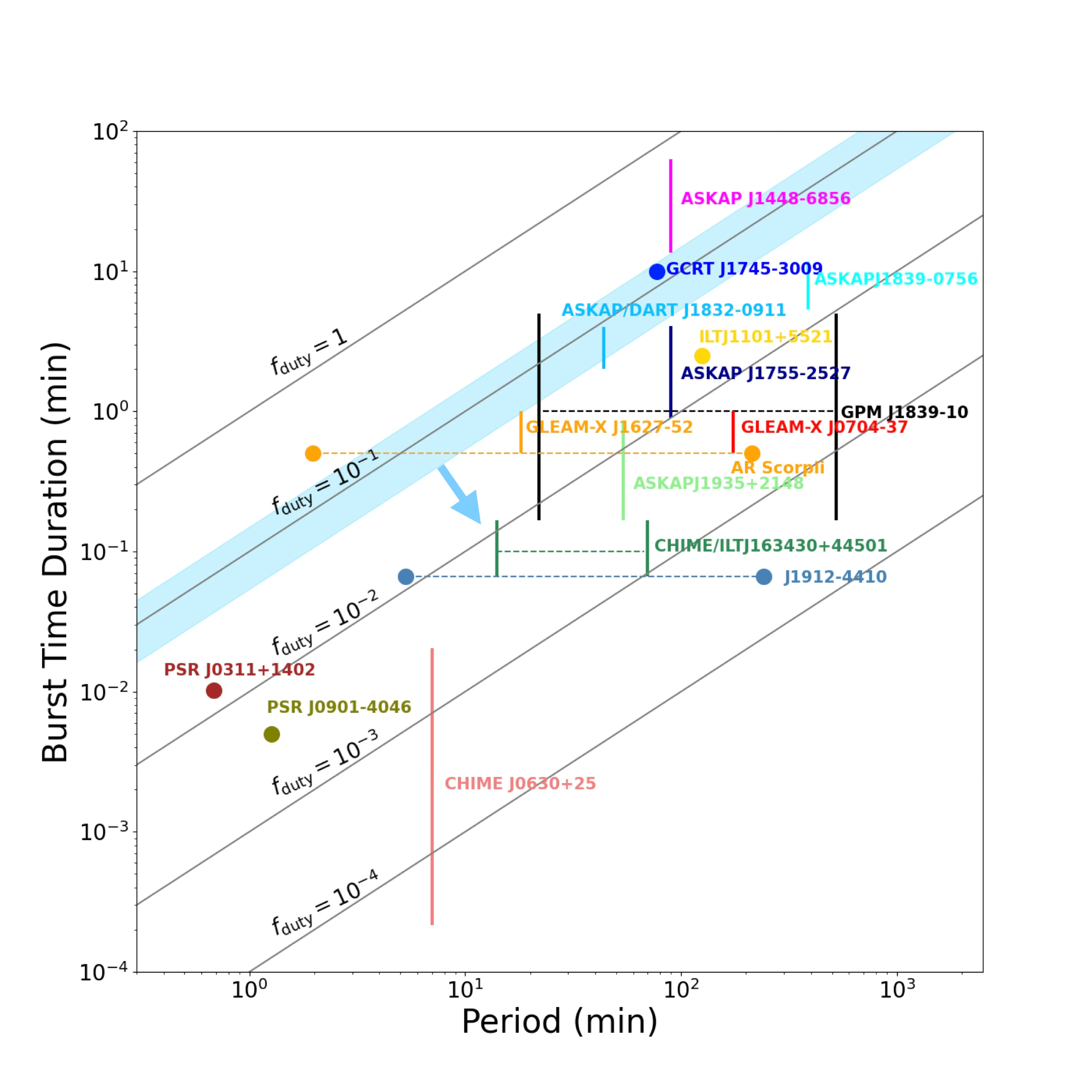}
    \caption{Duty cycle vs the period of LPRTs. When both beat and orbital periods are reported for a source, both periods are shown in the figure and connected by a dashed line to indicate that they correspond to the same source.
    The gray solid lines denote different duty cycles. The blue region denotes the predicted duty cycles for WD / NS + RD systems and $f_d=1$ is adopted. 
    The blue arrow indicates the parameter space below the blue region, where no mass transfer is expected to occur in binary systems.
    }
    \label{fig:duty_cycle}
\end{figure*}

The duty cycle is defined as the ratio of the pulse width to the period of LPRTs, which observationally ranges from $\sim 10^{-4} - 0.7$.
We present the duty cycle vs the period of LPRTs (AR Scorpii like objects are also included) in Figure~\ref{fig:duty_cycle}.
Within the binary scenario, the duration of an LPRT pulse is defined by the duration that takes the WD / NS to move across an effective interaction region of size $\sim 2R_{\rm RD}$.
The corresponding arc length is therefore $\Delta s=a\Delta\varphi$, and the burst duration is determined by the orbital sweeping time $\Delta t_{\rm LPRT}=a\Delta\varphi/v_{\rm orb}=P\Delta\varphi/2\pi$, where $v_{\rm orb}=2\pi a/P_{\rm orb}$. 
The angular width of the active window is $\Delta\varphi=2R_{\rm RD}/a$. 
Thus the corresponding duty cycle is
\begin{equation}\label{eq:f_duty}
f_{\rm duty}=\frac{\Delta t_{\rm LPRT}}{P_{\rm orb}}=\frac{R_{\rm RD}}{\pi a}.
\end{equation}
A necessary condition for the binary system to remain detached, i.e. without sustained mass transfer, is that the Roche lobe radius of the companion satisfies $R_{\rm MT}\geq R_{\rm RD}$. 
We parametrize the orbital separation from contact by introducing a dimensionless distance factor $f_d$ to express $R_{\rm MT}=f_d R_{\rm RD}$, where $f_d=1$ corresponds to marginal Roche lobe filling and $f_d>1$ describes detached systems.
We consider that LPRT binaries satisfy $f_d>1$ and obtain the upper limit on the duty cycle for WD / NS + RD binaries
\begin{equation}
f_{\rm duty}\leq \frac{r_{\rm L}}{\pi f_{d}}.
\end{equation}
Considering the upper limit of the duty cycle with $f_d=1$, the predicted duty cycle $f_{\rm duty}$ depends solely on the mass ratio $q_r$.
For the WD + RD and NS + RD systems, the mass ratio spans $q_r\sim 0.05 - 3$, which corresponds to a duty cycle range of $f_{\rm duty}\sim 5.4\times10^{-2} - 1.5\times10^{-1}$.
We adopt $f_d=1$ to illustrate this range as the blue shaded region in Figure~\ref{fig:duty_cycle}.
This region represents the upper limit of the duty cycle, since in realistic detached binaries $f_d$ can be greater than unity.
It is evident that the observed binary origin LPRTs and AR Scorpii-like objects predominantly lie below this blue region, consistent with the theoretical constraint given by Equation~(\ref{eq:f_duty}).

In terms of individual sources, the systems with confirmed binary counterparts, including AR Scorpii, J1912-4410, and ILT J1101+5521, are all located below the blue shaded region. GLEAM-X J0704-37, which includes the MD companion, and GPM J1839-10, which may have a companion, are also consistent with the predicted upper limit. 
These sources can therefore be naturally accommodated in the detached binary scenario.
GPM J1839-10 is an illustrative case.
The previously identified $\sim 22\ {\rm min}$ period corresponds to the beat period, the orbital period is $P_{\rm orb}\simeq 8.75\ {\rm hr}$ \citep{Horvath2026}.
Using the orbital period moves the source to the right in Figure~\ref{fig:duty_cycle}, placing it further below the predicted upper limit for detached binary systems. 
This strengthens its consistency with the
binary scenario.

\subsubsection{Energy Budget for Isolated WD / NS}

\begin{figure*}[]
\begin{center}
\setlength{\tabcolsep}{-5pt}
\begin{tabular}{ll}
\resizebox{97mm}{!}{\includegraphics[]{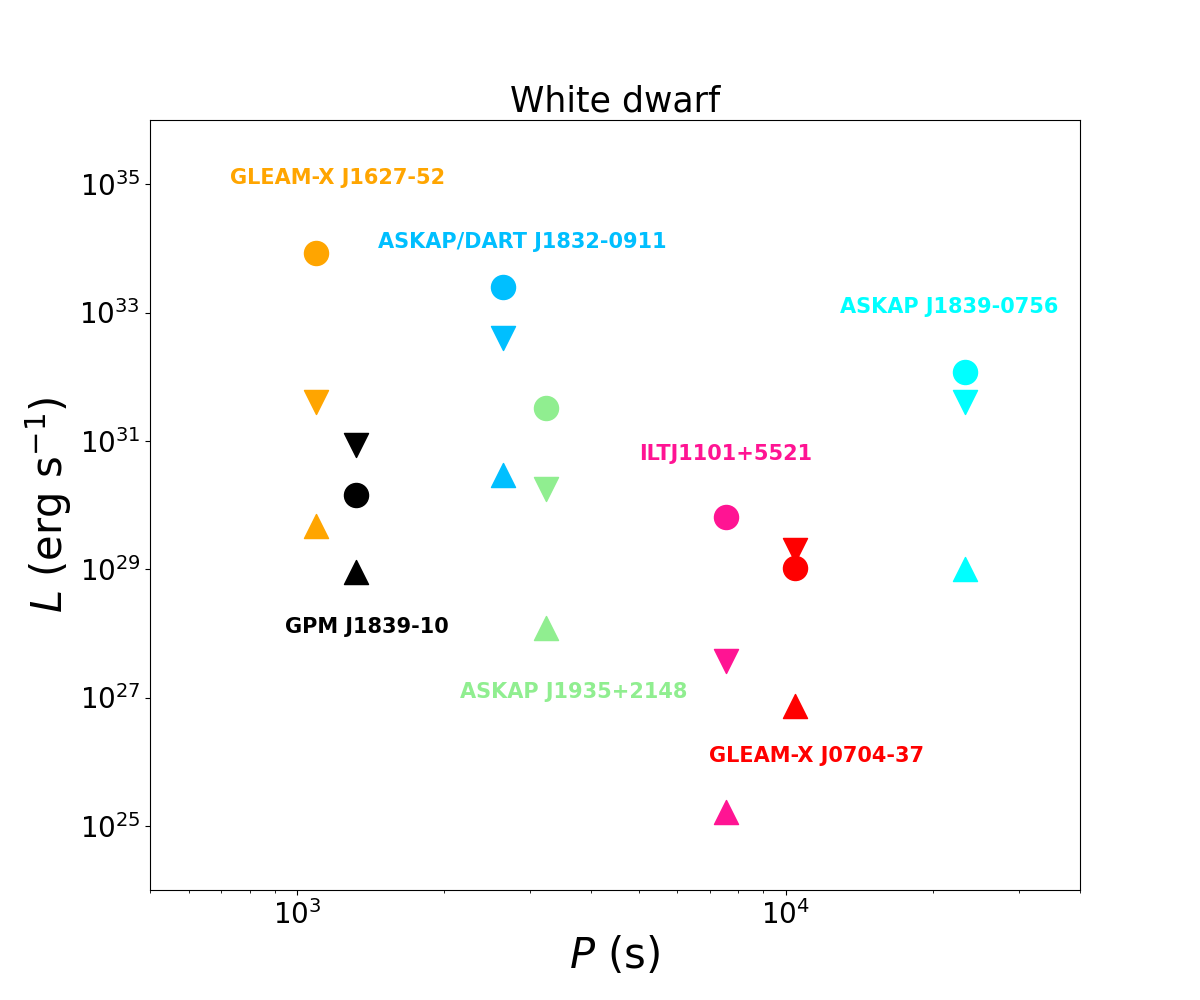}}&
\resizebox{97mm}{!}{\includegraphics[]{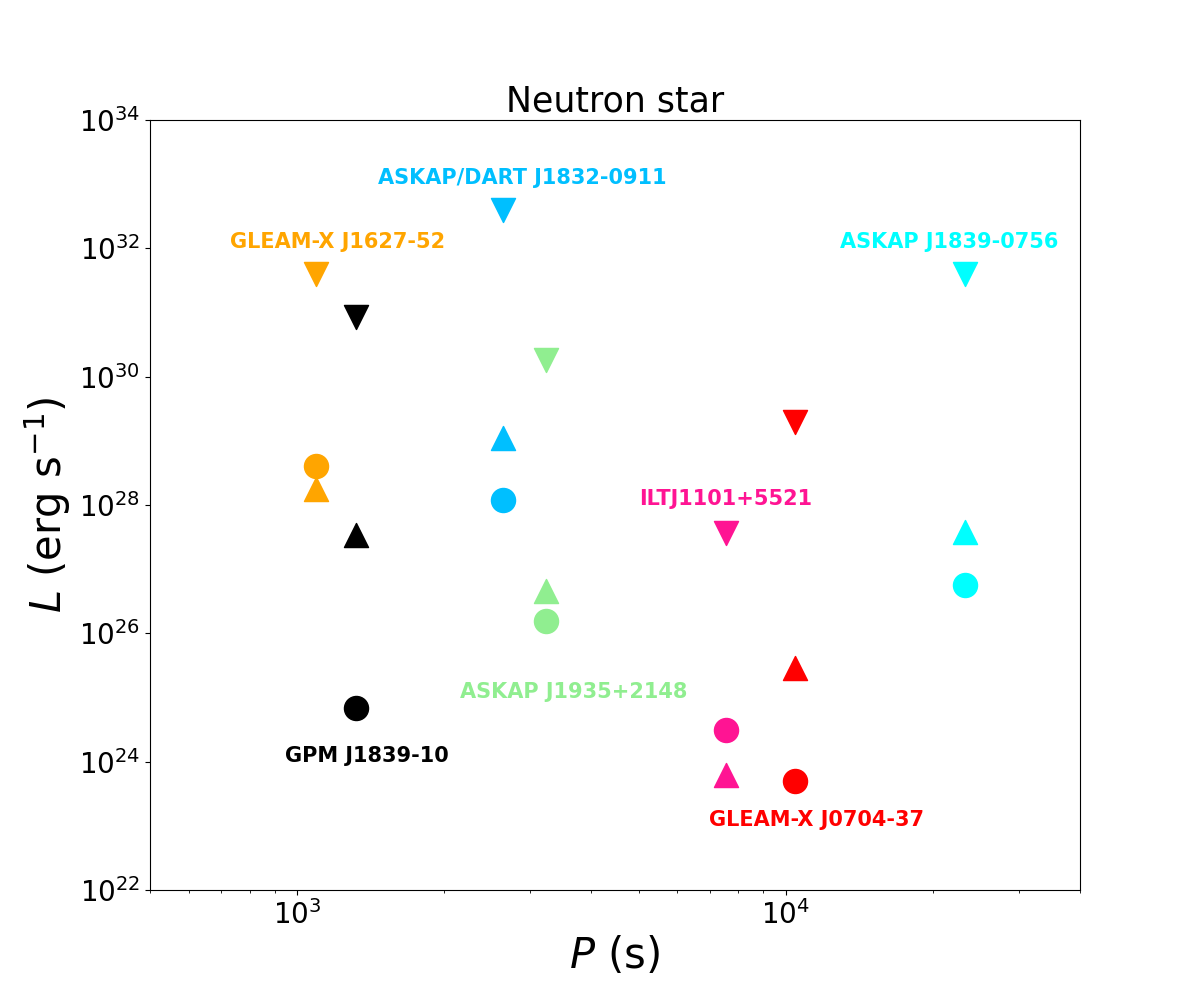}}
\end{tabular}
\caption{Comparison of the spin-down luminosity, beaming-corrected radio luminosity, and observed isotropic luminosity of several LPRT sources as a function of the observed radio emission period $P$.
The left and right panels correspond to the WD and NS cases, respectively. 
For each LPRT source, the dot shows the spin-down luminosity calculated using the observed period $P$ and the upper limit on the period derivative $\dot P$. 
The triangle represents the beaming-corrected luminosity and the inverted triangle represents the observed isotropic luminosity. 
The adopted parameters are $M_{\rm WD}=0.6M_\odot$, $R_{\rm WD}=0.01R_\odot$, $M_{\rm NS}=1.4M_\odot$, and $R_{\rm NS}=10^6\,{\rm cm}$.
}
\label{fig:contour_P_P_dot}
\end{center}
\end{figure*}

The observed isotropic luminosity of LPRTs ranges from $L_{\rm iso}\sim 10^{27}-10^{33} \ \rm erg \ s^{-1}$.
Assuming a typical duration of $\Delta t_{\rm LPRT}\sim 100 \ \rm s$,
the corresponding isotropic energy released during each period can be estimated as $E_{\rm iso}\sim10^{29}-10^{35} \ \rm erg$.
The spin rotation energy of the WD and NS can be calculated as
\begin{equation}
\begin{aligned}
E_{\rm rot,WD / NS}&=\frac{1}{2}I_{\star}\Omega_{\star}^2\\
&\simeq\left\{
\begin{aligned}
&(1.5\times10^{44} \ {\rm erg}) \ 
I_{\rm WD,50}\left(\frac{P_{\rm WD}}{60 \ \rm min}\right)^{-2}, \ {\rm WD}, \\
&(1.5\times10^{39} \ {\rm erg}) \ I_{\rm NS,45}\left(\frac{P_{\rm NS}}{60 \ \rm min}\right)^{-2}, \ {\rm NS},
\end{aligned}
\right.
\end{aligned}
\end{equation}
where $\Omega_{\star}$ denotes the spin angular velocity of the WD or NS and $I_\star$
is the moment of inertia. 
The magnetic field energy of the WD and NS can be calculated as
\begin{equation}
\begin{aligned}
E_{\rm mag, WD / NS}&\simeq \frac{1}{6}B_\star^2R_\star^3\\
&\simeq\left\{
\begin{aligned}
&(5.6\times10^{43} \ {\rm erg}) \ B_{\rm WD,9}^2\left(\frac{R_{\rm WD}}{0.01R_\sun}\right)^3, \ {\rm WD}, \\
&(1.7\times10^{41} \ {\rm erg}) \ B_{\rm NS,12}^2R_{\rm NS,6}^3, \ {\rm NS}.
\end{aligned}
\right.
\end{aligned}
\end{equation}
The crust quake energy of the NS can be estimated as
\begin{equation}\label{eq:E_Q}
E_{Q}\simeq \mu_s\epsilon_{\rm yield}^2\Delta L A\simeq(4\times10^{43} \ {\rm erg}) \ \mu_{30}\epsilon_{\rm yield,-1}^2\Delta L_{4.6}A_{11},
\end{equation}
where $\mu_s=10^{30} \ \rm erg \ cm^{-3}$ is chosen for the shear modulus of the deep crust, $\epsilon_{\rm yield}$ is the yield strain and quakes are expected to occur when $\epsilon_{\rm yield}\sim 0.1$ \citep{Horowitz&Kadau2009}. 
It is useful to estimate the characteristic duration over which a given energy reservoir can sustain the
observed LPRT emission. For a reservoir energy $E_{\rm res}$, the timescale can be estimated as
\begin{equation}
t\sim \frac{\eta E_{\rm res}}{f_b L_{\rm iso}},
\end{equation}
where $\eta<1$ is the efficiency to convert the reservoir energy into radio emission and $f_b=\Delta\Omega/4\pi$ is the beaming factor, so the true radiated luminosity is $f_bL_{\rm iso}$.
For the observed range $L_{\rm iso}\sim10^{27}-10^{33}\ {\rm erg\ s^{-1}}$, the above energy reservoirs correspond to characteristic sustaining timescales
\begin{equation}
t\sim\left\{
\begin{aligned}
&(2\times10^{3} - 5\times10^9 \ {\rm{yr}}) \ \eta f_b^{-1}, \ {\rm WD}, \\
&(5\times10^{-2} - 5\times10^6 \ {\rm yr}) \ \eta f_b^{-1}, \ {\rm NS},
\end{aligned}
\right.
\label{eq:Uph}
\end{equation}
depending on the adopted reservoir and source parameters.
Even the lower end of this range is much longer than the typical burst duration
$\Delta t_{\rm LPRT}\sim100\ {\rm s}$, which implies that the total available energy is not the limiting factor for producing individual LPRT bursts.

Since the central engines of LPRTs are expected to produce relatively steady radio emission, a natural energy channel is rotation spin-down power. 
Constraints can be derived from the spin period and its derivative of the WD / NS, assuming that the radio emission is powered by rotational energy loss. 
The spin-down power for WD / NS is
\begin{equation}
L_{\rm sd}=I_\star\Omega_\star\dot\Omega_\star=\frac{4\pi^2}{P_\star^3}I_\star\dot P_\star.
\end{equation}
The luminosity of radio emission after the solid angle correction can be written as $L_{\Delta\Omega}=f_bL_{\rm iso}$,
here $L_{\rm iso}$ is the observed isotropic luminosity and $\Delta\Omega/4\pi$ is responsible for collimation of the produced radiation into a solid angle $\Delta\Omega$.
The polar cap angle can be calculated as $\theta_{\rm pc,WD / NS}\simeq\sqrt{{R_\star}/{R_{\rm LC,WD / NS}}}$ and the solid angle of the radio emission region can be estimated as $\Delta\Omega\simeq\Delta\theta\Delta\phi\simeq 4\pi \theta_{\rm pc}$.

We compare the spin-down luminosity, observed isotropic luminosity, and beaming-corrected radio luminosity of several LPRT sources as a function of the period $P$ in Figure~\ref{fig:contour_P_P_dot}. 
For some LPRTs showing both an orbital period and a beat period, we use the shorter beat period, since it is associated with the WD / NS spin rather than the orbital motion.
The left and right panels show the results assuming a WD and NS central object, respectively. 
For each LPRT, the dot represents the spin-down luminosity calculated from the period $P$ and the observational upper limit on the period derivative $\dot P$. 
The inverted triangle and triangle represent the observed isotropic luminosity $L_{\rm iso}$ and the beaming-corrected luminosity $L_{\Delta\Omega}$, respectively.
If the spin-down luminosity is above $L_{\rm iso}$, the available spin-down power is sufficient to explain the observed radio emission without requiring beaming corrections. 
If it lies below $L_{\rm iso}$ but above $L_{\Delta\Omega}$, the spin-down power could still account for the emission provided that the beaming solid angle is sufficiently small. 
However, if the spin-down luminosity lies below both luminosity estimates, spin-down power is insufficient.
We note that the WD case generally gives a spin-down luminosity larger than the NS case because a WD has a much larger moment of inertia, and hence a larger reservoir of rotational energy. One can see that the spin-down power of an isolated WD is adequate to power LPRTs. However, the spin-down power of an isolated NS is generally insufficient to power the observed radio emission from very long period LPRTs. A different energy source (e.g. magnetic energy) is needed to power LPRTs if they are isolated NSs.

\subsection{Spectral Properties}

\subsubsection{Spectra Bandwidth}

Observations suggest that some LPRTs have relatively broad spectra with $\Delta\nu/\nu_0 > 1$ \citep{WangZT2025}, while others can exhibit much narrower bandwidths ($\Delta\nu/\nu_0 \ll 1$) \citep{Men2025}.
The continuous broad spectra implies that the intrinsic radiation mechanisms could be related to relativistic plasma emission. 
One pulse in GPM J1839-10 shows an extremely narrow bandwidth, and a down drifting substructure is also observed \citep{Men2025}, which implies that the emitted plasma may be moving from a high magnetic field region to a low field region if the radiation frequency is positively related to the background magnetic field.

To confront spectral data with models, we discuss three possible scenarios and their predicted spectral properties.
\begin{enumerate} 
\item When the central engine is an isolated WD or NS, charged particles are likely to be accelerated and radiate from the high to low magnetic field regions in the open field line zone, producing synchrotron or curvature radiation. 
Consequently, LPRTs are expected to exhibit a continuous and relatively broad spectra.
\item When the central engine is a WD / NS + RD system in which the role of RD only provides the orbital period, and the radiation is still produced near the WD or NS, there are no intrinsic differences from scenario (i).
\item When the central engine is an asynchronous WD / NS + RD system, the motion of the RD relative to the WD / NS magnetosphere can generate an electric potential drop via the unipolar induction effect, accelerating particles along the background magnetic field and producing radiation through the relativistic electron cyclotron maser mechanism \citep{Qu&Zhang2025}. 
In this case, the spectra could be narrower.
\end{enumerate}

\subsection{Polarization Properties}

\subsubsection{$B_\parallel$ in the medium surrounding LPRTs}\label{subsec:RM}

From the measured dispersion measure (DM) and Faraday rotation measure (RM), the component of the background magnetic field along the LOS for most LPRTs may be estimated as 
\begin{equation}
|B_\parallel|=\frac{2\pi m_e^2c^4}{e^3}\frac{|\rm RM|}{\rm DM}\approx 1.23{\rm \mu G} \ \left(\frac{|\rm RM|}{1 \ \rm rad \ m^{-2}}\right)\left(\frac{\rm DM}{1 \ \rm pc \ cm^{-3}}\right).
\end{equation}
For LPRTs with modest RM and DM values, this estimate gives a magnetic field strength comparable to that of the Galactic interstellar medium, suggesting that their surrounding environments are not strongly magnetized or dense. However, this is not necessarily true for all sources.
In particular, the source CHIME J0630+25 reported by \cite{DongFQ2025_421s} shows a relatively large RM, which may indicate a more strongly magnetized local environment around the central engine. 
Therefore, most LPRTs appear to reside in relatively clean environments, while high RM sources represent exceptional systems with significant local magneto-ionic contributions. 
It should also be pointed out that both the magnitude of RM and its temporal variation may become significant in binary systems where the companion contributes a non-negligible stellar wind.

In the following, we discuss the RM contribution from the stellar wind of a massive and low-mass companion. 
Consider the companion star mass $M_c$, radius $R_c$ and surface magnetic field strength $B_c$.
Wind velocity can be approximated by escape velocity $v_w\simeq(2GM_c/R_c)^{1/2}$.
We consider that the distance is the order of the binary separation, i.e. $r\simeq a$.
The electron density in the stellar wind can be estimated as
\begin{equation}
\begin{aligned}
n_w&\simeq \frac{\dot M}{4\pi m_pv_w a^2}\\
&\simeq\left\{
\begin{aligned}
&(1.8\times10^8 \ {\rm{cm^{-3}}}) \ \left(\frac{\dot M}{10^{-8} \ M_\sun \ {\rm yr}^{-1}}\right)\left(\frac{M_c}{50 \ M_\sun}\right)^{-1/2}\\
&\left(\frac{R_c}{10 \ R_\sun}\right)^{1/2}\left(\frac{M}{51.4M_\sun}\right)^{-2/3}\left(\frac{P_{\rm orb}}{24 \ \rm h}\right)^{-4/3}, \ {\rm NS + O-star},\\
&(2.3\times10^5 \ {\rm{cm^{-3}}}) \ \left(\frac{\dot M}{10^{-14} \ M_\sun \ {\rm yr}^{-1}}\right)\left(\frac{M_c}{0.2 \ M_\sun}\right)^{-1/2}\\
&\left(\frac{R_c}{0.6 \ R_\sun}\right)^{1/2}\left(\frac{M}{0.8M_\sun}\right)^{-2/3}\left(\frac{P_{\rm orb}}{100 \ \rm min}\right)^{-4/3}, \ {\rm WD + MD},\\
&(1.4\times10^{5} \ {\rm cm^{-3}}) \ \left(\frac{\dot M}{10^{-14} \ M_\sun \ {\rm yr}^{-1}}\right)\left(\frac{M_c}{0.2 \ M_\sun}\right)^{-1/2}\\
&\left(\frac{R_c}{0.6 \ R_\sun}\right)^{1/2}\left(\frac{M}{1.6M_\sun}\right)^{-2/3}\left(\frac{P_{\rm orb}}{100 \ \rm min}\right)^{-4/3}, \ {\rm NS + MD},
\end{aligned}
\right.
\end{aligned}
\end{equation}
where the mass loss rates $\dot M$ are normalized to $10^{-8} \ M_\sun \ {\rm yr}^{-1}$ and $10^{-14} \ M_\sun \ {\rm yr}^{-1}$ for O-stars and RD, respectively \citep{Puls1996,Muijres2012,Wood2002,Wood2005}.
The background magnetic field of the companion at distance $a$ is $B_{\rm bg}\simeq B_c (a/R_c)^{-3}$ when $a>R_c$.
We adopt $P_{\rm orb}=24 \ \rm {h}$ as a fiducial value for NS + O-star systems, since these systems have larger total masses and companion radii than WD / NS + MD systems. 
The RM contributed by the stellar wind can be estimated as
\begin{equation}
\begin{aligned}
|{\rm RM}|&\simeq \frac{e^3}{2\pi m_e^2c^4} n_w B_{\rm bg}a\\
&\simeq\left\{
\begin{aligned}
&(1.4\times10^7 \ {\rm rad \ m^{-2}}) \ B_c\left(\frac{\dot M}{10^{-8} \ M_\sun \ {\rm yr}^{-1}}\right)\left(\frac{M_c}{50 \ M_\sun}\right)^{-1/2}\\
&\left(\frac{R_c}{0.6 \ R_\sun}\right)^{7/2}\left(\frac{M}{51.4M_\sun}\right)^{-4/3}\left(\frac{P_{\rm orb}}{24 \ \rm h}\right)^{-8/3}, \ {\rm NS + O-star},\\
&(77.4 \ {\rm rad \ m^{-2}}) \ B_c\left(\frac{\dot M}{10^{-14} \ M_\sun \ {\rm yr}^{-1}}\right)\left(\frac{M_c}{0.2 \ M_\sun}\right)^{-1/2}\\
&\left(\frac{R_c}{0.6 \ R_\sun}\right)^{7/2}\left(\frac{M}{0.8M_\sun}\right)^{-4/3}\left(\frac{P_{\rm orb}}{100 \ \rm min}\right)^{-8/3}, \ {\rm WD + MD},\\
&(30.7 \ {\rm rad \ m^{-2}}) \ B_c\left(\frac{\dot M}{10^{-14} \ M_\sun \ {\rm yr}^{-1}}\right)\left(\frac{M_c}{0.2 \ M_\sun}\right)^{-1/2}\\
&\left(\frac{R_c}{0.6 \ R_\sun}\right)^{7/2}\left(\frac{M}{1.6M_\sun}\right)^{-4/3}\left(\frac{P_{\rm orb}}{100 \ \rm min}\right)^{-8/3}, \ {\rm NS + MD}.
\end{aligned}
\right.
\end{aligned}
\end{equation}
We conclude that extremely large RM and strong RM variations in some LPRTs may be associated with the stellar wind of a massive companion star and the associated compact object is a NS.
WD / NS systems with low-mass companions can in principle produce RM variations ranging from several tens up to even ten thousands, primarily determined by $\dot M$, $B_c$ and $P_{\rm orb}$.

\begin{figure}
\hspace*{-3mm}
\includegraphics[width=96mm]{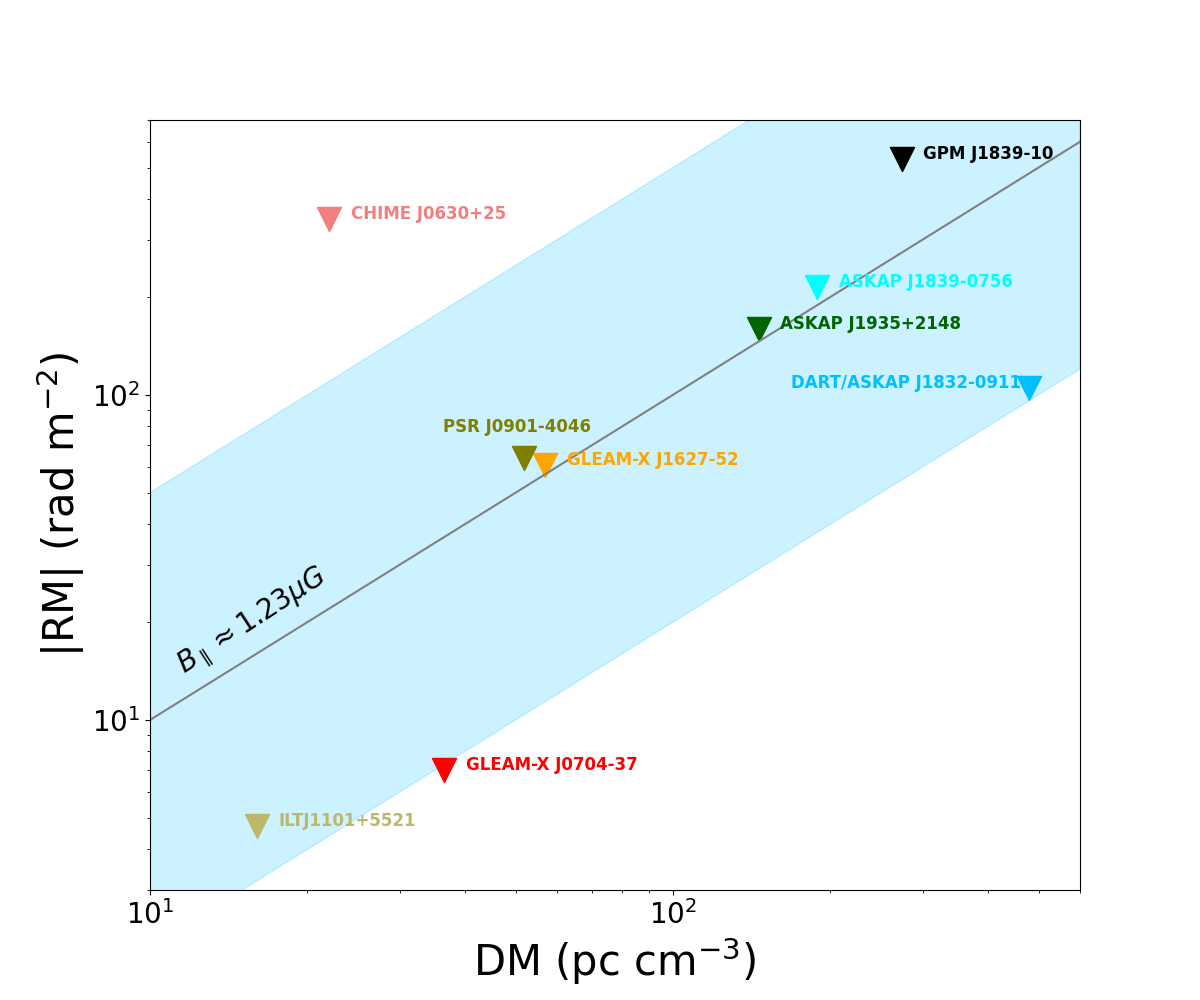}
\caption{Distribution of DM and RM for known LPRTs. The black dashed line denotes $B_\parallel=1.23\mu\rm G$.}
\label{fig:DM_RM}
\end{figure}

We present the distribution of DM and RM for the known LPRTs in Figure~\ref{fig:DM_RM}.
The black solid line corresponds to a magnetic field strength $B_{\parallel}\approx1.23 \ {\mu \rm G}$ along the LOS, when ${\rm RM}=1 \ \rm rad \ m^{-2}$ and ${\rm DM}=1 \ \rm pc \ cm^{-3}$. 
The upper and lower boundaries of the shaded blue region indicate magnetic field strengths that are a factor of five larger and smaller than this reference value, respectively.
One can see that most sources are found to reside in relatively clean environments, and the inferred background magnetic field strengths are generally consistent with the large scale magnetic field of the Milky Way.
One peculiar source, CHIMEJ0630+25, lies above the blue region, indicating a more complex magneto-ionic environment.
This suggests that the source may be associated with a binary system, in which the enhanced RM is likely contributed by the wind of the companion.

\subsubsection{Polarization properties vs. Intrinsic Radiation Mechanisms}

As discussed in Section~\ref{sec:observational results}, detections of linear and circular polarizations in LPRTs which are likely located in a clean environment. 
Thus intrinsic radiation mechanisms are required to generate two eigenmodes with a phase difference, leading to circular polarization. 
In the following, we briefly discuss several possible radiation mechanisms in isolated WD / NS and WD / NS + RD systems:
\begin{itemize}
\item In the emission site with an ordered magnetic field: 
Curvature radiation of a single charged particle can produce a high degree of circular polarization when observed at an off-beam viewing angle \citep{Rybicki&Lightman1979,Jackson1998}. 
In the presence of an ordered magnetic field, a high total polarization degree is expected from coherent curvature radiation by bunches with a mixture of linear and circular polarization components, which have been well studied in the context of radio pulsars and FRBs \citep{Gangadhara2010,Yang&zhang2018,Tong2022,Wang2022,Qu&Zhang2023}.
Both phase differences and varying polarization angles are required to produce circular polarization.
A mixture of linear and circular polarization is also expected.
\item A feature of non-relativistic ECME is its ability to produce nearly 100\% circular polarization when the LOS is aligned with the background magnetic field \citep{Melrose&Dulk1982}.
In the relativistic ECME, the inclusion of relativistic motion along the background magnetic field leads to high circular polarization when the LOS is aligned with the field direction, similar to the non-relativistic case. When the LOS lies along the $1/\gamma$ radiation cone, the emission can be 100\% linearly polarized \citep{Qu&Zhang2025}. In between, both linear and circular polarizations are expected. 
\end{itemize}

\section{Isolated Magnetic White Dwarf}\label{sec:Isolated magnetic white dwarf}

In this section, we investigate the predictions for LPRTs powered by isolated magnetic WDs. The magnetic field strengths of isolated magnetic WDs have been observed to span the range $10^3 – 10^9  \ \rm G$ \citep{Ferrario2015}. 
In the following, we adopt a fiducial spin period of $P_{\rm WD}=60 \ {\rm min}$ and a surface magnetic field strength of $B_{\rm WD}=10^9 \ {\rm G}$.
The electric force acting on a proton is much greater than the gravitational force, which justifies a magnetically dominated magnetosphere similar to that of pulsars.
The light cylinder of the WD can be calculated as $R_{\rm LC,WD}={cP}/{2\pi}\simeq(1.7\times10^{13} \ {\rm cm}) \ (P_{\rm WD}/60 \ {\rm min})$.

\subsection{Pair Production Condition}\label{subsec:WD pair production}

\subsubsection{Non-resonant Inverse Compton Scattering Channel}

The maximum unipolar electric potential drop across the polar cap of a WD can be estimated as
\begin{equation}
\begin{aligned}
\Phi_{\rm WD}&=\frac{2\pi^2B_{\rm WD}R_{\rm WD}^3}{c^2P_{\rm WD}^2}\\
&\simeq(5.7\times10^{8} \ {\rm statV}) \ 
\left(\frac{B_{\rm WD}}{10^9 \ {\rm G}}\right)\left(\frac{R_{\rm WD}}{0.01R_\sun}\right)^3\left(\frac{P_{\rm WD}}{60 \ \rm min}\right)^{-2}.
\end{aligned}
\end{equation}
The maximum Lorentz factor of the electrons accelerated in this potential drop can be calculated as
\begin{equation}\label{eq:gamma_max_WD}
\begin{aligned}
\gamma_{\rm max,WD}&=\frac{e\Phi_{\rm WD}}{m_e c^2}\\
&\simeq3.3\times10^{5} \ 
\left(\frac{B_{\rm WD}}{10^9 \ {\rm G}}\right)\left(\frac{R_{\rm WD}}{0.01R_\sun}\right)^3\left(\frac{P_{\rm WD}}{60 \ \rm min}\right)^{-2}.
\end{aligned}
\end{equation}
A key question is whether isolated magnetic WDs can generate the pair plasma required to sustain the pulsar-like emission processes responsible for LPRTs.
Based on Equation~(\ref{eq:gamma_max_WD}), we take a typical electron Lorentz factor $\gamma=10^5$ below.

Before discussing the non-resonant inverse Compton scattering channel, we first note that curvature radiation is not expected to be an efficient channel for producing pair-seeding photons in an isolated WD magnetosphere. 
For a dipolar WD magnetosphere, the curvature radius near the WD surface is very large $\rho\sim 4/3 \sqrt{R_{\rm WD}R_{\rm LC}}\sim (2\times10^{10} \ {\rm cm}) \ (R_{\rm WD}/0.01R_\sun)^{1/2}(P_{\rm WD}/{1 \ \rm min})^{1/2}$. 
Even if non-dipolar surface fields reduce the local curvature radius to $\rho\sim10^8 \ {\rm cm}$, 
the characteristic energy of curvature photons is
\begin{equation}
E_{\rm cur}=\frac{3}{2}\frac{\hbar c}{\rho}\gamma^3
\approx (300 \ {\rm eV}) \
\gamma_5^3\rho_8^{-1}.
\end{equation}
The resulting curvature photon energy remains below the energy required for efficient magnetic pair production. 
In the following, we focus on non-resonant inverse Compton scattering, which can upscatter thermal photons from the WD surface to much higher energies and is therefore the more favorable channel for pair-seeding photon production.

The surface temperature of a WD is taken as $T_{\rm WD}=(10^4 \ {\rm K}) \ T_{\rm WD,4}$,
thus the typical thermal photon energy can be estimated as $E_{\rm th,WD}\simeq2.8k_BT_{\rm WD}\simeq(3.9\times10^{-12} \ {\rm erg}) \ T_{\rm WD,4}$.
The typical gamma-ray photon via non-resonant ICS energy can be calculated as \citep{ZhangGil2005}
\begin{equation}
\begin{aligned}
E_{\gamma,\rm ICS}^{\rm NR}&={\rm min}(\gamma^2 2.8k_BT_{\rm WD},\gamma m_ec^2)\\
&\simeq{\rm min}(24 \ {\rm GeV} \ \gamma_5^2 T_{\rm WD,4},51 \ {\rm GeV} \ \gamma_5).
\end{aligned}
\end{equation}
The mean free path for an electron to produce one ICS gamma-ray photon can be estimated as
\begin{equation}
l_e^{\rm NR}=\frac{1}{n_{\rm ph,WD}\sigma_{\rm T}}\simeq(7.7\times10^{10} \ {\rm cm}) \ T_{\rm WD,4}^{-3},
\end{equation}
where $n_{\rm ph,WD}=a_{\rm rad}T_{\rm WD}^4/E_{\rm th}\propto T_{\rm WD}^3$ is the number density of thermal photons from the WD surface and $\sigma_{\rm T}$ is the Thomson cross section.
We note that for a typical WD surface temperature $T_{\rm WD}\sim10^4\ {\rm K}$, one has $l_e^{\rm NR}\gg R_{\rm WD}$, indicating that non-resonant ICS channel is inefficient for producing pairs. 
However, since $l_e^{\rm NR}\propto T_{\rm WD}^{-3}$, this mean free path can be significantly reduced if the polar-cap region is heated during a magnetically active episode to achieve $l_e^{\rm NR}\ll R_{\rm WD}$.
When $l_{\gamma B}^{\rm NR}\ll R_{\rm WD}$, the mean free path of the $\gamma$-ray before producing $e^\pm$ in the strong magnetic field via $\gamma-B$ process can be calculated as \citep{Erber1966,Ruderman1975,ZhangGil2005}
\begin{equation}
\begin{aligned}
l_{\gamma B}^{\rm NR}&=\chi \rho \left(\frac{B_c}{B_{\rm WD}}\right)\left(\frac{2m_ec^2}{E_\gamma}\right)={\rm max}(l_{\gamma B}^{\rm T},l_{\gamma B}^{\rm KN})\\
&\simeq {\rm max}(1.2\times10^8 \ {\rm cm} \ \rho_9 B_{\rm WD,9}^{-1}\gamma_5^{-2}T_{\rm WD,4}^{-1},\\
&~~~~~5.9\times10^7 \ {\rm cm} \ \rho_9 B_{\rm WD,9}^{-1}\gamma_5^{-1}),
\end{aligned}
\end{equation}
where $\chi=1/15$ is applied, $B_c=4.4\times10^{13} \ {\rm G}$ is the critical magnetic field.
The requirement $(l_e^{\rm NR}+l_{\gamma B}^{\rm NR})<R_{\rm WD}$ for copious pair production is not met in this long period rotating WD, which lies below the death line.
Thus we conclude that long period magnetic WDs might not be responsible for LPRTs.
However, as proposed by \cite{ZhangGil2005}, 
(i) during magnetically active events, magnetic reconnection would heat up the local magnetosphere to higher temperatures which could reduce $l_e^{\rm NR}$\footnote{Some LPRTs exhibit beat periods on the timescale of minutes, suggesting a possible shorter spin period. If these LPRTs are isolated magnetic WDs, the maximum Lorentz factor of particles can reach $\sim 10^7$ according to Equation~(\ref{eq:gamma_max_WD}). A high electron Lorentz factor can reduce $l_{\gamma B}^{\rm NR}$, but it does not affect $l_e^{\rm NR}$.},
(ii) the stronger surface magnetic field of WDs or more curved magnetic field lines could reduce $l_{\gamma B}^{\rm NR}$.
Under such conditions, the pair production condition $(l_e^{\rm NR}+l_{\gamma B}^{\rm NR})\ll R_{\rm WD}$ could be satisfied, effectively “turning on” the WD. 
However, this mechanism may not account for LPRTs that exhibit stable persistent activities.

To investigate whether isolated magnetic WDs can sustain pair production, we calculate the WD death line based on the non-resonant ICS channel.
We adopt the space charge-limited flow (SPLF) model for WDs and assume that pair formation must be completed within one stellar radius, i.e., the available gap height is taken to be $R_{\rm WD}$. 
For each spin period $P_{\rm WD}$, we determine the corresponding critical surface magnetic field and convert it into the $P-\dot P$ plot using the magnetic dipole spin-down relation $B_{\rm WD}=\left({3c^3I_{\rm WD}P_{\rm WD}\dot P_{\rm WD}}/{2\pi^2R_{\rm WD}^6}\right)^{1/2}$, where $B_{\rm WD}$ denotes the polar surface magnetic field.
The Lorentz factor of the primary electron is estimated from the maximum potential drop in the SCLF model, and is subsequently used to evaluate the characteristic energy of the non-resonant ICS photons and the corresponding magnetic pair production mean free path.

\begin{figure*}
\centering
\hspace*{-0.8cm}
\includegraphics[width=20.5 cm,height=13.5 cm]{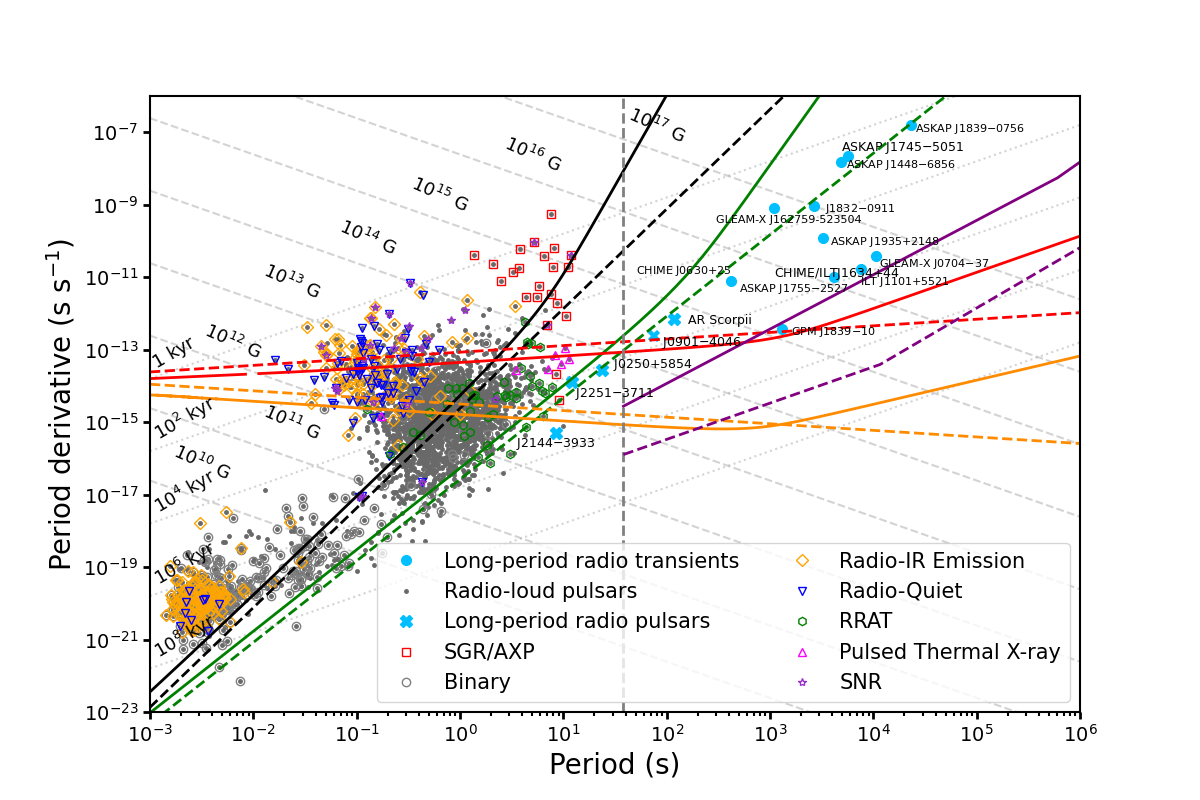}
    \caption{$P-\dot{P}$ diagram for pulsars and LPRTs based on the ATNF pulsar catalogue. 
    The various subclasses of pulsars are represented by the markers in the legend (SGR/AXP: soft gamma-ray repeaters/anomalous X-ray pulsars; RRAT: rotating radio transient; SNR: supernova remnant). 
    Blue circles denote LPRTs (summarized in Table~\ref{table}), blue crosses represent long-period radio pulsars. 
    Light-grey dashed and dotted lines indicate constant surface dipolar magnetic field strength and characteristic age, respectively. 
    The black dashed and solid curves show the analytical death line of the curvature radiation vacuum gap model with a dipolar magnetic field from \cite{Zhang2000} and numerical solution in this paper, respectively. 
    The green dashed and solid curves show the corresponding analytical and numerical death lines for the curvature radiation vacuum gap model with a multipolar magnetic field. 
    The red dashed and solid curves represent the analytical and numerical death lines for the resonant ICS vacuum gap model with a dipolar magnetic field, the orange dashed and solid curves correspond to the analytical and numerical resonant ICS vacuum gap model with a multipolar magnetic field. 
    The vertical grey dashed line marks the critical spin period of a WD. 
    The solid and dashed purple curves show the numerical WD death lines for dipolar and multipolar magnetic field configurations, respectively, assuming a hot WD with $T_{\rm WD}=5\times10^4 \ \rm K$; for cooler WDs (e.g., $T_{\rm WD}=10^4 \ \rm K)$, the WD death line is irrelevant because no pair production is expected \citep{ZhangGil2005}. }
\label{fig:death_line}
\end{figure*}

We adopt the WD surface temperature of $T_{\rm WD}=5\times10^4 \ {\rm K}$.
The two numerical WD death lines for dipolar (solid purple line) and multipolar (dashed purple line) magnetic field configurations are presented in Figure~\ref{fig:death_line}.
The vertical grey dashed line in Figure~\ref{fig:death_line} marks the critical spin period of a WD, below which the centrifugal force at the stellar surface exceeds gravity. 
Equating the gravitational and centrifugal accelerations, we have 
\begin{equation}
P_{\rm crit,WD}=2\pi\left(\frac{R_{\rm WD}^{3}}{{GM_{\rm WD}}}\right)^{1/2}.
\end{equation}
We apply the approximate WD mass-radius relation $R_{\rm WD}\simeq(1.2\times10^9 \ {\rm cm}) \ \left({Z}/{A}\right)^{5/3}\left({M_{\rm WD}}/{M_\odot}\right)^{-1/3}$, where $Z$ and $A$ denote the atomic number and mass number of the ions, respectively, and the critical period is $P_{\rm crit,WD}\simeq38 \ {\rm s}$ for $M_{\rm WD}=0.6M_\odot$ and $Z/A=1$.
This critical period represents the rotational break-up limit of a WD.

\subsubsection{Resonant Inverse Compton Scattering Channel}

In the following, we investigate resonant cyclotron scattering of thermal photons emitted from the WD surface as a possible mechanism for pair production.
We consider an electron moving along the background magnetic field with Lorentz factor $\gamma$, interacting with thermal photons from the WD surface. 
In the comoving frame of the electron, the resonance condition is satisfied when the energy of the incident thermal photon equals the electron cyclotron frequency.
Since the resonant scattering cross section $\sigma_{\rm res}\gg\sigma_{\rm T}$, we have $l_e^{\rm R}\simeq{1}/{n_{\gamma}\sigma_{\rm res}}\ll R_{\rm WD}$, where $n_\gamma$ denotes the number density of X-ray photons.
The critical electron Lorentz factor, below which the boosted gamma-ray has not yet entered the Klein–Nishina regime, can be estimated as
\begin{equation}
\begin{aligned}
\gamma_{\rm KN}&\simeq\frac{m_ec^2}{\hbar\omega_B}\\
&\simeq4.4\times10^4 \ \left(\frac{B_{\rm WD}}{10^9 \ {\rm G}}\right)^{-1}\left(\frac{r}{0.01R_\sun}\right)^3\left(\frac{R_{\rm WD}}{0.01R_\sun}\right)^{-3}.
\end{aligned}
\end{equation}
We take a typical electron Lorentz factor $\gamma=10^4$ below and the resonant scattering photon's frequency can be estimated as
\begin{equation}
\begin{aligned}
E_{\gamma,\rm ICS}^{\rm R}\simeq \gamma\hbar\omega_B&\simeq (1.2\times10^{-3} \ {\rm GeV}) \ \left(\frac{\gamma}{10^4}\right)\left(\frac{B_{\rm WD}}{10^9 \ {\rm G}}\right)\\
&\times\left(\frac{r}{0.01R_\sun}\right)^{-3}\left(\frac{R_{\rm WD}}{0.01R_\sun}\right)^{3}.
\end{aligned}
\end{equation}
Then the mean free path for the photon to attenuate via $\gamma-B$ process can be estimated as 
\begin{equation}
\begin{aligned}
l_{\gamma B}^{\rm R}=\chi \rho \left(\frac{B_c}{B_{\rm WD}}\right)&\left(\frac{2m_ec^2}{E_\gamma}\right)\simeq(2.6\times10^{12} \ {\rm cm}) \ \left(\frac{\rho}{10^9 \ {\rm cm}}\right)\\
&\times\left(\frac{B_{\rm WD}}{10^9 \ \rm G}\right)^{-2} \left(\frac{\gamma}{10^4}\right)^{-1},
\end{aligned}
\end{equation}
which is much greater than $R_{\rm WD}$.
Therefore, we conclude that the resonant ICS channel is less efficient than the non-resonant channel to produce pairs.

\section{Isolated Neutron Star}\label{sec:Isolated neutron star}

The pair formation front and polar-cap pair cascade driven by curvature radiation and ICS have been extensively studied for ordinary radio pulsars (e.g. \citep{Zhang&Harding2000,Zhang2000,Hibschman&Arons2001a,Hibschman&Arons2001b,Medin&Lai2010}), providing the theoretical basis for the calculations presented below.

\subsection{Pair Production Condition}

The maximum unipolar electric potential drop across the polar cap of a NS can be estimated as
\begin{equation}
\begin{aligned}
\Phi_{\rm NS}&=\frac{2\pi^2B_{\rm NS}R_{\rm NS}^3}{c^2P_{\rm NS}^2}\\
&\simeq(1.7\times10^{6} \ {\rm statV}) \ \left(\frac{B_{\rm NS}}{10^{15} \ {\rm G}}\right)\left(\frac{R_{\rm NS}}{10^6 \ {\rm cm}}\right)^3\left(\frac{P_{\rm NS}}{60 \ \rm min}\right)^{-2}.
\end{aligned}
\end{equation}
The maximum Lorentz factor of the electrons accelerated in this potential drop can be calculated as
\begin{equation}\label{eq:gamma_max_NS}
\begin{aligned}
\gamma_{\rm max,NS}&=\frac{e\Phi_{\rm NS}}{m_e c^2}\\
&\simeq9.9\times10^{2} \ \left(\frac{B_{\rm NS}}{10^{15} \ {\rm G}}\right)\left(\frac{R_{\rm NS}}{10^6 \ {\rm cm}}\right)^3\left(\frac{P_{\rm NS}}{60 \ \rm min}\right)^{-2}.
\end{aligned}
\end{equation}
We take the maximum Lorentz factor $\gamma=10^3$ by considering the NS with spin period $\lesssim 60 \ {\rm min}$ (see Equation~(\ref{eq:gamma_max_NS})).
In the NS case, both curvature radiation and inverse Compton scattering are standard channels for producing pairs in polar-cap cascade models \citep{Zhang&Harding2000}.
We therefore examine them separately below.

\subsubsection{Curvature radiation and $\gamma-B$ Pair Production Channel}

For a dipolar magnetic field, the curvature radius of the last open field line near the NS surface can be estimated as
\begin{equation}
\rho\simeq \frac{4}{3}\sqrt{R_{\rm NS}R_{\rm LC}}
\simeq (5.5\times10^9 \ {\rm cm}) \ R_{\rm NS,6}^{1/2}\left(\frac{P_{\rm NS}}{60 \ {\rm min}}\right)^{1/2}.
\end{equation}
For an optimistic estimate, we adopt a smaller curvature radius $\rho=10^9 \ {\rm cm}$, which may mimic locally more curved field lines than a pure dipole. 
The characteristic curvature photon energy is
\begin{equation}
E_{\rm cur}=\frac{3}{2}\frac{\hbar c}{\rho}\gamma^3
\simeq (3.0\times10^{-5} \ {\rm eV}) 
\gamma_3^3\rho_9^{-1}.
\end{equation}
This photon energy is far below the pair production threshold and $l_{\gamma B}^{\rm cur}\gg R_{\rm NS}$, which implies that curvature photons cannot trigger magnetic pair production. 
We also consider an optimistic multipolar field geometry, in which the near surface magnetic field lines are more strongly curved and the curvature radius could be comparable to the stellar radius, i.e. $\rho\sim R_{\rm NS}=10^6 \ {\rm cm}$, we have
\begin{equation}
E_{\rm cur}=\frac{3}{2}\frac{\hbar c}{R_{\rm NS}}\gamma^3\simeq(3.0\times10^{-2} \ {\rm eV}) \ \gamma_3^3 R_{\rm NS,6}^{-1}.
\end{equation}
Although this is larger than the dipolar estimate by several orders of magnitude, it is still far below the pair production threshold.

We numerically calculate the curvature radiation death lines for both dipolar (solid black line) and multipolar magnetic field geometries (solid green line). 
The corresponding two death lines are presented in Figure~\ref{fig:death_line}, together with the analytical estimates of \cite{Zhang2000} (dashed black line for dipolar and dashed green line for multipolar field configuration). 
The numerical solutions are generally consistent with the analytical results for ordinary pulsar periods.
As shown in Figure~\ref{fig:death_line}, even in the optimistic multipolar case, curvature photons cannot trigger magnetic pair production for most known LPRTs if the central engines are isolated magnetars. 
Therefore, we conclude that the curvature radiation channel cannot sustain pair cascade in isolated long-period magnetars.

\subsubsection{Non-resonant ICS and $\gamma-B$ Pair Production Channel}

The surface temperature of a NS is taken as $T_{\rm NS}=(10^6 \ {\rm K}) \ T_{\rm NS,6}$, thus the typical thermal photon energy is $E_{\rm th,NS}\simeq2.82k_BT_{\rm NS}\simeq(3.9\times10^{-10} \ {\rm erg}) \ T_{\rm NS,6}$.
The typical gamma-ray photon via non-resonant ICS energy gives
\begin{equation}
\begin{aligned}
E_{\gamma,\rm ICS}^{\rm NR}&={\rm min}(\gamma^22.8k_BT_{\rm NS},\gamma m_ec^2)\\
&\simeq{\rm min}(0.24 \ {\rm GeV} \ \gamma_3^2 T_{\rm NS,6},0.51 \ {\rm GeV} \ \gamma_3).
\end{aligned}
\end{equation}
The mean free path for an electron to produce one ICS gamma-ray photon can be estimated as
\begin{equation}\label{eq:NS l_e^NR}
l_e^{\rm NR}=\frac{1}{n_{\rm ph,WD}\sigma_{\rm T}}\simeq(7.7\times10^{4} \ {\rm cm}) \ T_{\rm NS,6}^{-3}\ll R_{\rm NS},
\end{equation}
where $n_{\rm ph,NS}=a_{\rm rad}T_{\rm NS}^4/E_{\rm th,NS}$ is the number density of thermal photons from NS surface.
One can see that the mean free path of non-resonant ICS is much shorter in NSs than in WDs, due to the hotter surface temperatures of NSs.
When $l_{\gamma B}^{\rm NR}\ll R_{\rm NS}$, the mean free path for the photon to attenuate via $\gamma-B$ process can be calculated as \citep{Erber1966,Ruderman1975,ZhangGil2005}
\begin{equation}\label{eq:NS l_gammaB^NR}
\begin{aligned}
l_{\gamma B}^{\rm NR}=\chi \rho \left(\frac{B_c}{B_{\rm NS}}\right)\left(\frac{2m_ec^2}{E_\gamma}\right)\simeq &{\rm max}(12.5 \ {\rm cm} \ \rho_6 B_{\rm NS,15}^{-1}\gamma_3^{-2}T_{\rm NS,6}^{-1},\\
&5.9 \ {\rm cm} \ \rho_6 B_{\rm NS,15}^{-1}\gamma_3^{-1}),
\end{aligned}
\end{equation}
where $\rho=10^{6} \ \rm cm$ is applied near the NS surface.
One can see that the requirement $(l_e^{\rm NR}+l_{\gamma B}^{\rm NR})<R_{\rm NS}$ for pair production is satisfied in the case of NSs, due to their hotter surfaces and stronger magnetic fields compared to WDs.

\subsubsection{Resonant Inverse Compton Scattering Channel}

The mean free path of an electron to scatter an X-ray photon $l_{e}^{\rm R}\simeq {1}/{n_{\gamma}\sigma_{\rm res}}$ could be much smaller than the NS radius, where $n_\gamma$ denotes the number density of X-ray photons.
The upscattered photons can produce pairs at a small distance from the scattering site compared to $l_e^{\rm R}$. 
The resonant ICS cross section in the electron comoving frame is given by $\sigma'_{\rm res}\simeq
2\pi^2{e^2\hbar}/{(m_ec)}
\delta(\epsilon'-\epsilon_B)$ \citep{Daugherty1978}, where $\epsilon_B=\hbar \omega_B$ is the gyration energy, $\epsilon$ and $\epsilon'$ denote the incident photon energies in the lab frame and comoving frame, respectively.
Here we adopt $\epsilon'\simeq\gamma\epsilon$and a characteristic thermal photon energy $\epsilon_{\rm res}\simeq2.82k_BT_{\rm NS}=\epsilon_B/\gamma$. 
The corresponding electron mean free path can be estimated as
\begin{equation}
\begin{aligned}
l_e^{\rm R}&\simeq\left[\frac{2\pi^2e^2\hbar}{m_ec}\int_0^\infty\frac{dn_{\rm ph}}{d\epsilon}\delta(\epsilon'-\epsilon_B)d\epsilon\right]^{-1}\\
&=
\left[
\frac{2\pi^2e^2\hbar}{m_ec\gamma}
\left.\frac{dn_{\rm ph}}{d\epsilon}\right|_{\epsilon=\epsilon_{\rm res}}
\right]^{-1}\approx 2\times10^3 \ \rm cm
\end{aligned}
\end{equation}
for $T_{\rm NS}=10^6$ K and $\gamma=10^4$, assuming a blackbody distribution for the thermal photon spectrum.
We find that $l_e^{\rm R}$ is shorter than the non-resonant ICS mean free path $l_e^{\rm NR}$.
Therefore, once the resonance condition is satisfied, resonant ICS proceeds much more rapidly than non-resonant ICS.

We numerically calculate the resonant ICS death lines for both dipolar (solid red line) and multipolar (solid orange line) magnetic field geometries, the numerical death lines are presented in Figure~\ref{fig:death_line}, together with the analytical estimates of \cite{Zhang2000}. 
The numerical calculations are in good agreement with the analytical solutions in the short period range.
As shown in Figure~\ref{fig:death_line}, most ordinary radio pulsars lie above the dipolar resonant ICS death line, consistent with the existence of active pair cascades.

We note that the pair production condition adopted here is less demanding than the screening condition considered by \citet{Harding&Muslimov2002}. 
They distinguished the pair-formation front, defined as the location where the first pairs are produced, from the larger pair multiplicity required for complete screening of the accelerating electric field.
For LPRTs which usually do not require extremely coherent conditions as pulsars, we do not require copious pairs capable of screening the parallel electric field, but only demand that pairs can be produced to generate observed radio emission.

\subsubsection{Photon Splitting and $\gamma-\gamma$ Pair Production Channel}

In the strong background magnetic field, which can be considered as the virtual photons to experience $\gamma+B\rightarrow \gamma+\gamma$. 
Magnetic photon splitting is a third-order QED process describing a single photon that splits into two lower energy photons. 
This process can be understood by considering the interaction between the photon virtual pairs and the background field and a vacuum polarization pair radiates photons in the lowest order.
Charge-parity (CP) invariance allows three polarization channels for photon splitting in QED: $\perp \rightarrow \parallel\parallel$, $\perp \rightarrow \perp\perp$, and $\parallel \rightarrow \perp\parallel$. 
However, as pointed out by \citet{Adler1971}, the requirement of energy and momentum conservation, in the weak linear vacuum dispersion, restricts the process to the $\perp \rightarrow \parallel\parallel$ mode below the pair production threshold. 
The other two CP-allowed channels are forbidden. 
If this is the case, then photons in the other two polarization channels must exist but will not undergo the splitting process. 
These photons can instead produce pairs via the $\gamma$–$B$ or $\gamma$–$\gamma$ processes. 
However, in the presence of strong vacuum dispersion, all three photon-splitting modes permitted by CP invariance in QED become allowed \citep{Baring&Harding1998,Baring&Harding2001}. 
In this case, photon splitting can proceed efficiently, leading to suppression of pair production.
In the following, we examine whether photon splitting can suppress pair production in long period rotating magnetars.

In the low energy limit that photon's energy is much smaller than the rest energy of electron, the approximate absorption coefficient of photon splitting are given by \citep{Baring&Harding2001,HuKun2019,HuKun2022}
\begin{equation}
{\alpha}_{\rm \perp \rightarrow \parallel \parallel}^{\rm sp}=\frac{1}{2}{\alpha}_{\rm \parallel \rightarrow \perp \parallel}^{\rm sp}=\frac{\alpha_f^3}{60\pi^2\lambdabar } \left(\frac{E_\gamma}{m_ec^2}\right)^5\left(\frac{B}{B_c}\right)^6\mathcal{M}_1^2\sin^6\theta_{B},
\end{equation}
and
\begin{equation}
{\alpha}_{\rm \perp \rightarrow \perp \perp}^{\rm sp}=\frac{\alpha_f^3}{60\pi^2\lambdabar } \left(\frac{E_\gamma}{m_ec^2}\right)^5\left(\frac{B}{B_c}\right)^6\mathcal{M}_2^2\sin^6\theta_{B},
\end{equation}
where the superscript ``sp" denotes photon splitting, $\alpha_f=e^2/(\hbar c)\simeq 1/137$ is the fine-structure constant.
The absorption coefficient of unpolarized photons can be written as \citep{Harding1997,HuKun2019,HuKun2022}
\begin{equation}
{\alpha}_{\rm ave}^{\rm sp}=\frac{\alpha_f^3}{120\pi^2\lambdabar}\left(\frac{E_\gamma}{m_ec^2}\right)^5\left(\frac{B}{B_c}\right)^6(3\mathcal{M}_1^2+\mathcal{M}_2^2)\sin^6\theta_{B}.
\end{equation}
The reaction amplitude coefficients $\mathcal{M}_1$ and $\mathcal{M}_2$ are defined as
\begin{equation}
\mathcal{M}_\sigma=\left(\frac{B}{B_c}\right)^{-4}\int_0^\infty \frac{ds}{s}e^{-\frac{B_c}{B}s}\Lambda_\sigma(s), \ \sigma=1,2,
\end{equation}
where $s$ denotes the path length of the photon and $\Lambda_\sigma$ were first identified by \cite{Adler1971} as
\begin{equation}
\Lambda_1(s)=\left(-\frac{3}{4s}+\frac{s}{6}\right)\frac{\cosh s}{\sinh s}+\frac{3+2s^2}{12\sinh^2s}+\frac{s \sinh s}{2 \sinh^3 s},
\end{equation}
and
\begin{equation}
\Lambda_2(s)=\frac{3}{4s}\frac{\cosh s}{\sinh s}+\frac{3-4s^2}{4\sinh^2s}-\frac{3s^2}{2\sinh^4s}.
\end{equation}
We note that $l_e^{\rm NR}\ll R_{\rm NS}$, thus in the limit of $B\gg B_c$, we have $\mathcal{M}_1\simeq B_c^3/(6B^3)$ and $\mathcal{M}_2\simeq B_c^4/(3B^4)$.
One can see that the absorption coefficient is independent on background magnetic field. 
For unpolarized photons, the mean free path of photon splitting can be estimated as \citep{Zhang2001}
\begin{equation}
\begin{aligned}
l_{\rm ave}^{\rm sp}&\simeq ({\alpha}_{\rm ave}^{\rm sp})^{-1}\simeq (3.7\times10^{6} \ {\rm cm}) \ f_{\Theta}^{-6/7}\left(\frac{\gamma}{10^3}\right)^{-10/7}\\
&\times\left(\frac{T_{\rm NS}}{10^6 \ \rm K}\right)^{-5/7}\left(\frac{P_{\rm NS}}{60 \ \rm min}\right)^{3/7}\left(\frac{r_e}{10^6 \ \rm cm}\right)^{3/7}>R_{\rm NS},
\end{aligned}
\end{equation}
where $f_\Theta$ is the ratio of the colatitude of the magnetic field line at the NS surface to the polar cap angle.
One can see that $(l_e^{\rm NR}+l_{\gamma B}^{\rm NR})\ll l_{\rm ave}^{\rm sp}$ (see Equations~(\ref{eq:NS l_e^NR}) and (\ref{eq:NS l_gammaB^NR})), indicating that photon splitting is insufficient to suppress the $\gamma-B$ pair production process.

The two-photon pair production $\gamma+\gamma\rightarrow e^{\pm}$ is well described in the framework of QED. 
The threshold condition of two-photon pair production can be written as $\nu_1 \nu_2 (1-\cos\theta_{\gamma\gamma})\geq 2 \left({m_e c^2}/{h}\right)^2$,
where $\theta_{\gamma\gamma}$ is the angle between the two photons moving direction. For the tail-on collision with $\theta_{\gamma\gamma}=0$, the kinetic condition cannot be satisfied and no pair production.
The absorption coefficient of $\gamma-\gamma$ pair production is given by \citep{Gould&Schreder1967,Zhang&Qiao1998,Zhang2001}
\begin{equation}
\begin{aligned}
\alpha_{\gamma\gamma}&=\frac{\alpha_f^2}{\pi\lambdabar}\left(\frac{k_B T_{\rm NS}}{m_ec^2}\right)^3F(E,s)\\
&\simeq(2.1\times10^{-6} \ {\rm cm^{-1}}) \ T_{\rm NS,6}^3F(\epsilon,s),
\end{aligned}
\end{equation}
where $\lambdabar=\hbar/(m_ec)\simeq3.86\times10^{-11} \ \rm cm$ is the Compton wavelength of the electron, the factor $F(\epsilon,s)=g(s)f(\epsilon)$ is defined, $\epsilon$ denotes the photon's energy in units of the electron’s rest mass. $g(s)\simeq(0.270-0.507\mu_c+0.237\mu_c^2)$ takes care of the nonisotropy of the soft photons with respect to the height \citep{Zhang&Qiao1998} and $\mu_c=\cos(\theta_c)=[1-R_{\rm NS}^2/(R_{\rm NS}+s)^2]^{1/2}$ is the maximum cosine of the impact angle between the two photons.
Consider $s\ll R_{\rm NS}$ and we have $g(s)\simeq0.27$.
The function $f(\epsilon)$ reaches the maximum value $\sim 1$ when $\epsilon_{\rm th}\sim m_ec^2/k_bT_{\rm NS}$.
The mean free path of $\gamma-\gamma$ process can be estimated as
\begin{equation}
l_{\gamma\gamma}\simeq \alpha_{\gamma\gamma}^{-1}\simeq (1.8\times10^6 \ {\rm cm}) \ \left(\frac{T_{\rm NS}}{10^6 \ \rm K}\right)^{-3},
\end{equation}
which is slightly larger than $R_{\rm NS}$.
If the seed photons are identified as non-resonant ICS photons, their energies are significantly higher than those of thermal photons. 
Then the factor $f(\epsilon)\sim(\pi^3/3)m_ec^2/k_bT_{\rm NS}\epsilon^{-1}\ln(0.117k_bT_{\rm NS}\epsilon)$ becomes relevant \citep{Gould&Schreder1967}, 
leading to an increase in the mean free path of two-photon pair production $l_{\gamma\gamma}$.
Therefore, the $\gamma-\gamma$ pair production process is less efficient than the $\gamma-B$ pair production process under the present conditions.

\section{WD / NS + RD Binary System}\label{sec:WD / NS-Red dwarf binary system}

To date, five LPRTs (e.g., AR Scorpii, J191213.72–441045.1, DSSJ230641.47+244055.8, ILTJ1101+5521, and GLEAM-X J0704–37) have been confirmed to be associated with WD–RD systems.
In this section, we examine the general predictions for the WD / NS + RD system.

\subsection{Beat Period}

The WD / NS + RD system can naturally account for LPRTs that exhibit either a single period or an additional beat period.
Consider the orbital angular velocity $\vec\Omega_{\rm orb}$ and the spin period velocity $\vec\Omega_{\rm WD / NS}$.
(i) If the binary is a polar system, i.e. $|\vec\Omega_{\rm WD / NS}|\simeq |\vec\Omega_{\rm orb}|$, then the unipolar inductor model is not applicable, and radiation can only be produced near the WD / NS. In this case, no beat period is expected.
(ii) If the binary is asynchronous, we have WD / NS generally spinning much faster than the orbital motion, i.e. $|\vec\Omega_{\rm WD / NS}|\gg|\vec\Omega_{\rm orb}|$.
When the WD / NS is an aligned rotator with $\vec\Omega_{\rm WD / NS}\parallel \vec m$, here $\vec m$ is the magnetic axis of the WD / NS. 
The beat angular velocity can be defined as \citep{Qu&Zhang2025,Yang2026}
\begin{equation}
\begin{aligned}
\vec\Omega_{\rm beat}&=\vec\Omega_{\rm WD / NS}-\vec\Omega_{\rm orb}\\
&\approx \vec\Omega_{\rm WD / NS}, \ \vec\Omega_{\rm WD / NS}\gg \vec\Omega_{\rm orb}.
\end{aligned}
\end{equation}
The magnitude of the beat angular velocity is the difference between $|\vec\Omega_{\rm orb}|$ and $|\vec\Omega_{\rm WD/NS}|$ when the two vectors point in the same direction, and it becomes the sum of the two magnitudes when their directions are opposite.

\subsection{Spin-up and Spin-down}\label{sec:spin-up-down}

Recently, CHIME/ILT J1634+44 has been suggested to exhibit two possible characteristic periods: a longer period that may be associated with orbital motion and a shorter period that could correspond to a beat period.
This source also shows a significantly negative period derivative of $\dot P\sim -9\times10^{-12} \ {\rm s \ s^{-1}}$, associated with the shorter period \citep{DongFQ2025,Bloot2025}.
We note that the spin-up can be achieved in the binary system due to accretion under specific conditions.
In the following, we define several characteristic radii relevant to a binary system, including the separation distance of the binary, Alfv\'en radius and corotating radius. 
For subsequent calculations, we adopt the period $P_\star\sim 14 \ \rm min$ detected in CHIME/ILT J1634+44 as the spin period of the WD / NS.
Here we emphasize that the accretion introduced in the following is intended to describe the torque acting on the compact object during possible accreting episodes, rather than the condition under which coherent radio emission is produced. 
In the detached binary phase, where the RD cannot fill its Roche lobe and sustained Roche overflow is absent. However, the RD may still lose mass through a stellar wind, and the WD / NS can capture a small fraction of this wind.
In an eccentric binary, the mass transfer rate may vary strongly with the orbital phase and accretion could occur near the periastron.
Coherent radio emission is expected to be observable only during relatively detached phases, when the accretion flow has largely ceased and the magnetosphere becomes sufficiently clean. 
The spin evolution discussed in this subsection should be interpreted as the episodic torque history of the compact object, while the coherent radio emission phase corresponds to a different part of the orbit.
The light cylinder radius of a WD / NS is defined as
\begin{equation}
R_{\rm LC}=\frac{cP_{\rm WD / NS}}{2\pi}\simeq(4.0\times10^{12} \ {\rm cm}) \ \left(\frac{P_{\rm WD / NS}}{14 \ \rm min}\right).
\end{equation}
The separation distance of the WD / NS + RD is
\begin{equation}
\begin{aligned}
a&=(GM)^{1/3}\left(\frac{P_{\rm orb}}{2\pi}\right)^{2/3}\\
&\simeq\left\{
\begin{aligned}
&(3.6\times10^{10} \ {\rm cm}) \ \left(\frac{M}{0.8M_\sun}\right)^{1/3}\left(\frac{P_{\rm orb}}{70 \ \rm min}\right)^{2/3}, \ {\rm WD}, \\ 
&(4.6\times10^{10} \ {\rm cm}) \ \left(\frac{M}{1.6M_\sun}\right)^{1/3}\left(\frac{P_{\rm orb}}{70 \ \rm min}\right)^{2/3}, \ {\rm NS}.
\end{aligned}
\right.
\end{aligned}
\end{equation}
One can see that the entire binary system lies within the light cylinder radius of the WD / NS. 
The Alfv\'en radius $R_A$ is defined as the radial distance from the WD / NS at which the magnetic energy density equals the kinetic energy density of the outflow plasma. 
The mass loss rate of the RD is $\dot M\sim 10^{-15}-10^{-12} \ M_\sun \ {\rm yr}^{-1}$ \citep{Wood2002,Wood2005} and we take a typical value of $\dot M=10^{-13} \ M_\sun \ {\rm yr}^{-1}$ below.
The Alfv\'en radius can be calculated as \citep{Bhattacharya1991}
\begin{equation}
\begin{aligned}
R_{A}&=\left(\frac{B_\star^4R_\star^{12}}{GM_\star\dot M_{\rm RD}^2}\right)^{1/7}\\
&\simeq\left\{
\begin{aligned}
&(1.7\times10^{11} \ {\rm cm}) \ 
\left(\frac{B_{\rm WD}}{10^6 \ {\rm G}}\right)^{4/7}\left(\frac{R_{\rm WD}}{0.01R_\sun}\right)^{12/7}\\
&\times\left(\frac{M_{\rm WD}}{0.6M_\sun}\right)^{-1/7}\left(\frac{\dot M_{\rm RD}}{10^{-13} \ M_\sun \ {\rm yr}^{-1}}\right)^{-2/7}, \ {\rm WD}, \\ 
&(2.8\times10^{11} \ {\rm cm}) \ \left(\frac{B_{\rm NS}}{10^{15} \ {\rm G}}\right)^{4/7}\left(\frac{R_{\rm NS}}{10^6 \ \rm cm}\right)^{12/7}\\
&\times\left(\frac{M_{\rm NS}}{1.4M_\sun}\right)^{-1/7}\left(\frac{\dot M_{\rm RD}}{10^{-13} \ M_\sun \ {\rm yr}^{-1}}\right)^{-2/7}, \ {\rm NS}.
\end{aligned}
\right.
\end{aligned}
\end{equation}
The corotating radius $R_{\rm co}$ is defined as the radius at which the local Keplerian angular velocity equals the angular velocity of the WD / NS, which can be calculated as
\begin{equation}
\begin{aligned}
R_{\rm co}&=\left(\frac{GM_\star}{\Omega_{\rm WD / NS}^2}\right)^{1/3}\\
&\simeq\left\{
\begin{aligned}
&(1.1\times10^{10} \ {\rm cm}) \ \left(\frac{M_{\rm WD}}{0.6M_\sun}\right)^{1/3}\left(\frac{P_{\rm WD}}{14 \ \rm min}\right)^{2/3}, \ {\rm WD}, \\ 
&(1.5\times10^{10} \ {\rm cm}) \ \left(\frac{M_{\rm NS}}{1.4M_\sun}\right)^{1/3}\left(\frac{P_{\rm NS}}{14 \ \rm min}\right)^{2/3}, \ {\rm NS}.
\end{aligned}
\right.
\end{aligned}
\end{equation}
One can see that the $R_A\gg R_{\rm co}$ for typical WDs and NSs with strong magnetic fields. 
The inflowing plasma is forced to rotate faster than the local Keplerian speed due to magnetic torques and is eventually ejected by centrifugal forces. 
This process is known as the propeller effect, which extracts angular momentum from the WD / NS and causes it to spin down.
Let $\Omega_{\rm K}(R_A)$ denote the Keplerian angular velocity at the Alfv\'en radius $R_A$.
From angular momentum conservation, we obtain
\begin{equation}
I_{\rm WD / NS}\frac{d\Omega_{\rm WD / NS}}{dt}=-\dot M_{\rm RD} R_A^2[\Omega_{\rm WD / NS}-\Omega_{\rm K}(R_A)],
\end{equation}
which gives a period derivative as
\begin{equation}
\begin{aligned}
\dot P&=\frac{\dot M_{\rm RD}R_A^2P}{I}\left[1-\frac{\Omega_{\rm K}(R_A)P}{2\pi}\right]\\
&\simeq\left\{
\begin{aligned}
&(6.5\times10^{-13} \ {\rm s \ s^{-1}}) \ \left(\frac{B_{\rm WD}}{10^6 \ {\rm G}}\right)^{8/7}\left(\frac{R_{\rm WD}}{0.01R_\sun}\right)^{10/7}\\
&\times\left(\frac{M_{\rm WD}}{0.6M_\sun}\right)^{-9/7}\left(\frac{\dot M_{\rm RD}}{10^{-13} \ M_\sun \ {\rm yr}^{-1}}\right)^{3/7}\left(\frac{P_{\rm WD}}{14 \ \rm min}\right), \ {\rm WD}, \\ 
&(3.7\times10^{-7} \ {\rm s \ s^{-1}}) \ \left(\frac{B_{\rm NS}}{10^{15} \ {\rm G}}\right)^{8/7}\left(\frac{R_{\rm NS}}{10^6 \ \rm cm}\right)^{10/7}\\
&\times\left(\frac{M_{\rm NS}}{1.4M_\sun}\right)^{-9/7}\left(\frac{\dot M_{\rm RD}}{10^{-13} \ M_\sun \ {\rm yr}^{-1}}\right)^{3/7}\left(\frac{P_{\rm NS}}{14 \ \rm min}\right), \ {\rm NS}.
\end{aligned}
\right.
\end{aligned}
\end{equation}

In contrast, for an NS with a typical surface magnetic field strength $B_{\rm NS}=10^{12} \ \rm G$, the system can enter the spin-up regime with $R_A\ll R_{\rm co}$ if accretion occurs.
The plasma is moving along magnetic field lines and accreted onto the magnetic poles of the NS, transferring angular momentum from the disk to the NS and resulting in a spin-up.
The angular momentum conservation gives
\begin{equation}
I_{\rm WD / NS}\frac{d\Omega_{\rm WD / NS}}{dt}=\dot M_{\rm RD} R_A^2\Omega_{\rm K}(R_A),
\end{equation}
which gives the period derivative as
\begin{equation}
\begin{aligned}
\dot P&=-\frac{P^2}{2\pi I}\dot M_{\rm RD}R_A^2\Omega_{\rm K}(R_A)\\
&\simeq(-6.4\times10^{-10} \ {\rm s \ s^{-1}}) \ \left(\frac{B_{\rm NS}}{10^{12} \ {\rm G}}\right)^{2/7}\left(\frac{R_{\rm NS}}{10^6 \ \rm cm}\right)^{-8/7}\\
&\times\left(\frac{M_{\rm NS}}{1.4M_\sun}\right)^{-4/7}\left(\frac{\dot M_{\rm RD}}{10^{-13} \ M_\sun \ {\rm yr}^{-1}}\right)^{6/7}\left(\frac{P_{\rm NS}}{14 \ \rm min}\right)^2
\end{aligned}
\end{equation}
for NS + RD systems.
For the WD + RD scenario, accretion driven spin-up is difficult to realize since the Alfv\'en radius can be larger than the corotating radius for a magnetized WD. 
Therefore, a WD + RD system is unlikely to explain the observed negative period derivative $\dot P\sim -9\times10^{-12}\ {\rm s\ s^{-1}}$ in CHIME/ILT J1634+44 unless the system undergoes a much stronger accretion episode. However, such an accretion state may suppress coherent radio emission.
In contrast, an accreting NS with a moderate surface magnetic field can more easily satisfy $R_A<R_{\rm co}$, allowing material to accrete onto the surface and exert a positive torque. 
This favors a NS central engine if the negative $\dot P$ is interpreted as accretion driven spin-up during the episodic mass transfer phase.

It should be pointed out that the above estimates assume that the transferred material reaches the magnetosphere of the compact object. 
This condition does not need to hold throughout the entire orbit phase. 
If the binary has a non-negligible eccentricity, the interaction between the companion outflow and the WD / NS magnetosphere can be highly phase dependent. 
Near the periastron, the enhanced density of the inflowing material may lead to either accretion or a propeller phase, thus producing spin-up or spin-down torques. 
However, such a dense plasma environment is unfavorable for the escape of coherent radio emission. 
At larger orbital separations, the accretion rate can drop substantially, allowing the magnetosphere to become cleaner. 
The coherent radio emission mechanism operates in a detached phase. 
In this picture, the accreting phase is responsible for the angular momentum evolution, and the observed coherent radio emission can be produced during non-accreting or weakly accreting phases.

\subsection{Unipolar Induction Model}\label{subsec:unipolar}

We consider that the low magnetized RD moves with respect to the WD / NS magnetosphere with $\vec \Omega_{\rm WD / NS}$ and $\vec \Omega_{\rm orb}$ aligned. 
An electric potential drop of the magnitude of
$\Phi\simeq 2R_c|\vec E|$ can be produced due to the unipolar induction effect.  
The induced electric field strength is $|\vec E|=\left|{\vec v}\times\vec B_\star\right|/c,$ where $\vec v = |\vec\Omega_{\rm WD / NS} - \vec\Omega_{\rm orb}| \times  \vec r = \Delta\Omega  r \hat{\phi}$ is the velocity of the companion star (RD) relative to the WD / NS magnetosphere, $\hat{\phi}$ is the unit vector along the azimuthal direction, $B_\star$ is the background magnetic field of the WD / NS.
We define $\zeta$ to quantify the degree of asynchronicity in this binary system as
\begin{equation}
\Delta\Omega=|\Omega_{\rm orb}-\Omega_{\rm WD / NS}|=\zeta\Omega_{\rm orb}.
\end{equation}
We can then obtain the total voltage when the two stars are at a separation of $a$ as
\begin{equation}\label{eq:Phi_binary}
\begin{aligned}
\Phi&\simeq \frac{2B_\star R_\star^3 R_{\rm RD}\zeta}{ca^2}\Omega_{\rm orb}\\
&\simeq\left\{
\begin{aligned}
&(1.6\times10^{8} \ {\rm statvolt}) \ \zeta\left(\frac{B_{\rm WD}}{10^6 \ \rm G}\right)\left(\frac{R_{\rm WD}}{0.01R_\sun}\right)^3\left(\frac{R_{\rm RD}}{0.2R_\sun}\right)\\
&\times\left(\frac{M}{0.8M_{\sun}}\right)^{-2/3}\left(\frac{P_{\rm orb}}{100 \ \rm min}\right)^{-7/3}, \ {\rm WD + RD}, \\ 
&(2.9\times10^{8} \ {\rm statvolt}) \ \zeta\left(\frac{B_{\rm NS}}{10^{15} \ \rm G}\right)\left(\frac{R_{\rm NS}}{10^6 \ \rm cm}\right)^3\left(\frac{R_{\rm RD}}{0.2R_\sun}\right)\\
&\times\left(\frac{M}{1.6M_{\sun}}\right)^{-2/3}\left(\frac{P_{\rm orb}}{100 \ \rm min}\right)^{-7/3}, \ {\rm NS + RD}.
\end{aligned}
\right.
\end{aligned}
\end{equation}
We note that the maximum unipolar potential of the RD
\begin{equation}
\begin{aligned}
\Phi_{\rm RD}&=\frac{2\pi^2 B_{\rm RD}R_{\rm RD}^3}{c^2P_{\rm RD}^2}\simeq (1.6\times10^{6} \ {\rm statvolt}) \ \left(\frac{B_{\rm RD}}{10^3 \ \rm G}\right)\\
&\times\left(\frac{R_{\rm RD}}{0.2R_\sun}\right)^3\left(\frac{P_{\rm RD}}{100 \ \rm min}\right)^{-2}
\end{aligned}
\end{equation}
is negligible compared with Equation~(\ref{eq:Phi_binary}).
Therefore, the maximum Lorentz factor achievable by an electron accelerated by the electric potential in the WD / NS–RD system can be estimated as
\begin{equation}\label{eq:gamma_binary_max}
\begin{aligned}
\gamma_{\rm max}&=\frac{q\Phi}{m_ec^2}\\
&\simeq\left\{
\begin{aligned}
&9.1\times10^{4} \ \zeta\left(\frac{B_{\rm WD}}{10^6 \ \rm G}\right)\left(\frac{R_{\rm WD}}{0.01R_\sun}\right)^3\left(\frac{R_{\rm RD}}{0.2R_\sun}\right)\\
&\times\left(\frac{M}{0.8M_{\sun}}\right)^{-2/3}\left(\frac{P_{\rm orb}}{100 \ \rm min}\right)^{-7/3}, \ {\rm WD + RD}, \\ 
&1.7\times10^{5} \ \zeta\left(\frac{B_{\rm NS}}{10^{15} \ \rm G}\right)\left(\frac{R_{\rm NS}}{10^6 \ \rm cm}\right)^3\left(\frac{R_{\rm RD}}{0.2R_\sun}\right)\\
&\times\left(\frac{M}{1.6M_{\sun}}\right)^{-2/3}\left(\frac{P_{\rm orb}}{100 \ \rm min}\right)^{-7/3}, \ {\rm NS + RD}.
\end{aligned}
\right.
\end{aligned}
\end{equation}

Notice that the star can spin faster than the orbit.   
The current in the circuit can be calculated as
${\Phi}/{2\mathcal{R}_{\rm mag}}$, where
$\mathcal{R}_{\rm mag}={4\pi}/{c}$
is the resistance of the magnetosphere. 
The total electric power dissipation rate of the binary system can be estimated as \citep{Lai2012,Piro2012,Qu&Zhang2025}
\begin{equation}
\begin{aligned}
\Dot{E}&_{\rm diss}=\frac{2\Phi^2}{\mathcal{R}_{\rm mag}}\\
&\simeq\left\{
\begin{aligned}
&(5.8\times10^{25} \ {\rm erg \ s^{-1}}) \ \zeta^2\left(\frac{B_{\rm WD}}{10^6 \ \rm G}\right)^2\left(\frac{R_{\rm WD}}{0.01R_\sun}\right)^6\left(\frac{R_{\rm RD}}{0.2R_\sun}\right)^2\\
&\times\left(\frac{M}{0.8M_{\sun}}\right)^{-4/3}\left(\frac{P_{\rm orb}}{100 \ \rm min}\right)^{-14/3}, \ {\rm WD + RD}, \\ 
&(2.0\times10^{26} \ {\rm erg \ s^{-1}}) \ \zeta^2\left(\frac{B_{\rm NS}}{10^{15} \ \rm G}\right)^2\left(\frac{R_{\rm NS}}{10^6 \ \rm cm}\right)^6\left(\frac{R_{\rm RD}}{0.2R_\sun}\right)^2\\
&\times\left(\frac{M}{1.6M_{\sun}}\right)^{-4/3}\left(\frac{P_{\rm orb}}{100 \ \rm min}\right)^{-14/3}, \ {\rm NS + RD}.
\end{aligned}
\right.
\end{aligned}
\end{equation}
The emission power generated by unipolar induction is sufficient to account for ILTJ1101+5521, a confirmed WD + RD system \citep{Ruiter2025,Rodriguez2025}, even under the assumption of a relatively low surface magnetic field for the WD \citep{Qu&Zhang2025}.
Recently, GPM J1839–10 has been confirmed to exhibit both a beat period ($\sim22 \ \rm min$) and an orbital period ($\sim 8.75 \ \rm h$) \citep{Horvath2026}, resulting in a value of $\zeta\simeq23$ and such a luminosity $\sim 10^{30} \ \rm erg \ s^{-1}$ in this LPRT is reachable.

\subsection{Validity of the Unipolar Induction Model}

\begin{figure*}[]
\begin{center}
\setlength{\tabcolsep}{-18pt}
\begin{tabular}{ll}
\resizebox{107mm}{!}{\includegraphics[]{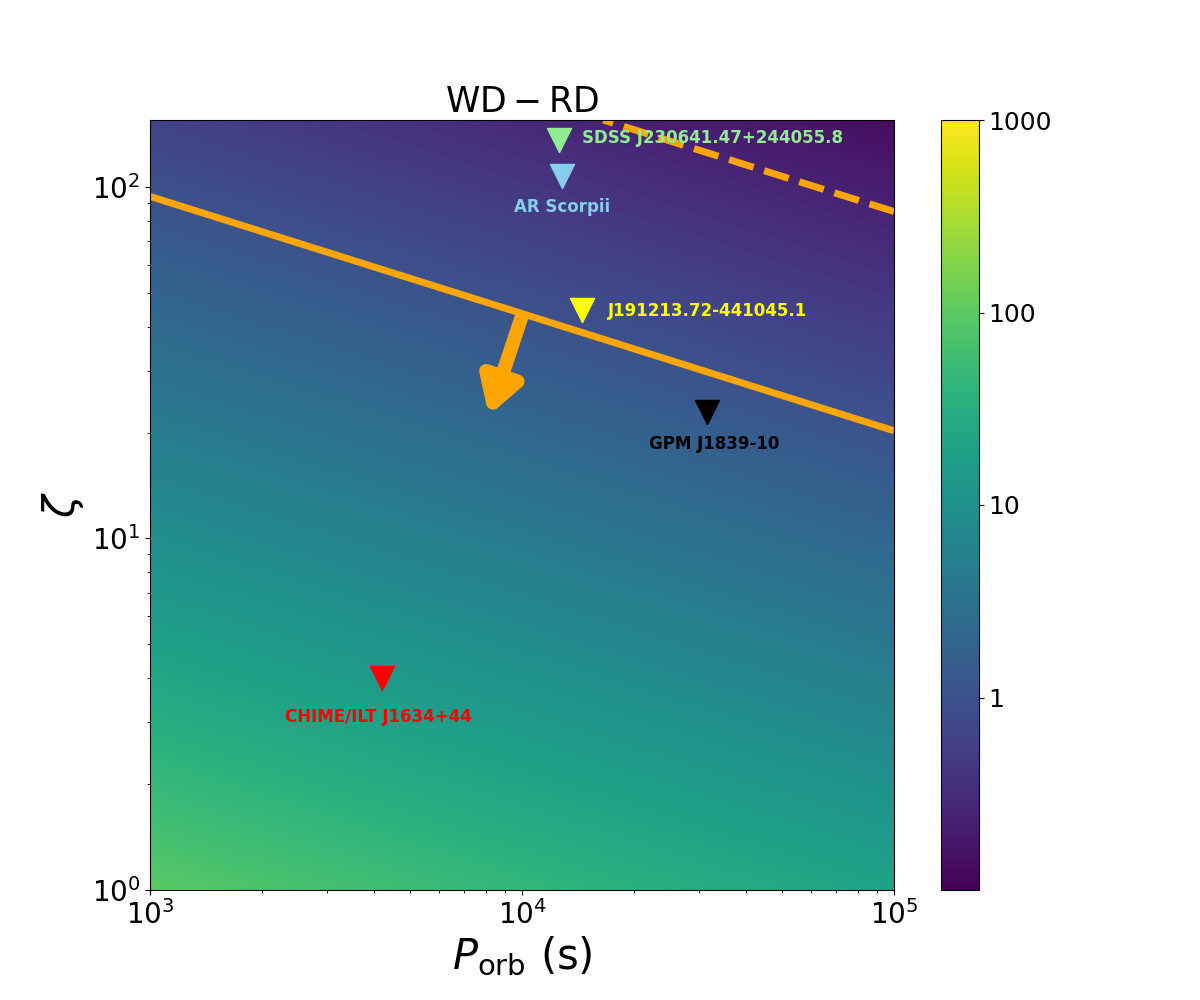}}&
\resizebox{107mm}{!}{\includegraphics[]{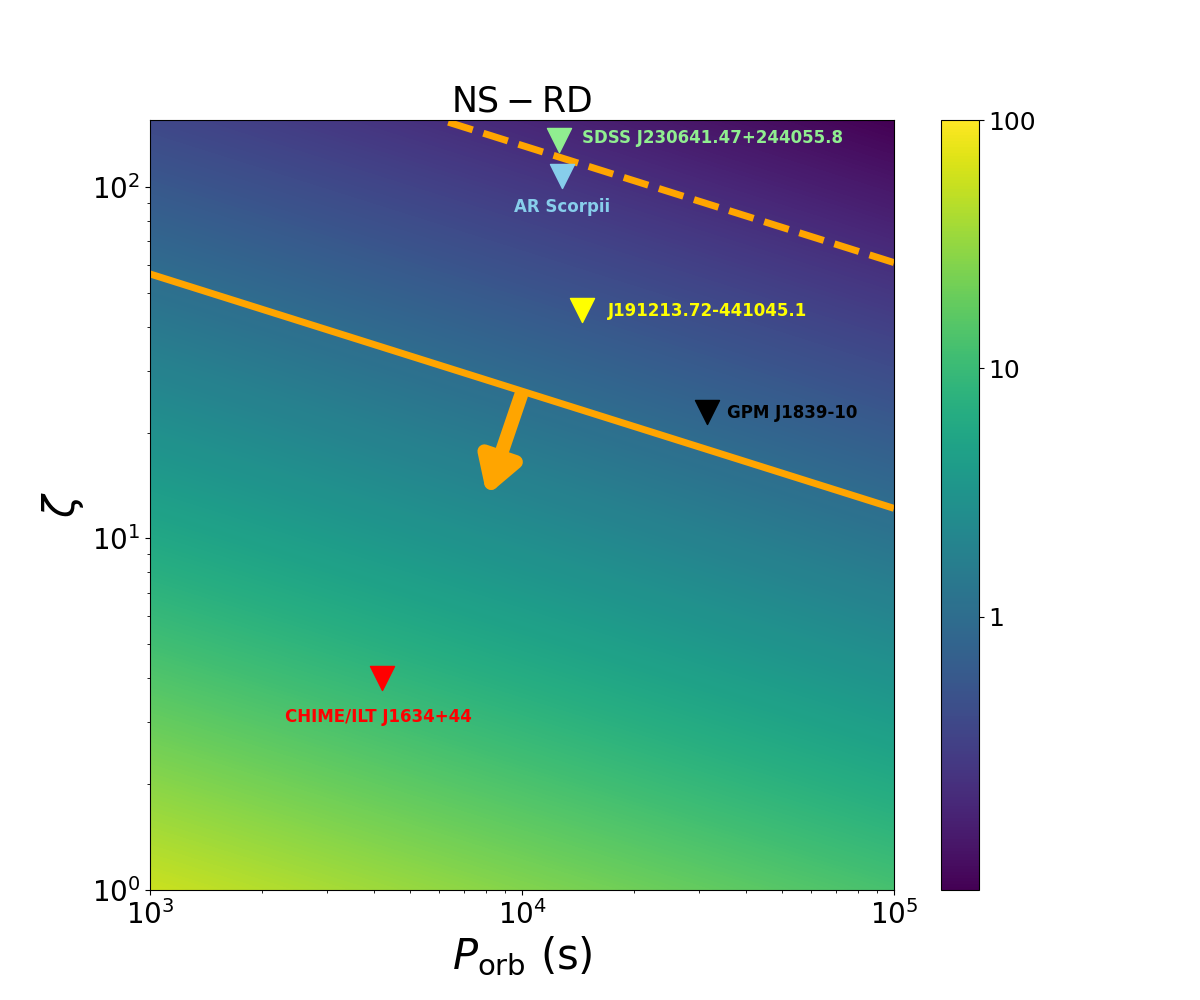}}
\end{tabular}
\caption{The value of $f_{\rm unipolar}$ as a function of orbital period $P_{\rm orb}$ and $\zeta$ for WD + RD (left panel) and NS + RD (right panel).
The orange solid line denotes $f_{\rm unipolar}=1$ for both panels.
The orange arrow denotes the regime ($f_{\rm unipolar}>1$) where the steady state current can be maintained in the unipolar induction model.
Following parameters are adopted: 
WD mass $M_{\rm WD}=0.6M_\sun$,
NS mass $M_{\rm NS}=1.4M_\sun$, and RD mass $M_{\rm RD}=0.1M_\sun$.
}
\label{fig:f_unipolar}
\end{center}
\end{figure*}

For a steady-state current system to be maintained, two conditions are required \citep{Drell1965,Chanmugam&Dulk1983}:
(i) The current is not large enough to induce a comparable magnetic field to background magnetic field.
(ii) The background magnetic field connectivity between the RD and the WD / NS must persist over the Alfv\'en waves round trip crossing timescale $\sim 2 \tau_{A}\sim 2a/v_{A}$, the Alfv\'en speed $v_{A}$ for a relativistic cold MHD fluid can be estimated as
\begin{equation}
v_{A}=\sqrt{\frac{\sigma}{\sigma+1}}c\simeq c, \ \sigma=\frac{B_{\rm bg}^2}{4\pi\gamma_0 n_0 m c^2}\gg 1,
\end{equation}
where $\sigma$ is the magnetization describing the ratio of Poynting to kinetic energy flux.
During the round trip crossing time, the relative motion between the companion and the rotating magnetic field lines causes a displacement of length $2a\Delta\Omega\tau_{A}$.
We define a parameter $f_{\rm unipolar}$ to describe the validity of the unipolar induction model as
\begin{equation}\label{eq:f_unipolar}
f_{\rm unipolar}=\frac{R_{\rm RD}}{2a\Delta\Omega\tau_{A}}=\frac{R_{\rm RD} c}{2a^2\Delta\Omega},
\end{equation}
where $\tau_{A}=a/c$ is applied.
The physical meaning of $f_{\rm unipolar}$ can be understood by comparing the displacement of the magnetic field lines relative to the companion during the Alfv\'en wave crossing time with the size of the companion. 
When $f_{\rm unipolar}>1$, the displacement $2a\Delta\Omega\tau_A$ is smaller than the radius of the companion and a steady current circuit can be established. The large displacement prevents the maintenance of a steady current system for $f_{\rm unipolar}<1$.

From Equation~(\ref{eq:f_unipolar}), one can see that the worst case scenario in which a steady-state current system cannot be maintained in the unipolar induction model corresponds to a larger total mass and hence a larger binary semimajor axis, resulting in a smaller value of $f_{\rm unipolar}$. 
For a rough order-of-magnitude estimate, we fix the typical masses of the WD and NS to be $0.6M_\sun$ and $1.4M_{\rm sun}$, respectively. 
Since the RD mass ranges from approximately $\sim 0.1M_{\sun} - 0.5 M_{\sun}$, we adopt the minimum value of $0.1M_\sun$ to represent this worst case condition, thereby minimizing $f_{\rm unipolar}$.

We present the value of $f_{\rm unipolar}$ as a function of orbital period $P_{\rm orb}$ and $\zeta$ for WD + RD (left panel) and NS + RD (right panel) in Figure~\ref{fig:f_unipolar}.
The solid orange line in both panels denotes $f_{\rm unipolar}=1$.
The parameter space below the solid orange line (orange arrow) denotes $f_{\rm unipolar}>1$, where the unipolar induction model can sustain a closed current circuit.
We note that the WD + RD system spans a slightly larger parameter space than the NS + RD system to satisfy $f_{\rm unipolar}>1$ due to its lower total mass and smaller semimajor axis, which increases $f_{\rm unipolar}$.
In both panels of Figure~\ref{fig:f_unipolar}, we also present the values of $P_{\rm orb}$ and $\zeta$ for the five observed LPRTs that exhibit both orbital and beat periods.
One can see that the three LPRTs confirmed through optical observations to host WD pulsars, with radiation likely originating from the pulsars and exhibiting relatively broad and continuous spectra, lie well above the orange solid line in the WD + RD system.

In Figure~\ref{fig:f_unipolar}, we also present the orange dashed line denoting $f_{\rm unipolar}=1$ for $M_{\rm RD}=0.6M_{\sun}$, which corresponds to the maximum assumed RD mass and represents the most optimistic case for explaining the largest possible number of LPRTs.
If it is the case, one can see that the three LPRTs associated with WD pulsars can be still explained in the context of the unipolar induction model.

\subsection{Magnetospheric Interaction and Reconnection–Driven Model}

As the orbital separation increases or as the surface magnetic field of the companion becomes sufficiently strong, the magnetic field of the WD / NS evaluated at the companion surface may fall below the intrinsic field of the companion. 
In this case, the system transitions from the unipolar induction regime to a magnetospheric interaction regime in which magnetic reconnection becomes the dominant channel for energy dissipation.
This scenario has been discussed in \citet{Yang2026}. 
Here we revisit this scenario in the WD + low-mass companion framework and focus on its energetic viability for LPRTs.
In particular, we estimate the location of the interaction layer from the balance of the two dipolar magnetic fields and evaluate the reconnection power with an effective interaction area constrained by the observed burst duty cycle. 
This allows us to directly compare the available reconnection power with the observed radio luminosities of LPRTs and to assess under which conditions magnetospheric interaction can be an important emission channel.

We define the magnetospheric interaction regime by the condition that the WD / NS magnetic field at the companion surface is weaker than the intrinsic companion field, so that the companion maintains its own closed magnetosphere and the two stellar magnetospheres interact through an interface where magnetic reconnection develops.
The critical binary separation can be defined as
\begin{equation}
\frac{B_{\rm WD / NS} R_{\rm WD / NS}^3}{a_{\rm crit}^3}=B_{c} \ \Rightarrow \ a_{\rm crit}=R_{\rm WD / NS}
\left( \frac{B_{\rm WD / NS}}{B_{c}} \right)^{1/3}.
\end{equation}
For $a \lesssim a_{\rm crit}$, the WD / NS magnetic field dominates at the companion surface and the system approaches the unipolar--induction regime (see Sec~\ref{subsec:unipolar}).
For $a \gtrsim a_{\rm crit}$, the companion preserves its own closed magnetosphere, and the magnetic fields of the two stars come into direct contact, forming an interaction layer where magnetic reconnection naturally occurs.
This condition for the onset of magnetospheric interaction can be equivalently expressed in terms of the orbital period as
\begin{equation}\label{eq:P_crit}
P_{\rm orb} \gtrsim P_{\rm crit}
= 2\pi\left[\frac{a_{\rm crit}^3}{G(M_{\rm WD / NS}+M_c)}\right]^{1/2}.
\end{equation}
The magnetic moments of the WD / NS and the companion are taken as $\mu_{\rm WD / NS} = B_{\rm WD / NS} R_{\rm WD / NS}^3$ and $\mu_{c} = B_{c} R_{c}^3$, respectively.
Along the line connecting the two stars, with the WD / NS located at $r=0$ and the companion at $r=a$, the corresponding dipole magnetic fields are approximated by
\begin{equation}
B_{\rm WD / NS}(r)=\frac{\mu_{\rm WD / NS}}{r^3}, \ B_{c}(a-r)=\frac{\mu_{c}}{(a-r)^3}.
\end{equation}
We define the magnetic moment ratio as $q_\mu={\mu_{c}}/{\mu_{\rm WD / NS}}$.
The background magnetic fields of the WD / NS and companion balance with each other implies the position of the interaction point as $r_{\rm int}={a}/{(1+q_\mu^{1/3})}$.
The characteristic field in the interaction region can be estimated as
\begin{equation}
B_{\rm int}=\frac{\mu_{\rm WD / NS}}{a^3}\left(1+q_\mu^{1/3}\right)^3.
\label{eq:B_int_def}
\end{equation}
The magnetic reconnection power can be estimated as
\begin{equation}\label{eq:P_rec}
\begin{aligned}
P_{\rm rec}&\simeq \epsilon_{\rm rec}\frac{B_{\rm int}^2}{8\pi}A_{\rm int}v_{\rm rel}\\
&\simeq\epsilon_{\rm rec}\frac{\mu_{\rm WD / NS}^2}{a^6}\left(1+q_\mu^{1/3}\right)^6v_{\rm rel}\pi r_{\rm int}^2\tan^2\theta_{\rm rec}\\
&\leq\epsilon_{\rm rec}\frac{\mu_{\rm WD / NS}^2}{a^6}\left(1+q_\mu^{1/3}\right)^6v_{\rm rel}\pi r_{\rm int}^2,
\end{aligned}
\end{equation}
where $\epsilon_{\rm rec}$ is the reconnection efficiency and $v_{\rm rel}=\Delta\Omega r_{\rm int}$ is the relative motion velocity.

\begin{figure*}[]
\begin{center}
\setlength{\tabcolsep}{-18pt}
\begin{tabular}{ll}
\resizebox{107mm}{!}{\includegraphics[]{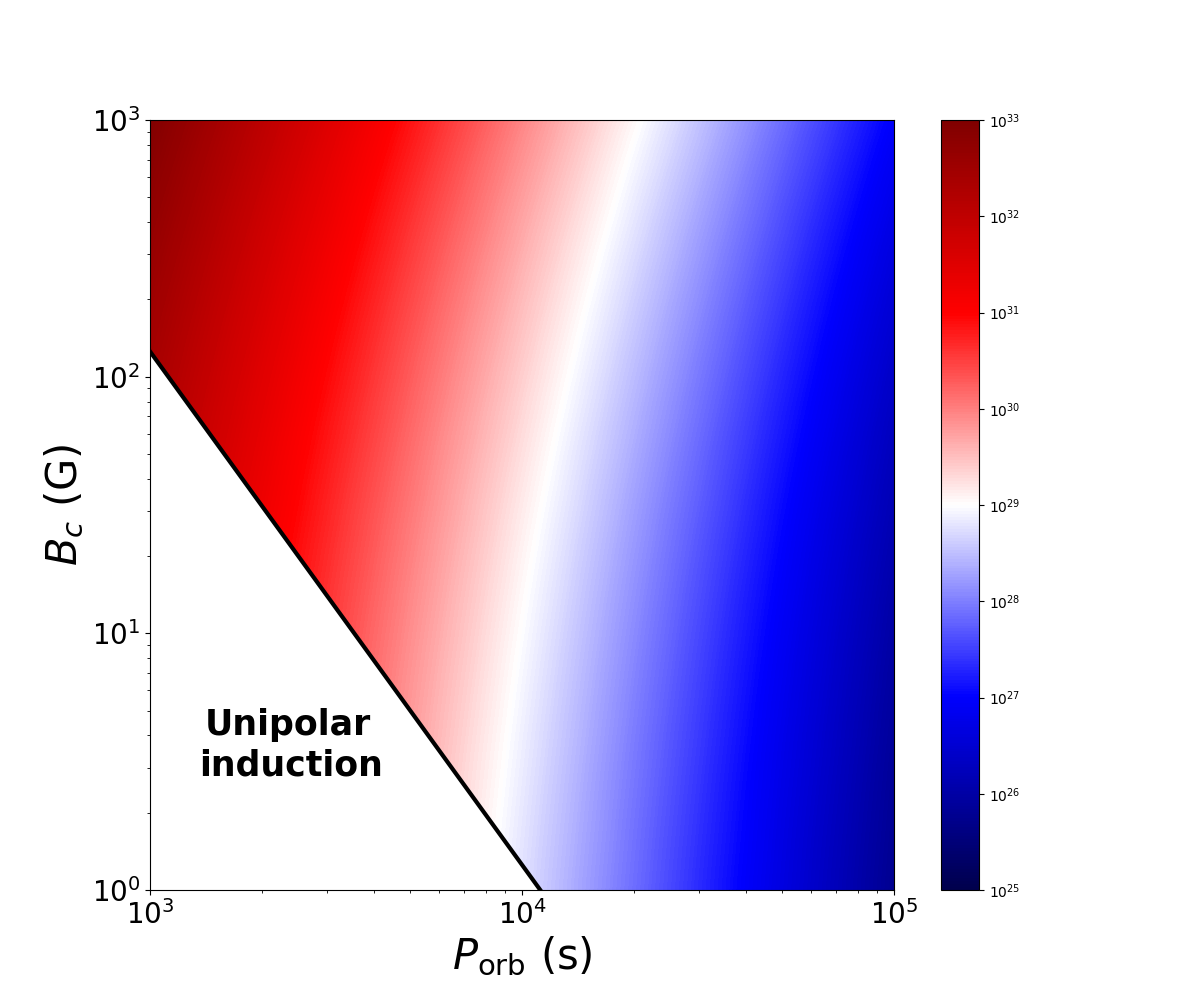}}&
\resizebox{107mm}{!}{\includegraphics[]{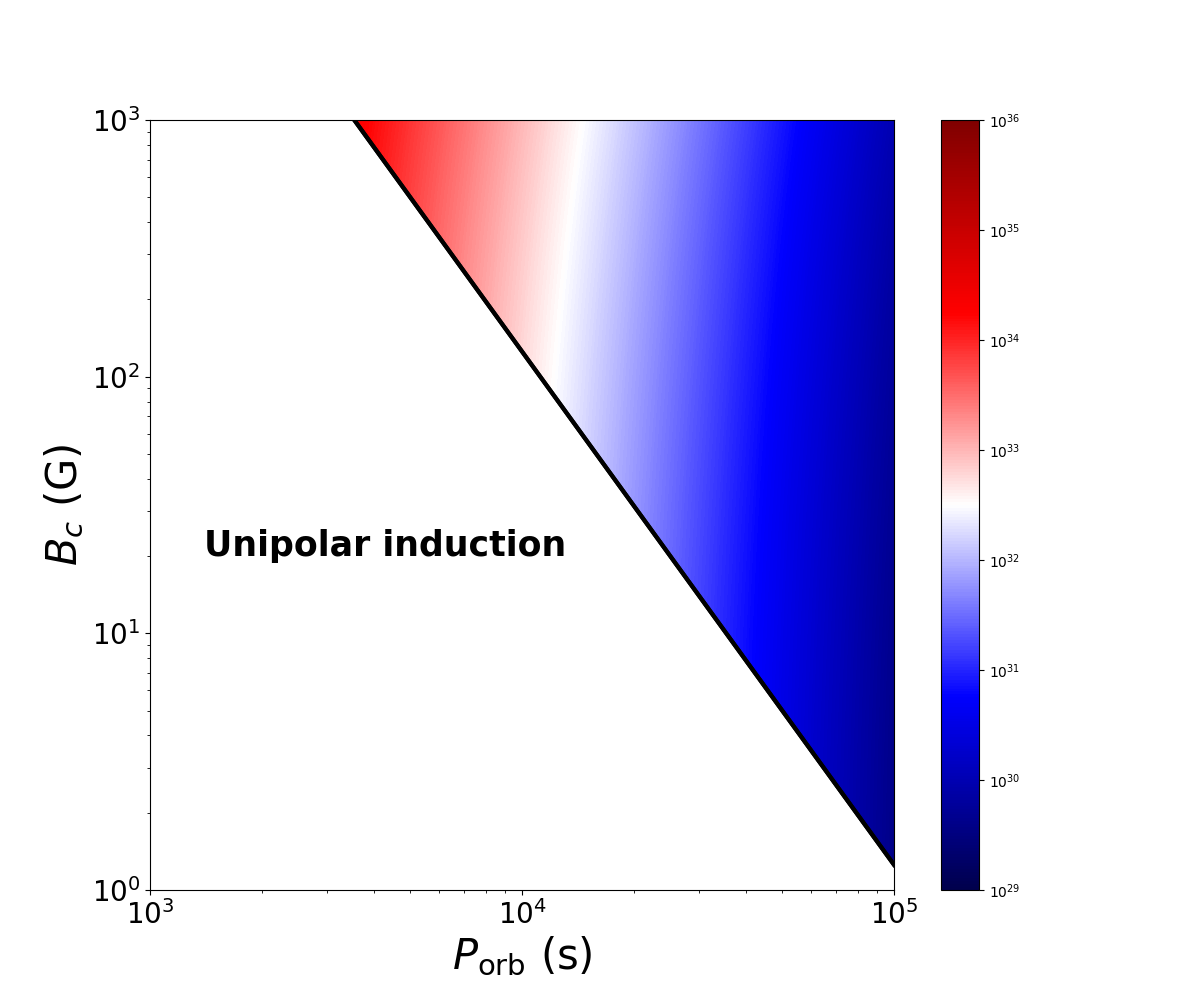}}
\end{tabular}
\caption{Reconnection power as a function of orbital period $P_{\rm orb}$ and surface magnetic field strength $B_c$ of the low-mass companion for $B_{\rm WD}=10^6 \ \rm G$ (left panel) and $B_{\rm WD}=10^8 \ \rm G$ (right panel), respectively. 
Following parameters are adopted: 
WD mass $M_{\rm WD}=0.6M_\sun$ and RD mass $M_{\rm RD}=0.2M_\sun$.
The reconnection efficiency $\epsilon_{\rm rec}=0.1$ is adopted.
}
\label{fig:P_rec}
\end{center}
\end{figure*}

We present the magnetic reconnection power as a function of orbital period $P_{\rm orb}$ and surface magnetic field strength $B_c$ of the low-mass companion in Figure~\ref{fig:P_rec}, for $B_{\rm WD} = 10^6 \ \rm G$ and $10^8 \ \rm G$ in the left and right panels, respectively.
The white regions in both panels denote $P_{\rm orb}<P_{\rm crit}$ (Equation~(\ref{eq:P_crit})) where unipolar induction starts to operate.
One can see that the magnetospheric interaction is likely to operate for low magnetized WD / NS or when the companion has a strong magnetic field strength.

There are several factors that may hinder the detectability of magnetospheric interaction in such systems. 
First, the magnetic field strength of the companion star may not be sufficiently large. 
Although MDs can host surface magnetic fields of order $\sim 10^2 - 10^3 \ \rm G$, most RM observations of LPRTs are typically small (Section~\ref{subsec:RM}). 
The inferred magnetic field strength along LOS ($B_\parallel \sim {\rm RM/DM}$) is consistent with the Galactic background, suggesting that the local magnetized environment around the source is not strongly enhanced. 
This implies that the companion’s magnetosphere may be relatively weak, limiting the efficiency of magnetospheric interaction.
Second, magnetospheric interaction tends to operate in wider binaries (i.e., larger $P_{\rm orb}$), where the magnetic field of WD / NS does not directly penetrate the companion’s surface field and unipolar induction cannot operate. 
In this regime, the two magnetospheres can interact and enable magnetic reconnection. 
However, the larger orbital separation also leads to a significant reduction in the available reconnection power, as it decreases rapidly with distance (Equation~(\ref{eq:P_rec})). 
As a result, the expected luminosity may fall below the typical radio luminosity of LPRTs, making such systems intrinsically more difficult to detect.

\subsection{Orbital Modulation with Emission Originating from the WD / NS}

If the radiation is generated in the immediate vicinity of the WD / NS, i.e., close to its magnetic moment, the companion star does not participate directly in the emission process. 
This situation naturally arises in wide binaries, where the orbital separation is sufficiently large that both the unipolar induction power and the reconnection power are far too weak to account for the observed luminosities of LPRTs.
In such systems, orbital modulation of the observed emission does not imply that the companion contributes energetically to the radiation. 
Instead, the modulation can result purely from geometric effects associated with the rotation of the WD / NS magnetic axis relative to the orbital plane and the LOS.
An especially diagnostic case occurs when both the beat period and the orbital period are detected. 
If radio emission appears only during several consecutive beat cycles within each orbital period, while no detectable emission is present during the remainder of the orbit, this strongly suggests that the magnetic moment of the WD / NS lies close to the orbital plane. 
In this configuration, the beamed emission region intersects the LOS only during a restricted range of orbital phases, producing the observed orbital modulation without requiring any direct contribution from the companion.

\subsection{Relativistic Electron Cyclotron Maser}

The gyromagnetic resonance condition gives the emission frequency of the relativistic electron cyclotron maser as \citep{Wu&Lee1979,Melrose2017}
\begin{equation}
\begin{aligned}
\omega_{\rm ECM}&=\frac{s\omega_B}{\gamma\left(1-\beta_\parallel\cos\theta_v\right)}\\
&=s\frac{\omega_B}{\gamma}+k_\parallel v_\parallel, \ s=0,\pm 1,\pm 2 \dots,
\end{aligned}
\end{equation}
where $s$ is the harmonic number, $\omega_B=eB/m_ec$ is the gyration frequency, $k_\parallel$ and $\beta_\parallel={v_\parallel}/{c}$ denote the wave vector and normalized velocity of the electron along the background magnetic field. 
We consider the case of $s=1$. 
As discussed in Section~\ref{sec:physical constraints}, the background magnetic field cannot be too strong, as this would make it difficult to explain the typical observed frequencies of LPRTs. 
In the following, we normalize the WD and NS surface magnetic fields to $B_{\rm WD}=10^6 \ \rm G$ and $B_{\rm NS}=10^{12} \ \rm G$, respectively.
The emission frequency of relativistic version of ECME can be estimated as
\begin{equation}
\begin{aligned}
\nu_{\rm ECM}&={\cal D} \nu_B
\simeq \gamma\nu_B\\
&\simeq\left\{
\begin{aligned}
&(5.9\times10^{8} \ {\rm Hz}) \ \left(\frac{\gamma}{5}\right)\left(\frac{B_{\rm WD}}{10^6 \ \rm G}\right)\left(\frac{r}{2\times10^{10} \ \rm cm}\right)\\
&\times\left(\frac{R_{\rm WD}}{0.01R_\sun}\right)^3, \ {\rm WD + RD}, \\ 
&(5.2\times10^{8} \ {\rm Hz}) \ \left(\frac{\gamma}{5}\right)\left(\frac{B_{\rm NS}}{10^{12} \ \rm G}\right)\left(\frac{r}{3\times10^{9} \ \rm cm}\right)\\
&\times\left(\frac{R_{\rm NS}}{10^6 \ \rm cm}\right)^3, \ {\rm NS + RD}.
\end{aligned}
\right.
\end{aligned}
\end{equation}
where the Doppler factor ${\cal D}\simeq \gamma$ when $\theta_v \sim 1/\gamma$ and $\nu_B=\omega_B/2\pi$.

We note that cyclotron emission has a characteristic frequency, 
thus the ECME mechanism is also expected to produce narrow spectra. 
In the unipolar induction scenario, electrons are accelerated from the RD to the WD / NS by an electric field that is aligned with the background magnetic field. 
As a result, the electrons move relativistically along the field lines, but remain non-relativistic in the comoving frame of the bulk motion. 
Consequently, the emission spectra of relativistic ECME remains narrow.
Relativistic ECME can lead to spectral broadening if the Lorentz factor has a finite spread around a central value and the local magnetic field varies along the electron trajectory.

Polarization properties: 
In the comoving frame, radiation is fully linearly polarized when the viewing angle is $\theta_v'=\pi/2$, and fully circularly polarized when $\theta_v'=0$.
The viewing angle transforms as $\sin\theta_v'={\cal D}\sin\theta_v$, where $\gamma$ is the Lorentz factor of the electron moving along the background magnetic field.
Therefore, an observer at $\theta_v\sim 1/\gamma$ sees nearly 100\% linear polarization, while at $\theta_v=0$, the emission appears to be nearly 100\% circularly polarized, and the radiation is elliptically polarized at intermediate angles $0<\theta_v<1/\gamma$ \citep{Qu&Zhang2025}.
We conclude that non-relativistic ECME primarily produces highly circularly polarized emission. 
In contrast, relativistic ECME can generate either highly circular or highly linear polarization within the same LPRT, depending on the viewing geometry.

For completeness, we briefly discuss the case with $\zeta \gtrsim 0$. In this near-polar configuration, the parallel voltage approaches zero, resulting in a strong suppression of particle acceleration. As a consequence, particles stream non-relativistically along the background magnetic field and emit via non-relativistic ECME. This mechanism is expected to produce significant circular polarization.

\section{High Energy Counterpart}\label{sec:high_energy}

\subsection{X-ray Emission Powered by Magnetar Magnetic Energy}\label{subsec:B-field decay}

\begin{figure*}[]
\begin{center}
\setlength{\tabcolsep}{-5pt}
\begin{tabular}{ll}
\resizebox{93mm}{!}{\includegraphics[]{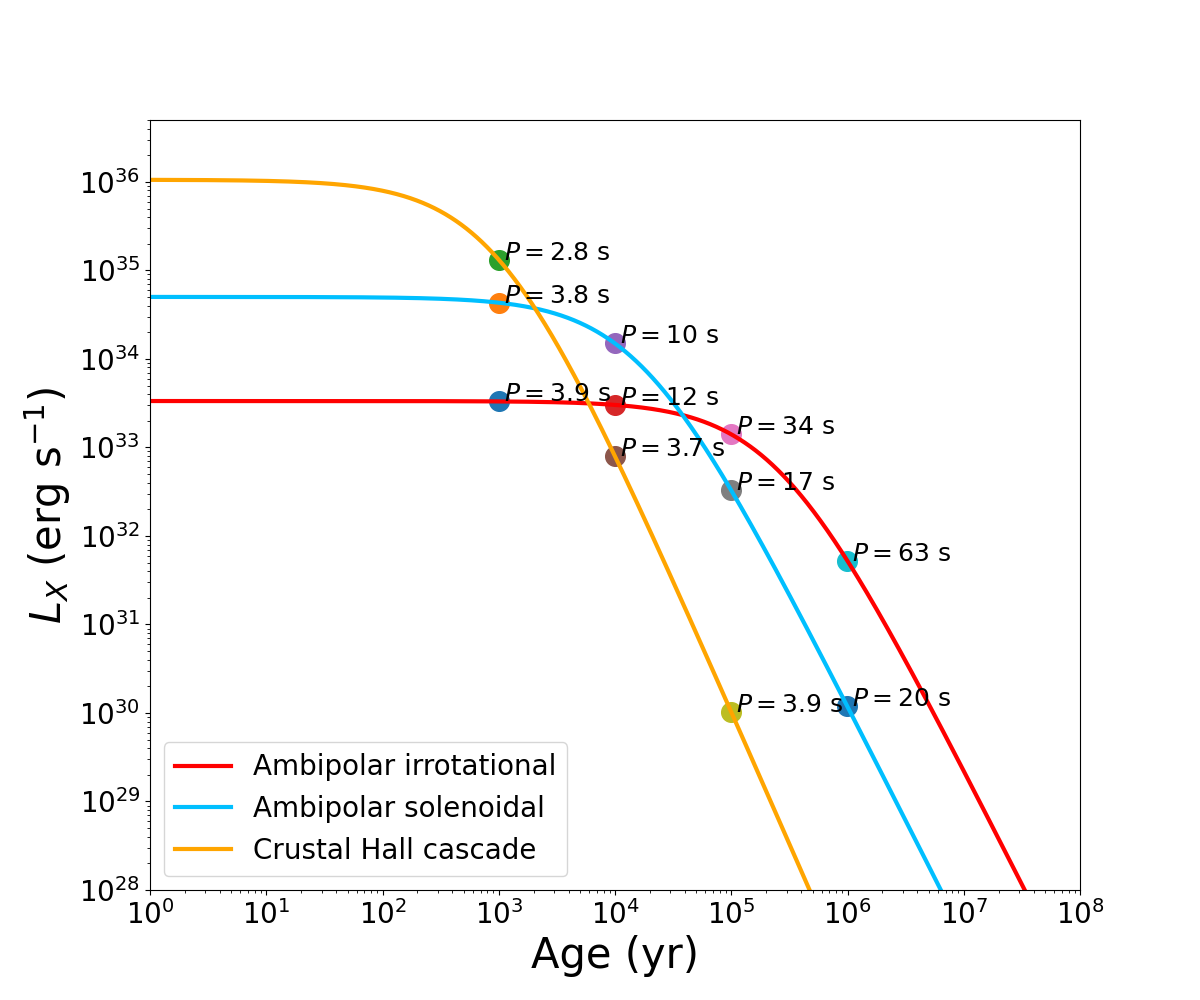}}&
\resizebox{93mm}{!}{\includegraphics[]{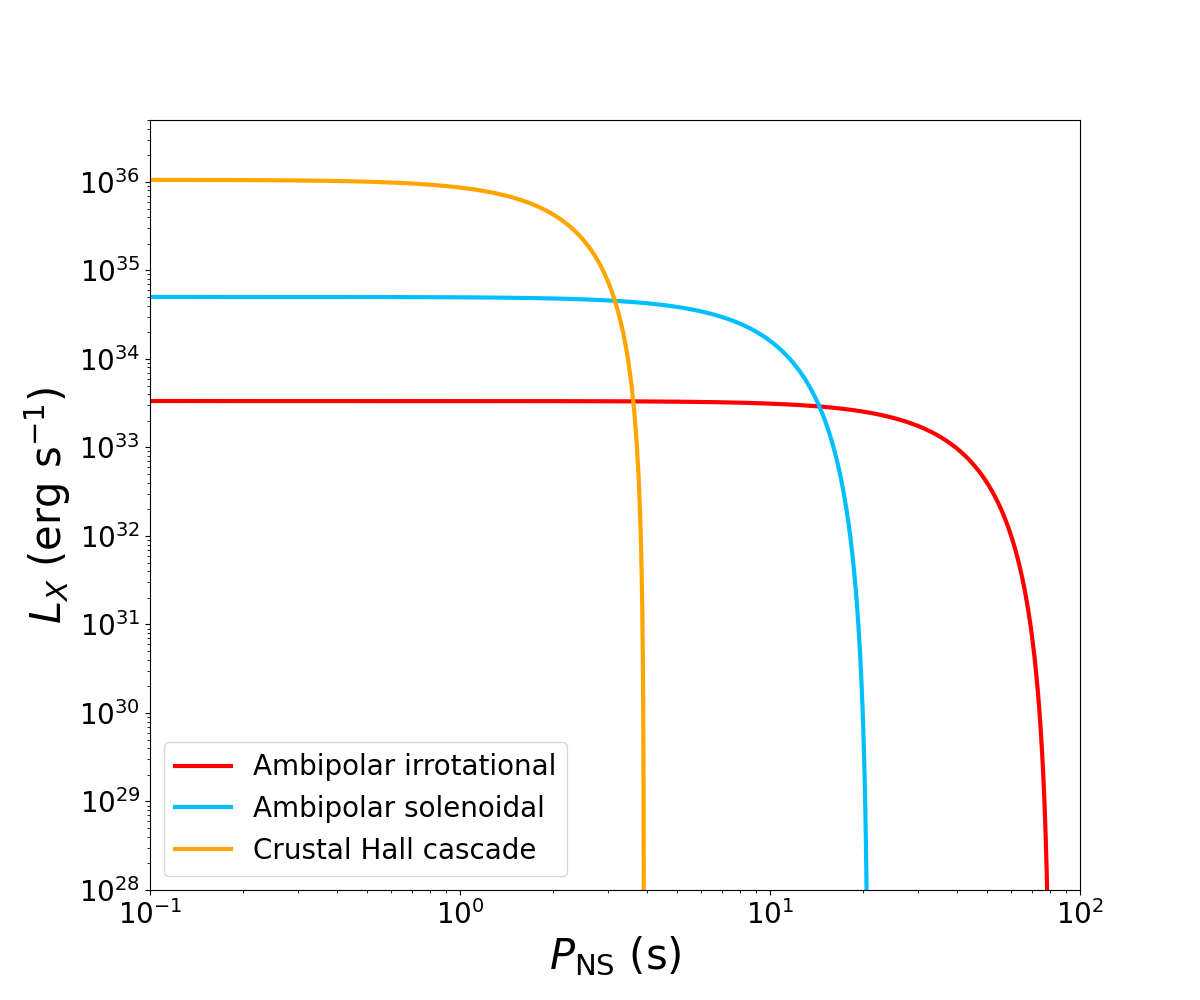}}
\end{tabular}
\caption{
Estimated X-ray luminosity powered by magnetic-field decay in the isolated magnetar scenario.
The left panel shows $L_X$ as a function of the magnetar age, while the right panel shows the corresponding parametric tracks in the $L_X-P$ plane.
The three curves correspond to different magnetic-field decay channels: ambipolar diffusion in the irrotational mode, ambipolar diffusion in the solenoidal mode, and crustal Hall cascade.
The marked points on the left panel indicate representative spin periods along each evolutionary track.
Following parameters are adopted: $B_{\rm NS,0}=10^{15}\ {\rm G}$, $P_0=0.1 \ {\rm s}$, $R_{\rm NS}=10^6 \ {\rm cm}$, $I_{\rm NS}=10^{45} \ {\rm g \ cm^2}$ and $\eta_X=0.1$.
}
\label{fig:L_X}
\end{center}
\end{figure*}

Several LPRTs have been reported to show X-ray counterparts. 
If these sources are isolated magnetars, their X-ray emission might be powered by magnetic energy dissipation rather than by spin-down power, as commonly discussed in the context of magnetars and anomalous X-ray pulsars \citep{Thompson&Duncan1996,Colpi2000}. 
We therefore estimate the X-ray luminosity that can be supplied by magnetic-field decay.
The magnetic energy stored in the dipolar field of a magnetar can be estimated as
\begin{equation}
E_B\simeq \frac{B_{\rm NS}^2R_{\rm NS}^3}{6}\simeq(1.7\times10^{47} \ {\rm erg}) \ B_{\rm NS,15}^2R_{\rm NS,6}^3.
\end{equation}
Following the phenomenological decay law $d\tilde B/d\tilde t=-a\tilde B^{1+\alpha}$ adopted by \citet{Colpi2000}, where $\tilde B=B/(10^{13}{\rm G})$ and $\tilde t=t/(10^6{\rm yr})$, the magnetic field evolves as $\tilde B(\tilde t)=\tilde B_{\rm NS,0}/(1+a\alpha \tilde B_{\rm NS,0}^\alpha \tilde t)^{1/\alpha}$. 
$B_{\rm NS,0}$ denotes the initial magnetic field strength of the magnetar.
The magnetic field decay of isolated neutron stars may proceed through several physical channels. 
Ambipolar diffusion refers to the drift of the magnetic field and charged particles relative to the background neutrons in the stellar core, and its charged-particle flux can be decomposed into solenoidal and irrotational components \citep{Goldreich&Reisenegger1992}. 
On the other hand, Hall drift is associated with the Hall component of the electric field. 
It does not directly dissipate magnetic energy by itself, but it can transfer magnetic energy to smaller spatial scales where Ohmic dissipation becomes more efficient, leading to a Hall-cascade like decay in the crust \citep{Goldreich&Reisenegger1992}. 
Following \citet{Colpi2000}, we consider three decay channels: ambipolar diffusion in the irrotational mode with $a=0.01$ and $\alpha=5/4$; ambipolar diffusion in the solenoidal mode with $a=0.15$ and $\alpha=5/4$; crustal Hall cascade with $a=10$ and $\alpha=1$. 
Assuming magnetic dipole braking, the spin period evolves as \citep{Colpi2000}
\begin{equation}
P^2(t)=P_0^2+
\frac{2K\alpha B_{\rm NS,0}^2t_B}{2-\alpha}
\left[
1-\left(1+\frac{t}{t_B}\right)^{(\alpha-2)/\alpha}
\right],
\end{equation}
where $K={2\pi^2R_{\rm NS}^6}/{3c^3I_{\rm NS}}$, $P_0$ is the initial period of the magnetar and the characteristic magnetic field decay timescale is defined as $t_B=(10^6{\rm yr})\tilde t_B$ and $\tilde t_B= {1}/{a\alpha \tilde B_{\rm NS,0}^\alpha}$.
We assume that a fraction $\eta_X$ of the dissipated magnetic energy is converted into X-ray radiation. 
The corresponding X-ray luminosity can be estimated as
\begin{equation}
L_X=-\eta_X \dot E_B=
\eta_X \frac{R_{\rm NS}^3}{3}a_{\rm cgs}B^{2+\alpha},
\end{equation}
where $a_{\rm cgs}=
{a}/[{(10^6{\rm yr})(10^{13}{\rm G})^\alpha}]$.
The late-time ($t\gg t_B$) behavior is $L_X\propto t^{-2.6}$ for ambipolar diffusion with $\alpha=5/4$, and $L_X\propto t^{-3}$ for the Hall-cascade case with $\alpha=1$.

We present the estimated X-ray luminosity powered by magnetic field decay as a function of magnetar age and spin period in the left panel and right panel of Figure~\ref{fig:L_X}.
The luminosity decreases rapidly as the magnetic field decays. 
The Hall cascade channel gives the largest early luminosity with the adopted parameters because of its faster initial magnetic field decay, it also declines more steeply at late times compared with ambipolar diffusion channels.
In contrast, ambipolar diffusion gives a lower initial luminosity but can sustain detectable X-ray emission for a longer time. 
The right panel of Figure~\ref{fig:L_X} demonstrates that the same magnetic field decay responsible for the X-ray luminosity also drives spin evolution through magnetic dipole braking.
We note that the period in the right panel denotes the magnetar spin period, not necessarily the observed LPRT period, which may correspond to the orbital or beat period in binary systems.
One can see that young magnetars with $t\sim 10^3-10^4 \ \rm yr$ can have $L_X\sim10^{33}-10^{35} \ {\rm erg \ s^{-1}}$ and older systems become rapidly faint.

\subsection{X-ray Emission Powered by Accretion in NS + RD System}\label{subsec:NS+RD}

X-ray emission might also be powered by accretion in a binary system. 
If matter from the companion is captured by the compact object and eventually reaches its surface or inner magnetosphere, the X-ray luminosity due to accretion can be estimated as $L_X^{\rm acc}\simeq \eta_{\rm acc}\dot M_{\rm acc} c^2$, where $\dot M_{\rm acc}$ is the mass accretion rate and $\eta_{\rm acc}$ is the radiative efficiency. 
We note that $\dot M_{\rm acc}$ is not equal to the total mass loss rate of the RD companion for the wind accretion case, which can be estimated as $\dot M_{\rm acc}\simeq\pi R_A^2 \rho_w v_w$, where the wind density is evaluated as $\rho_w\simeq{\dot M_{\rm RD}}/
{4\pi (a-R_A)^2v_w}$.
The relative velocity between the NS and the wind is considered to be comparable to the wind velocity. 
Then we have
\begin{equation}
\frac{x^{11/2}}{(1-x)^2}=\frac{4}{\dot M_{\rm RD}}
\left(
\frac{B_{\rm NS}^4R_{\rm NS}^{12}}
{GM_{\rm NS}a^7}
\right)^{1/2},
\end{equation}
where $x=R_A/a$ is defined.
The efficiency can be estimated as
\begin{equation}
\eta_{\rm acc,NS}\sim \frac{GM_{\rm NS}}{R_{\rm NS}c^2}\sim0.2\left(\frac{M_{\rm NS}}{1.4M_\odot}\right)
R_{\rm NS,6}^{-1}, \ {R_A<R_{\rm co}},
\end{equation}
and
\begin{equation}
\eta_{\rm acc,A}\sim \frac{GM_{\rm NS}}{R_{A}c^2}\sim10^{-5}, \ {R_A>R_{\rm co}}.
\end{equation}
We take $B_{\rm NS}=10^{12} \ \rm G$, $P_{\rm orb}=100 \ \rm min$ and $\dot M_{\rm RD}=10^{-13} \ M_\sun \ {\rm yr}^{-1}$ below, and the corresponding X-ray luminosity can be estimated as
\begin{equation}
\begin{aligned}
L_{X,{\rm NS}}^{\rm acc}&=\eta_X\eta_{\rm acc,NS} \dot M_{\rm acc}c^2\\
&\simeq3.5\times10^{30} \ {\rm erg \ s^{-1}} \ \eta_{X,-1}, \ R_A<R_{\rm co}
\end{aligned}
\end{equation}
when the accreted material can reach the NS surface, and
\begin{equation}
\begin{aligned}
L_{X,{\rm NS}}^{\rm acc}&=\eta_X\eta_{\rm acc,A} \dot M_{\rm acc}c^2\\
&\simeq2.3\times10^{26} \ {\rm erg \ s^{-1}} \ \eta_{X,-1}, \ R_A>R_{\rm co}
\end{aligned}
\end{equation}
when accretion onto the NS surface is inhibited by the propeller effect.
These estimates suggest that wind accretion from a low-mass RD companion can provide a modest X-ray luminosity $L_{X,{\rm NS}}^{\rm acc}\sim10^{27}-10^{31} \ \rm erg \ s^{-1}$, but it is generally difficult to reach the observed luminosities of some X-ray detected LPRTs, unless the binary system enters a stronger mass transfer phase and the accretion rate is significantly higher than the fiducial value.

\

We summarize the main conclusions of this section in Table~\ref{table:X-ray}. 
For the isolated magnetar scenario (Section~\ref{subsec:B-field decay}), magnetic field decay can provide a relatively high X-ray luminosity, $L_X\sim10^{33}-10^{35}\ {\rm erg\ s^{-1}}$, for young systems with ages of $t\sim10^3-10^4\ {\rm yr}$. 
This channel is therefore able to account for the brighter X-ray counterparts associated with LPRTs. 
For NS + RD systems (Section~\ref{subsec:NS+RD}), weak wind accretion can produce a modest X-ray luminosity, typically $L_X^{\rm acc}\sim10^{26}-10^{30}\ {\rm erg\ s^{-1}}$ for the fiducial parameters, unless the system enters a stronger episodic accretion phase. 
The WD + RD channel is expected to be even fainter in X-rays, with a characteristic luminosity of $L_X\sim10^{23}-10^{27}\ {\rm erg\ s^{-1}}$ calculated in \cite{Qu&Zhang2025}. 
Therefore, a relatively bright X-ray counterpart with $L_X\gtrsim10^{33}\ {\rm erg\ s^{-1}}$ would favor a young magnetar-like central engine, whereas faint X-ray counterparts or non-detections are more consistent with isolated WD or WD + RD systems.

\begin{table}[t]
\centering
\caption{
Estimated X-ray luminosities from different physical models.
Model A: X-ray emission powered by isolated magnetar magnetic-field decay (Section~\ref{subsec:B-field decay}).
Model B: isolated WD.
Model C: X-ray emission powered by accretion in an NS + RD system (Section~\ref{subsec:NS+RD}).
Model D: X-ray emission produced in a WD + RD system \citep{Qu&Zhang2025}.
}
\setlength{\tabcolsep}{6pt}
\begin{tabular}{c|c}
\hline
Model & $L_{\rm X} \ ({\rm erg\ s^{-1}})$ \\
\hline
A  & $10^{33}-10^{35}$  \\
\hline
B & Not applicable \\
\hline
C & $10^{26}-10^{30}$ \\
\hline
D & $10^{23}-10^{27}$ \\
\hline
\end{tabular}
\label{table:X-ray}
\end{table}

\section{Propagation Effects}\label{sec:propagation effects}

Radio emissions from LPRT sources undergo various propagation effects. 
The LPRT source GPM J1839+10 exhibits a range of emission properties, many of which are also observed in repeating FRBs, such as Faraday conversion \citep{Men2025}. 
Resonant cyclotron absorption has been reported in GPM J1839+10 \citep{Men2025,Men2026}, although this process has not been observed in FRBs. 
In this section, we discuss three propagation effects in LPRTs and investigate their implications for central engines.

\subsection{Resonant Cyclotron Absorption}

\begin{figure*}
\begin{center}
\begin{tabular}{ll}
\resizebox{90mm}{!}{\includegraphics[]{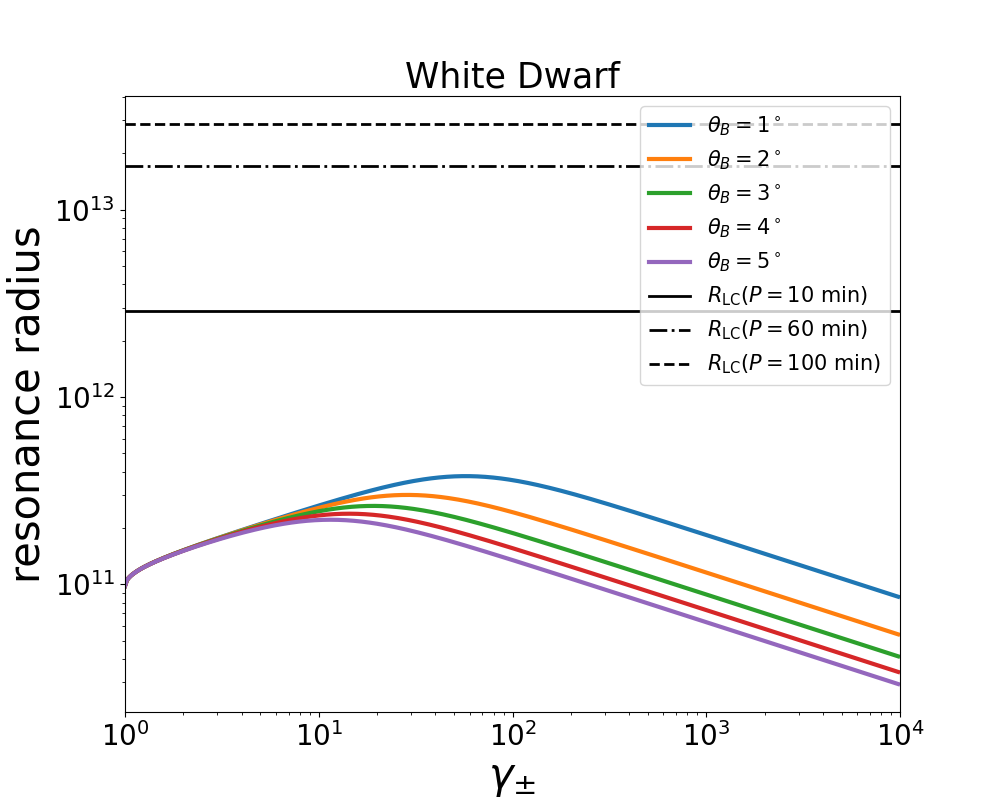}}&
\resizebox{90mm}{!}{\includegraphics[]{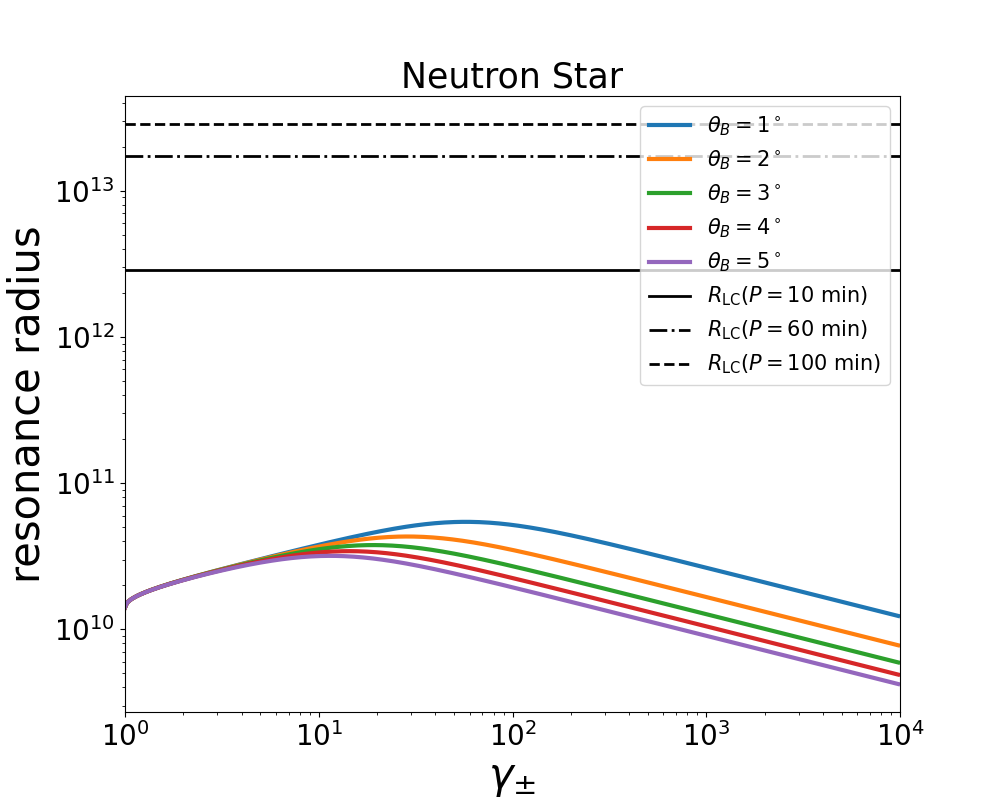}}
\end{tabular}
\caption{The cyclotron resonance radius as a function of pair plasma Lorentz factor for different angles $\theta_B$ between the wave vector and the background magnetic field from $1^\circ$ to $5^\circ$ and for WD with $B_{\rm WD}=10^{9}$ G (left panel) and NS with $B_{\rm NS}=10^{15}$ G (right panel), respectively. 
The horizontal black dashed, dot dashed and solid lines denote the radius of light cylinder for three spin periods of WD and NS. 
Following parameters are adopted: WD radius $R_{\rm WD}=0.01R_\sun$, NS radius $R_{\rm NS}=10^6$ cm, radio waves frequency of LPRTs $\nu_{\rm LPRT}=10^9$ Hz. 
One can see that the resonance condition is well within the light cylinder for the WD and NS.}
\label{fig:resonance radius}
\end{center}
\end{figure*}

To trigger cyclotron resonance, electrons or positrons must be able to transition between two energy levels separated by the characteristic frequency of the radio emission from LPRTs, measured in the electron rest frame\footnote{Unlike in LPRTs, resonant cyclotron absorption is likely not relevant in FRBs \citep{Qu&Zhang2023}, because FRB waves have large amplitudes that rapidly accelerate electrons to a large Lorentz factor, thus disrupting the resonance condition in the comoving frame.}.
Resonant cyclotron absorption of left- and right-hand circularly polarized radio waves can produce net circular polarization if the emission is initially 100\% linearly polarized and the electrons and positrons exhibit asymmetric distributions in Lorentz factor and number density.
The resonant cyclotron condition can be written as 
\begin{equation}\label{eq:resonant cyclotron absorption}
\omega'=\gamma_{\pm}\omega(1-\beta\cos\theta_B)=\omega_B,
\end{equation}
where $\omega'$ is the angular frequency of radio waves in the comoving frame of leptons, $\gamma_+$ and $\gamma_-$ are the Lorentz factors of positrons and electrons, respectively, $\theta_B$ is the angle between the momentum of radio waves and the background magnetic field at the resonance radius.
The resonance condition gives the cyclotron resonance absorption radius, which can be written as
\begin{equation}
\begin{aligned}
R_{\rm res,\pm}=\left[\frac{e B_\star R_\star^3}{m_ec\gamma_{\pm}\omega(1-\beta\cos\theta_B)}\right]^{1/3},
\end{aligned}
\end{equation}
where $B_\star$ and $R_\star$ denote the surface magnetic field strength and radius of WD and NS, respectively.
We present the cyclotron resonance radius as a function of $\gamma_\pm$ in Figure~\ref{fig:resonance radius}.
The horizontal black dashed, dot-dashed, and solid lines indicate the light cylinder radii for spin periods of $P = 100\ \rm min$, $P = 60\ \rm min$, and $P = 10\ \rm min$ for WDs and NSs.
It can be seen that the resonance radii lie within the magnetospheres of both the WD and NS.
In the following, we discuss two possible scenarios in which resonant cyclotron absorption may occur:

(i) In the context of an isolated WD or NS, if radio waves are produced at an emission radius where the electron gyration frequency exceeds the radio wave frequency in the comoving frame of leptons, the condition for resonant cyclotron absorption can be satisfied. 
In such cases, the observed absorption features may be used to infer the emission radius, assuming the central engine is an isolated WD or NS.

(ii) In the context of a WD / NS–RD binary system: 
if the radio emission originates from the WD or NS, the situation is largely the same as in case (i), except that the magnetic field of the RD can also contribute to resonant cyclotron absorption under specific wave propagation configurations, particularly when the radio waves pass near the RD.
If the radio emission originates from the RD in the context of the unipolar induction model, the radio waves are produced at an emission radius where their angular frequency matches the electron gyration frequency in the comoving frame. 
These waves then propagate toward the WD / NS and subsequently away from it, passing through regions where the decreasing background magnetic field can satisfy the condition for resonant cyclotron absorption.

Therefore, we conclude that the detection of resonant cyclotron absorption alone cannot distinguish between different types of central engines.

We note that resonant cyclotron absorption discussed here should not be confused with the ECME. 
The ECME corresponds to negative absorption and requires a non-thermal, anisotropic particle distribution. 
Resonant cyclotron absorption is a propagation effect and corresponds to positive absorption along the ray path. 
These two processes do not need to occur in the same region. 
In the binary scenario, coherent radio waves may be generated by ECME in a localized source region and then propagate into another region with a comparable magnetic field strength, where the resonance condition is satisfied and net cyclotron absorption occurs. 
If radio emission is produced by a mechanism other than ECME, resonant cyclotron absorption can still occur independently as long as the wave encounters a region satisfying the resonance condition.

\subsection{Faraday Conversion}\label{subsec:faraday conversion}

Only one LPRT (GPM J1839–10) has been reported to have Faraday conversion \citep{Men2025}.
Faraday conversion via field reversal, where the magnetic field component along the LOS is nearly equal to zero, i.e. $\vec k\perp\vec B_{\rm bg}$, is considered to be responsible for linear and circular conversion process. 
In such a region, the waves eigenmodes are considered to be X and O-modes where $\omega\gg\omega_p$.
In the magnetized pair plasma in the limit $\omega \ll \omega_B$, i.e. inside the magnetosphere of the WD or NS, and under the assumption that the plasma is moving relativistically along the background magnetic field with the bulk Lorentz factor $\gamma$, we have
\begin{equation}
\Delta k_{\rm XO}'\simeq\frac{\omega_p^2}{\gamma^3\omega^2}{\cal D}^2\sin^2\theta_B\simeq\frac{\omega_p^2}{\gamma^5\omega^2}\sin^2\theta_B,
\end{equation}
where we assume $\theta_B>1/\gamma$ and ${\cal D}\simeq 1/\gamma$ for the last approximation. 
It should be pointed out that the phase difference scales with $\gamma^{-4}$ in the open field line region where the plasma is likely relativistically streaming (see Section~\ref{sec:Isolated magnetic white dwarf} and \ref{sec:Isolated neutron star} for calculations on maximum Lorentz factors for WDs and NSs) and the radio waves are nearly propagating along the background magnetic field, and the frequency of radio waves is also much greater than the plasma frequency.
The corresponding phase difference in the lab frame can be expressed as
\begin{equation}
\begin{aligned}
\Delta\phi_{\rm XO}&\simeq \frac{\omega_p^2}{\gamma^5\omega^2}\sin^2\theta_B \frac{L}{\cal D}\simeq \frac{\omega_p^2}{\gamma^4\omega^2}\sin^2\theta_B L\\
&\simeq\left\{
\begin{aligned}
&3.1\times10^{-9}  \ \kappa\sin^2\theta_B\left(\frac{\gamma}{10^2}\right)^{-4}\left(\frac{L}{100R_{\rm WD}}\right)\\
&\times\left(\frac{P_{\rm WD}}{21 \ \rm min}\right)^{-1}\left(\frac{\nu}{10^9 \ \rm Hz}\right)^{-2}\left(\frac{B_{\rm WD}}{10^9 \ \rm G}\right)\\
&\times\left(\frac{r}{100R_{\rm WD}}\right)^{-3}\left(\frac{R_{\rm WD}}{0.01R_\sun}\right)^3, \ {\rm WD}, \\ 
&4.4\times10^{-6}  \ \kappa\sin^2\theta_B\left(\frac{\gamma}{10^2}\right)^{-4}\left(\frac{L}{100R_{\rm WD}}\right)\\
&\times\left(\frac{P_{\rm NS}}{21 \ \rm min}\right)^{-1}\left(\frac{\nu}{10^9 \ \rm Hz}\right)^{-2}\left(\frac{B_{\rm NS}}{10^{15} \ \rm G}\right)\\
&\times\left(\frac{r}{100R_{\rm NS}}\right)^{-3}\left(\frac{R_{\rm NS}}{10^6 \ \rm cm}\right)^3, \ {\rm NS}.
\end{aligned}
\right.
\end{aligned}
\end{equation}
where $L$ denotes the characteristic length scale of the magnetic field reversal region.
The phase difference is much smaller than unity.
Thus we conclude that the Faraday conversion is not significant within the WD / NS magnetosphere.

In the context of the binary system, consider the non-relativistic electron-ion plasma, the radio waves can satisfy $\omega\gg\omega_B$ and the wave vector difference of the two eigenmodes at $\theta=\pi/2$ is $\Delta k_{\rm XO}=-{\omega_p^2\omega_B^2}/{2c\omega^3}$.
We note that the reversal region lies closer to the RD, and the background magnetic field is dominated by the RD rather than the WD or NS.
The wind density of the RD can be estimated as
\begin{equation}
\begin{aligned}
n_w\simeq\frac{\dot M}{4\pi m_pv_wr^2}&\simeq(3.0\times10^4 \ {\rm cm^{-3}}) \ \left(\frac{\dot M}{10^{-13} \ M_\sun \ {\rm yr}^{-1}}\right)\\
&\times\left(\frac{v_w}{10^9 \ \rm cm \ s^{-1}}\right)^{-1}\left(\frac{r}{10^{11} \ \rm cm}\right)^{-2}.
\end{aligned}
\end{equation}
Thus the phase difference can be estimated as
\begin{equation}
\begin{aligned}
\Delta\phi_{\rm XO}&\simeq \frac{\omega_p^2\omega_B^2L}{2c\omega^3}\\
&\simeq1.4  \ \kappa\left(\frac{L}{10^{11} \ \rm cm}\right)\left(\frac{\nu}{10^9 \ \rm Hz}\right)^{-3}\left(\frac{v_w}{10^9 \ \rm cm \ s^{-1}}\right)^{-1}\\
&\times\left(\frac{r}{10^{11} \ \rm cm}\right)^{-5}\left(\frac{\dot M}{10^{-13} \ M_\sun \ {\rm yr}^{-1}}\right)\left(\frac{B_{\rm RD}}{10^3 \ \rm G}\right)\left(\frac{R_{\rm RD}}{0.2R_\sun}\right)^3.
\end{aligned}
\end{equation}

It should be pointed out that the phase difference is independent of the properties of the WD or NS, since the radio waves undergo Faraday conversion in a field reversal region close to the RD, where the plasma density and magnetic field are primarily determined by the RD environment.
Our estimate suggests that a binary system naturally provides favorable conditions for efficient Faraday conversion.
However, Faraday conversion is not unique to the binary scenario. 
Similar polarization conversion may also arise in isolated magnetars through propagation effects in the near source environment \citep{Lower2024}. 
Therefore, the detection of Faraday conversion alone cannot be regarded as definitive evidence of a binary origin. 
Additional observational diagnostics are required to distinguish between isolated and binary progenitor scenarios (see Figure~\ref{fig:classification}).

\subsection{Scintillation}

Some burst spectra observed in LPRTs appear to be narrow, which may be attributed to scintillation.
To determine whether scintillation can account for all narrow spectra observed in LPRTs, assuming a broad intrinsic bandwidth, one must quantify the probability of scintillation.
This probability depends on the signal-to-noise ratio ($S_N$), the telescope's detection bandwidth, and the scintillation bandwidth.
The interstellar medium in the Milky Way can also cause radio waves of LPRTs phase variations via scintillation. 
The empirical scintillation bandwidth for the Milky Way is given by $\delta\nu_{\rm scint,MW}\simeq(4 \ {\rm MHz}) \ |\sin{b}|^{6/5}\nu_9^{4.4}$ \citep{Cordes&Chatterjee2019}, where $b$ is the latitude.
The Galactic Latitude of CHIME/ILT J1634+44 is $b\sim 43^\circ$, which gives $\delta\nu_{\rm scint,MW}\sim 2.5 \ \rm MHz$.

The probability density function of the radio waves amplitude $A(\nu)$ due to scintillation can be described by Rayleigh distribution \citep{LynePulsar}.
Let us break up the observing band $\nu_1-\nu_2$ into $N=(\nu_2-\nu_1)/\delta\nu_{\rm scint}$ channels. To explain an observed narrow spectrum as a result of scintillation where the observed flux is below the detection threshold flux $f_{\rm min} \equiv f_0 \alpha$ over all frequencies except between $\nu_{1a}$ and $\nu_{2b}$, the observed flux is $f_{\rm obs}=S_N f_{\rm min}=S_N\alpha f_0$.
The maximum probability of the narrow bandwidth spectrum due to scintillation can be written as \citep{KQZ2024}
\begin{equation}
P_{\rm obs}(\alpha_{\rm max}) = \left({n_1\over n_1 + n_2 S_N}\right)^{n_1} \times \left({n_2 S_N\over n_1 + n_2 S_N} \right)^{n_2 S_N},
 \label{scint-prob4}
\end{equation}
where 
\begin{equation}
\alpha_{\rm max}=-\ln\left( {n_2 S_N\over n_1 + n_2 S_N} \right),
\end{equation}
and
\begin{equation}
n_1 \equiv { \nu_2 - \nu_1 + \nu_{1a} - \nu_{2b} \over \delta\nu_{\rm scint} } \ \  \& \ \  n_2 \equiv { \nu_{2b} - \nu_{1a}\over \delta\nu_{\rm scint} }.
\end{equation}
In the following, we discuss one concrete example of narrow LPRT spectra (e.g., LPRT CHIME/ILT J1634+44) and present the probability of making the spectra narrow due to the scintillation effect.
For LPRT CHIME/ILT J1634+44, we take $S_N=8.5$ as a typical value \citep{DongFQ2025}, the radio signal is detected in a frequency bin of size $\delta\nu_{\rm scint}$, and the rest are consistent with noise.
The dynamic spectra for a portion of the first burst of CHIME J1634+44 seen by VLA/realfast seem to have scintillation \citep{DongFQ2025}. 
We assume that a peak at $\sim 1.4$ GHz, the scintillation bandwidth is $\delta\nu_{\rm scint,MW}\sim 15 \ \rm MHz$ and the bandwidth is $\sim 60 \ \rm MHz$.
The probability can be estimated as $P_{\rm obs} \sim 10^{-5}$, which is too small for scintillation to be a possible mechanism for the narrow spectra.

\section{LPRT engines and relations with other phenomena}\label{sec:application}

The current LPRT data seem to suggest that there are at least two types of engines: one related to WD / NS + RD binaries and another related to isolated compact objects (NSs or WDs). No single source system can interpret all the observed LPRT phenomenology. 

In Section~\ref{subsec:classification}, we present an observational classification scheme for central engines of LPRTs.
We then explore their potential connections with CVs and WD pulsars in Sections~\ref{subsec:link_CV} and \ref{subsec:link_WD_Pulsars}, 
respectively.

\subsection{Central Engine Classification of LPRTs}\label{subsec:classification}

\begin{figure*}
    \includegraphics[width=18 cm,height=14.5 cm]{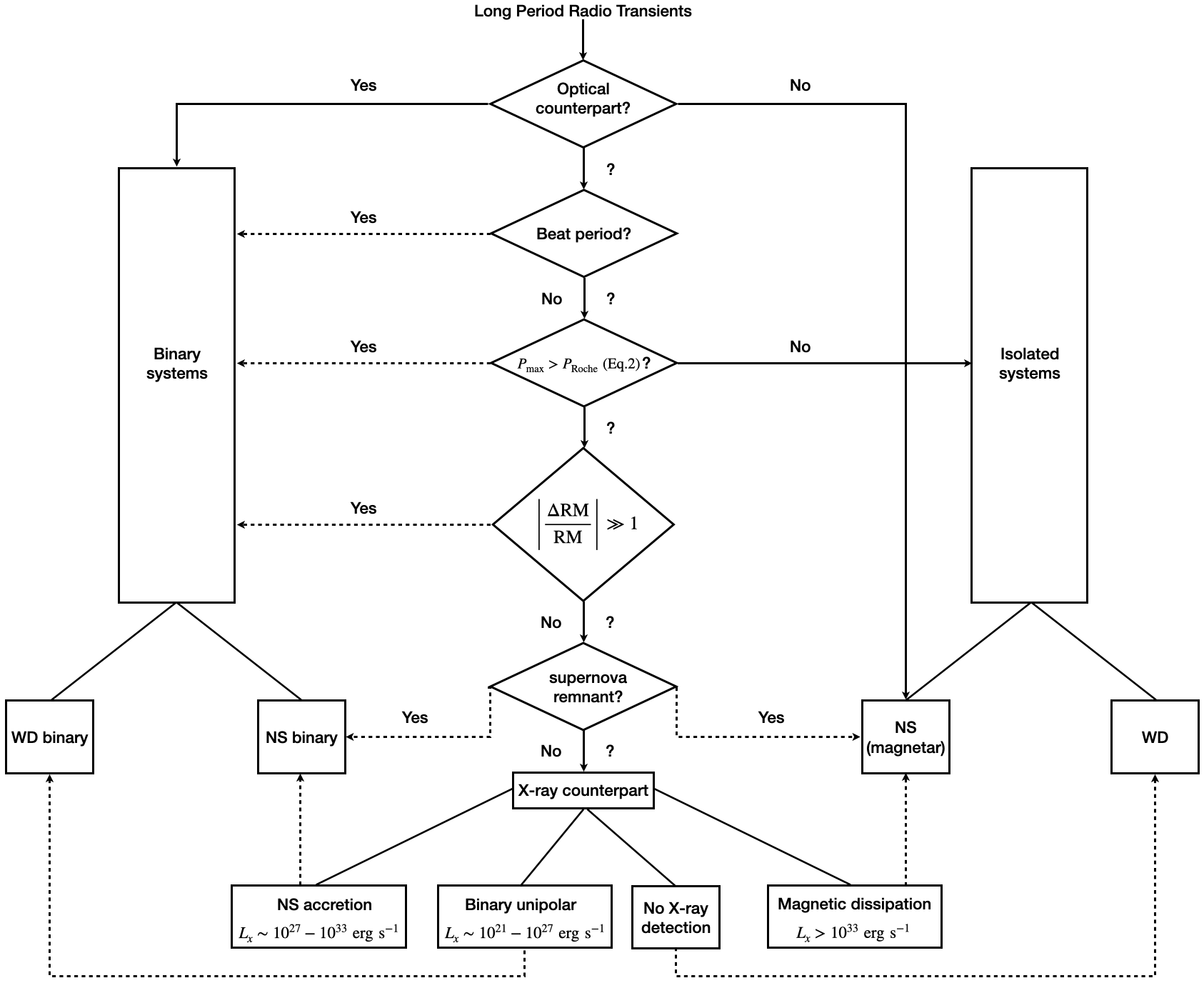}
    \caption{Recommended procedure to associate an LPRT with a possible central-engine category. 
    Multiple observational criteria are applied, including the optical counterpart, beat period detection, period constraint, significant variable RM, supernova remnant association, and X-ray counterpart. 
    Question marks denote cases in which no information is available to judge the validity of the criterion. 
    Dashed arrows indicate criteria that have a strong indication, while solid arrows show the robust conclusions or main decision paths.
    }
    \label{fig:classification}
\end{figure*}

In Figure~\ref{fig:classification}, we present an observational flow chart to diagnose the possible central engines of LPRTs. 
The most direct evidence for a binary origin is the identification of an optical counterpart as the companion star of the LPRT source. 
In this case, the system might be classified as a WD / NS + RD binary and further diagnostics are still needed to distinguish whether the compact object is a WD or an NS. 
If no optical counterpart is detected, or if the current observations are not deep enough to rule one out, other criteria should be considered.

A useful timing criterion is the detection of a beat period, which strongly favors an asynchronous binary system. 
Nevertheless, a beat period may also arise from precession of an isolated compact object, and therefore this criterion should be combined with other observational evidence before drawing a firm conclusion.
Another constraint comes from comparing the observed period with the Roche limit constraint for a WD / NS + RD binary system (Equation~(\ref{eq:P_Roche})). 
If the observed period is shorter than $P_{\rm Roche}$, a binary system cannot exist.
If the observed period is too long to be accommodated by an isolated rotator, a binary interpretation is favored. 
Propagation effects can also provide important evidence. 
A large variable RM with $|\Delta{\rm RM}/{\rm RM}|\gg1$ (Section~\ref{subsec:RM}), or the detection of Faraday conversion (Section~\ref{subsec:faraday conversion}) suggests that the radio waves propagate through a dense and magnetized local environment, which is expected in binary systems.

An association with a supernova remnant would strongly support a young NS or magnetar-related origin. 
The detected X-ray counterpart provides a useful discriminator among different scenarios (Section~\ref{sec:high_energy}). 
A relatively bright X-ray counterpart with $L_X\gtrsim 10^{33} \ {\rm erg \ s^{-1}}$ would favor NS / magnetar-related systems. 
A modest luminosity with $L_X\sim10^{27}-10^{33}\ {\rm erg\ s^{-1}}$ may be produced by accretion onto an NS in a binary system. 
A lower luminosity with $L_X\sim10^{21}-10^{27}\ {\rm erg \ s^{-1}}$ is more consistent with unipolar induction models. 
Non-detection of an X-ray counterpart would not uniquely determine the engine but would be more consistent with WD-related scenarios.

\subsection{Links Between LPRTs and CVs}\label{subsec:link_CV}

LPRTs may represent an evolutionary phase that bridges WD pulsars and magnetic CVs, which are mass-transferring systems and particularly include polars and intermediate polars (IPs). 
CVs typically have orbital periods in the range of $\sim 80 \ {\rm min} - 10 \ {\rm h}$, with an observed period gap between $\sim 2.45 - 3.2 \rm h$ \citep{Spruit&Ritter1983,Schreiber2024}.
In the context of polars with $\zeta = 0$, the WD magnetic field is strong enough to prevent the formation of an accretion disk, and the binary system is typically synchronized.  
IPs are typically asynchronous systems ($\zeta > 0$), in which a magnetic WD accretes material from a Roche lobe filling binary companion. 
Neither polars nor IPs are known to exhibit highly coherent radio emission, 
probably because ongoing accretion quenches such emission, as in radio pulsars \citep{Papitto2013}.

Unlike IPs, some LPRTs with long periods are interpreted as detached asynchronous WD / NS + RD systems ($\zeta>0$) without accretion in the radio emission phase, and thus ideal conditions for coherent radio emission to survive. 
The induced voltage drives the acceleration of charged particles to radiate via the relativistic ECME \citep{Qu&Zhang2025}.
The key distinction between LPRTs and CVs in this picture is the accretion state. 
Accretion can strongly modify the magnetospheric environment and make it difficult for coherent radio waves to be generated. 
By contrast, the LPRT phase corresponds to a detached or weakly interacting pre-CV stage, in which the companion has not yet developed sustained Roche-lobe overflow. 
The absence of strong accretion allows a relatively clean magnetospheric interaction region to survive.

This transitional picture finds observational support: For most CVs, highly circularly polarized radio emission have been detected which is the polarization feature of the non-relativistic ECME \citep{Barrett2020,Melrose&Dulk1982}, due to the parameter $\zeta$ being not large enough to efficiently accelerate particles.
In contrast, unipolar systems may not have entered Roche lobe overflow, placing them before the onset of accretion that are attributed to CVs. 
We also note the detection of two LPRTs (e.g., GLEAM-X J0704–37 and ILTJ1101+5521) within the CVs' period gap, which implies that some asynchronous binaries can contribute coherent radio emission via unipolar induction model.

The X-ray transient source EP240309a/EP J115415.8-501810 was first detected by Einstein Probe (EP) \citep{Ling2024ATel,XiaoY2025}.
Subsequent optical observations confirmed it as a CV of the IP type with a 238.2 s spinning WD in a $\sim 3.76 \ \rm hr$ orbit which is revealed by optical observations.
The spin period of the WD and the orbital period were revealed to be 3.97 min and 3.76 h \citep{Buckley2024ATel,Potter2024}. 
The X-ray emission from this source is thermal with luminosity $\sim2\times10^{32} \ \rm erg \ s^{-1}$.
This raises the question of why no coherent radio emission has been detected from this source. 
A natural explanation is that EP240309a is already in an accreting IP phase and the coherent radio emission is quenched. 
EP-discovered systems may remain fundamentally CVs in the future, their detection via high energy transient surveys provides a new approach for identifying and studying magnetic WDs. 
However, not all CVs are expected to be radio-loud or display characteristic features of LPRTs.

\subsection{Links Between LPRTs and WD Pulsars}\label{subsec:link_WD_Pulsars}

Three so-called LPRTs (AR Scorpii, J191213.72-441045.1 and SDSS J230641.47+244055.8) are different from others and their central engines are believed to be WD pulsars in a binary system \citep{Marsh2016,Pelisoli2023,Segura2025}.
While these sources share certain phenomenological similarities with some LPRTs, including periodic radio emission, X-ray emission counterpart, and the presence of both orbital and beat periods, they also exhibit several notable differences that challenge a unified interpretation:
(i) All three WD pulsar binary systems show a distinct beat period arising from the interaction between the WD's spin and the orbital motion, which is an observational feature not commonly seen in the broader LPRT population. 
(ii) Their radio luminosities are typically lower than those of other LPRTs exhibiting coherent radio emission. 
AR Scorpii and J191213.72-441045.1 have peak radio luminosities of $\sim10^{26} \ \rm erg \ s^{-1}$ at $1$-GHz. 
This value is even seven order of magnitudes below the brightest radio emissions in some LPRTs (e.g., $\sim 10^{33} \ \rm erg \ s^{-1}$ in ASKAP/DART J1832-0911).
(iii) These sources display relatively low levels of linear polarization, in contrast to some LPRTs that show extremely high linear polarization degrees (also some LPRTs show extremely high circular polarization degrees). This discrepancy raises questions about the underlying emission mechanisms and magnetospheric conditions in WD pulsars compared to those in other LPRTs.

An additional factor that may contribute to the apparent difference between AR Scorpii-like systems and LPRTs is the viewing geometry. 
AR Scorpii-like systems are all asynchronous WD + MD binaries, similar in this respect to the detached binary systems invoked for some LPRTs.
The relativistic ECME is expected to be strongly beamed. 
Therefore, the absence of bright coherent radio bursts in AR Scorpii-like systems may indicate that the ECME beam does not sweep the LOS. 
In this case, the observed emission would be dominated by the broader and less strongly beamed incoherent component.
This interpretation is also consistent with the broad band emission observed from AR Scorpii-like systems. 
Such continuous emission is likely associated with synchrotron radiation from electrons accelerated in an extended interaction region, such as a bow shock between the WD and MD. 
In some LPRTs the LOS may sweep the relativistic radiation beam leading to much brighter emission, while the underlying incoherent component may be too faint or overwhelmed by the coherent bursts to be easily identified.

\section{Conclusions and Discussions}\label{sec:conclusions}

In this work, we have investigated the physical origins of LPRTs by combining observational constraints with theoretical models for isolated compact objects and compact binaries. 
Our main conclusions are as follows.

\begin{itemize}
\item LPRTs are naturally divided into two classes by their observed periods. 
Motivated by the Roche-limit and mass transfer constraints (Equation~(\ref{eq:P_Roche}) \& (\ref{eq:P_MT})), we propose that short period LPRTs are more likely powered by isolated NS rather than WD, whereas longer-period sources are naturally explained by detached compact objects (NS or WD) + RD binary systems. 
This provides a physically motivated framework for interpreting the increasing diversity of the LPRT population.
\item Among isolated compact objects, NSs are more promising than WDs as steady coherent radio emitters. 
We find that isolated WDs generally cannot initiate pair production under typical conditions due to their low surface temperatures (Section~\ref{sec:Isolated magnetic white dwarf}). 
In contrast, slowly rotating NSs, especially magnetars, can remain marginally active through inverse-Compton-driven pair cascades (Section~\ref{sec:Isolated neutron star}), although very long period NSs require an additional energy source beyond rotational spin-down.
\item Detached binary systems provide a natural explanation for long period LPRTs. 
When the magnetic field of the compact object dominates that of the companion, asynchronous WD / NS + RD systems can generate electric potentials via unipolar induction, accelerating particles, and producing coherent radio emission through relativistic ECME. 
At larger binary separations or the companion has stronger surface magnetic field strength, where unipolar induction becomes inefficient, coherent radio emission may be powered by magnetospheric interaction via magnetic reconnection. 
These two scenarios naturally account for the observed orbital and beat period modulations of long-period sources.
\item We summarize our results in the diagnostic flow chart shown in
Figure~\ref{fig:classification}, which provides observational criteria to identify the central engines of LPRTs.
First, the detection of an optical counterpart provides the most direct evidence for the presence of a binary companion.
In the absence of an identified optical counterpart, the detection of two periods including a beat period would support a binary system.
In the absence of a beat period, one can compare the observed period with the Roche-limit period $P_{\rm Roche}$ (Equation~(\ref{eq:P_Roche})) and the mass transfer limit (Equation~(\ref{eq:P_MT})). 
A shorter observed period than this would favor an isolated system.
A large fractional variation of RM with $|\Delta{\rm RM}/{\rm RM}|\gg1$ would support a binary interpretation thanks to the contribution from the companion (Section~\ref{subsec:RM}).
Finally, the association with a supernova remnant provides strong evidence for an NS origin.
\item Multi-wavelength observations provide diagnostics of the central engine. 
Bright X-ray counterparts favor magnetar-related systems, either isolated or in binaries, whereas WD-related channels are expected to produce faint or no detectable X-ray emission. 
Propagation effects further constrain the plasma environment and emission geometry, providing complementary diagnostics to distinguish different progenitor channels. 
\end{itemize}

In summary, our study suggests that LPRTs are a diverse population that likely includes isolated NS and WD / NS + low-mass companion binary systems. 
By combining multiple observational criteria, we propose a physically motivated classification framework that can be used to identify the central engines of LPRTs (Figure~\ref{fig:classification}).

\section*{Acknowledgements}
We thank Ziteng Wang, Manisha Caleb, Iris de Ruiter, Pawan Kumar, Yunpeng Men, Myles Sherman, Wenbin Lu and Ruichong Hu for helpful discussion.
We thank Kaya Mori and Matthew Lundy for organizing the LPRT workshop at Columbia and all attendees for insightful discussions.
YQ's work is supported by the Nevada Center for Astrophysics University of Nevada, Las Vegas, and the Research Council of Finland Centre of Excellence in Neutron-Star Physics (project 374063).

\bibliography{example}{}
\bibliographystyle{aasjournal}

\end{document}